\def\rn{\noindent\parshape 2 0truecm 8.8truecm 0.3truecm 8.5truecm}
\def\nn#1 #2{#1, #2.}				% Name with 1 initial
\def\nnn#1 #2 #3{#1, #2. #3.}			% Name with 2 initials
\def\nnnn#1 #2 #3 #4{#1, #2. #3. #4.}		% Name with 3 initials
\def\nnnnn#1 #2 #3 #4 #5{#1, #2. #3. #4. #5.}	% Name with 4 initials
\def\dualand{, \&\hbox{ }}				% Lower case "and" already in use.
\def\multiand{, \&\hbox{ }}				% Lower case "and" already in use.
\def\rg#1;#2;#3;#4;#5;#6 {\par\rn#1 #2, {\it #3}, {\bf #4}, #5 (``#6'') \par}
\def\rf#1;#2;#3;#4;#5 {\par\rn#1 #2, {\it #3}, {\bf #4}, #5\par}
\def\rfbook#1;#2;#3;#4;#5 {{\frenchspacing\par\rn#1 #2, {\it #3} (#4: #5)\par}}
\def\rfproc#1;#2;#3;#4;#5;#6 {{\frenchspacing\par\rn#1 #2, in {\it #3}, ed. #4 (#5: #6)\par}}
\def\rfprep#1;#2;#3  {{\par\rn#1 #2, #3\par}}
\def\rfprepp#1;#2;#3 {{\par\rn#1 #2, #3\par}}

\def\K{{\rm K}}

\def\mK{{\rm \mu K}}
\def\muK{{\rm \mu K}}
\def\MJy{{\rm MJy}}

\def\sr{{\rm sr}}
\def\MJysr{\MJy/\sr}

\def\GHz{{\rm GHz}}
\def\specint{{I_\nu}}
\def\bfzero{{\bf 0}}

\def\Ob{\Omega_b}

\def\Ol{\Omega_\Lambda}
\def\Om{\Omega_m}
\def\On{\Omega_\nu}

\newcommand{\Omhh}{\Omega_m h^2}
\newcommand{\Obhh}{\Omega_b h^2}
\newcommand{\Onhh}{\Omega_\nu h^2}
\newcommand{\yp}{Y_P}
\newcommand{\ns}{n_S}
\newcommand{\nt}{n_T}
\newcommand{\ts}{T/S}

\def\expec#1{\langle#1\rangle}
\def\bexpec#1{\left\langle#1\right\rangle}

\def\etal{{\frenchspacing\it et al.}}
\def\ie{{\frenchspacing\it i.e.}}
\def\eg{{\frenchspacing\it e.g.}}
\def\etc{{\frenchspacing\it etc.}}
\def\rms{rms}
\def\beq#1{\begin{equation}\label{#1}}
\def\eeq{\end{equation}}
\def\beqa#1{\begin{eqnarray}\label{#1}}
\def\eeqa{\end{eqnarray}}
\def\eq#1{equation~(\ref{#1})}
\def\Eq#1{Equation~(\ref{#1})}
\def\eqn#1{~(\ref{#1})}

\newcommand{\beeq}{\begin{equation}} 
\newcommand{\beeqa}{\begin{eqnarray}}

\def\fig#1{Figure~\ref{#1}}
\def\Fig#1{Figure~\ref{#1}}

\def\sec#1{\S\ref{#1}}
\def\Sec#1{\S\ref{#1}}

\def\bs{\hskip-0.2truecm}

\def\spose#1{\hbox to 0pt{#1\hss}}
\def\simlt{\mathrel{\spose{\lower 3pt\hbox{$\mathchar"218$}}
     \raise 2.0pt\hbox{$\mathchar"13C$}}}
\def\simgt{\mathrel{\spose{\lower 3pt\hbox{$\mathchar"218$}}
     \raise 2.0pt\hbox{$\mathchar"13E$}}}
\def\simpropto{\mathrel{\spose{\lower 3pt\hbox{$\mathchar"218$}}
     \raise 2.0pt\hbox{$\propto$}}}

\def\ed{\end{document}}

\def\coh{\xi}
\def\Da{\Delta\alpha}
\def\freq{f}	% In case we change our minds again...
\def\freqq{{f'}}	% In case we change our minds again...
\def\nfreq{F}
\def\kk{{(k)}}
\def\l{\ell}
\def\lfac{\left({2\l+1\over 4\pi}\right)}
\def\Cltot{C_{\l\rm(tot)}}
\def\fsky{f_{\rm sky}}

\def\a{{\bf a}}

\def\e{{\bf e}}

\def\n{{\bf n}}
\def\vp{{\bf p}}
\def\r{{\bf r}}
\def\w{{\bf w}}
\def\x{{\bf x}}
\def\y{{\bf y}}

\def\xT{{\bf x}^T}
\def\xE{{\bf x}^E}
\def\xB{{\bf x}^B}
\def\xt{\tilde{\x}}
\def\at{\tilde{a}}

\def\bzero{{\bf 0}}
\font\bfmath=cmmib10
\def\err{\hbox{\bfmath\char'042}}	% Bold-face $\varepsilon$
\def\rh{\widehat{\r}}
\def\A{{\bf A}}

\def\C{{\bf C}}
\def\E{{\bf E}}
\def\F{{\bf F}}

\def\I{{\bf I}}
\def\J{{\bf J}}
\def\M{{\bf M}}
\def\N{{\bf N}}

\def\Q{{\bf Q}}
\def\R{{\bf R}}

\def\Sig{{\bf\Sigma}}

\newcommand{\calC}{{\mathcal{C}}}
\newcommand{\calA}{{\mathcal{A}}}
\newcommand{\calW}{{\mathcal{W}}}
\newcommand{\calL}{{\mathcal{L}}}
\newcommand{\calN}{{\mathcal{N}}}
\newcommand{\calNa}{{\mathcal{N}}_a}

\newcommand{\bfq}{{\mathbf{q}}}
\newcommand{\bfr}{{\mathbf{r}}}
\newcommand{\bfs}{{\mathbf{s}}}

\newcommand{\bfR}{{\mathbf{R}}}
\newcommand{\bfC}{{\mathbf{C}}}
\newcommand{\kfid}{k_{\rm fid}}

\def\tr{\hbox{tr}\,}
\def\lm{_{\l m}}
\def\alm{a\lm}
\def\Cl{C_\l}
\def\Ct{\tilde{C}_\l}
\def\Nt{\tilde{N}_\l}
\def\bCt{\tilde{\C}}

\def\ith{i^{th}}

\def\kth{k^{th}}

\def\second{2^{nd}}

\def\alphabar{{\bar\alpha}}
\def\Tcmb{T_{\rm cmb}}
\def\conv{c} % Conversion from thermodynamic to antenna temperature
\def\convv{c_*} % Conversion from antenna temperature to brightness
\def\nus{\nu_*}
\def\p{p}
\def\DF{{\rm DF}}
\def\sigmaconf{\sigma_{\rm conf}}
\def\sigman{\sigma_{\rm n}}
\def\sigmaps{\sigma_{\rm ps}}

\newcommand{\tableskip}{\tablevspace{3pt}}
\newlength{\tskip}
\newcommand{\colskip}{@{\hspace{0.1in}}}
\newcommand{\colskippp}{@{\hspace{0.3in}}}
\newcommand{\colskipp}{@{\hspace{0.5in}}}
\newlength{\colwidth}
\newlength{\idlwidth}
\newlength{\smwidth}

\newcommand{\boom}{{\it Boomerang}}
\newcommand{\map}{{\it MAP}}
\newcommand{\planck}{{\it Planck}}

\newcommand{\eV}{{\rm\ eV}}

\newcommand{\impc}{{\rm\,Mpc}^{-1}}
\newcommand{\kmsmpc}{{\rm\ km\ s^{-1}\ Mpc^{-1}}}

\documentstyle[emulateapj,danonecolfloat]{article}
\begin{document}
% \hbadness=2000  %I don't want to hear about underfull hboxes
\twocolumn[%%% Begin front material

%%%%%%%%%%%%%%%%%%%%%%%%%%%%%

%\tighten
%\eqsecnum
%\received{4 August 1988}
%\accepted{23 September 1988}
\journalid{337}{15 January 1989}
\articleid{11}{14}

\submitted{Submitted to ApJ May 20, 1999; accepted August 26}
%\submitted{\today. To be submitted to ApJ.}

\title{Foregrounds and Forecasts for the Cosmic Microwave Background}

\author % [Tegmark et al.]
  {Max Tegmark$^a$\altaffilmark{1},
  Daniel J.~Eisenstein$^a$, 
  Wayne Hu$^a$\altaffilmark{2} 
  and
  Angelica de Oliveira-Costa$^{a,b}$\\
  $^a$Institute for Advanced Study, Princeton, NJ 08540; max,eisenste,whu,angelica@ias.edu\\
  $^b$Princeton University, Department of Physics, Princeton, NJ 08544\\
}
\date{Submitted 1998, August 31}
%\date{Accepted 1993 December 11. Received 1993 March 17}
%\pubyear{1998}

\begin{abstract}
One of the main challenges facing upcoming CMB experiments
will be to distinguish the cosmological signal from 
foreground contamination.
We present a comprehensive treatment of this problem
and study how foregrounds degrade the accuracy
with which the \boom, \map\ and \planck\ experiments can measure 
cosmological parameters.
Our foreground model
includes not only the
normalization, frequency dependence and scale dependence 
for each physical component, 
but also variations in frequency dependence across the sky.
When estimating how accurately cosmological parameters can be measured,
we include the important complication that foreground model parameters
(we use about 500) must be simultaneously measured from the data as well. 
Our results are quite encouraging: despite all these complications, 
precision measurements of most cosmological parameters are degraded by
less than a factor of 2 for our main foreground model and
by less than a factor of 5 in our most pessimistic scenario.
Parameters measured though large-angle polarization signals 
suffer more degradation: up to 5 in the main model and 25 in 
the pessimistic case.
The foregrounds that are potentially most damaging and therefore 
most in need of further study are vibrating dust emission and 
point sources, especially those in the radio frequencies.
It is well-known that $E$ and $B$ polarization contain 
valuable information about reionization and gravity waves, 
respectively. However, 
the cross-correlation between polarized and unpolarized foregrounds also 
deserves further study, as we find that it carries the bulk of the 
polarization information about most other cosmological parameters.
\end{abstract}

%\preprint{IASSNS-AST 97/666}
%\date{\today}
%\date{Submitted November 21, 1996; accepted February 26, 1997}
]%%% End front material

%%%%%%%%%%%%%%%%%%%%%%%%%%%%%%%%%%%%%%%%%%%%%%
%%%%%%%%%%%%%%%%%%%%%%%%%%%%%%%%%%%%%%%%%%%%%%

\altaffiltext{1}{Hubble Fellow}
\altaffiltext{2}{Alfred P. Sloan Fellow}

\section{Introduction}
\label{IntroSec}

Our ability to measure cosmological parameters using the 
cosmic microwave background (CMB) will only be as good as our
understanding of microwave foregrounds,
\eg, synchrotron, free-free and dust emission from our own
Galaxy and extragalactic objects.
For this reason, the recent dramatic progress in the CMB field has stimulated
much work on modeling foregrounds and on algorithms 
for removing them.

Early work on the subject 
(Lubin \& Smoot 1991; 
Bennett {\etal} 1992, 1994;
Brandt {\etal} 1994; 
Dodelson \& Stebbins 1994)
focused on the frequency dependence of foregrounds and how this
could be used to discriminate them from CMB.
Work done for the Phase A study of the \planck\ satellite mission
(Tegmark \& Efstathiou 1996, hereafter TE96; 
Bouchet {\etal} 1996) showed that the
scale dependence of foregrounds was also important, often
being quite different from the roughly scale-free CMB fluctuations, 
and that a multifrequency version of Wiener filtering could 
take advantage of this to improve foreground removal.

% The realization that CMB polarization measurements can substantially
% enhance cosmological information 
The growing interest in CMB polarization,
driven by the combination of theoretical utility
(Kamionkowski {\etal} 1997; Zaldarriaga \& Seljak 1997; Hu \& White 1997)
and experimental feasibility (Staggs {\etal} 1999),
has spurred the modeling of foreground polarization signals
(\eg, Keating {\etal} 1998; Zaldarriaga 1998).
Such models have been further refined for both the \map\ mission
(Refregier {\etal} 1998) and the final \planck\ science case 
(AAO 1998), 
% Bouchet {\etal} (1998) and 
much of which is reviewed in 
Bouchet \& Gispert (1999, hereafter BG99)
and de Zotti {\etal} (1999). 

Yet another complication is that the frequency dependence of 
foregrounds generally varies slightly across the sky. This can be
modeled as each foreground having two or more subcomponents
(TE96; AAO 1998; BG99) or more generally by introducing the 
notion of frequency coherence (Tegmark 1998, hereafter T98).

The purpose of the present paper is to assess the impact of foregrounds
on CMB experiments, including all of the above-mentioned complications. 
This is important for two reasons, 
aside from a general desire to have realistic expectations for future
CMB experiments:
\begin{enumerate}
\item
It helps identify which foregrounds are most 
damaging and therefore most in need of further study. 
\item 
It is useful for optimizing future missions
and for assessing the science impact of design changes 
to, \eg, \planck.
\end{enumerate}
Such a comprehensive analysis is quite timely, since our knowledge
of foregrounds has undergone somewhat of a phase transition
during the last year or two: whereas earlier foreground 
models were quite speculative, generally based on extrapolations 
from lower or higher frequencies, foregrounds
have now been convincingly detected and quantified 
at CMB frequencies by CMB experiments such as 
COBE DMR (Kogut {\etal} 1996, hereafter K96), MAX (Lim {\etal} 1996), 
Saskatoon (de Oliveira-Costa {\etal} 1997), 
OVRO (Leitch {\etal} 1997), 
the 19~GHz survey (de Oliveira-Costa {\etal} 1998) 
and Tenerife (de Oliveira-Costa {\etal} 1999a).

This paper extends prior work in a number of ways.
The treatment of spectral variations is more general than 
in the work for the \planck\ proposal (TE96; Bersanelli {\etal} 1996;
Bouchet {\etal} 1999; BG99) and in Knox (1999, hereafter K99). 
It propagates the effect of 
foregrounds all the way through to the measurement of cosmological parameters,
which has not been previously done except for
a limited parameter set (Prunet {\etal} 1998a). 
Finally, it % 
quantifies the degradation caused by the need to 
measure the statistical properties of the foregrounds directly 
from the CMB data, jointly with the CMB parameters.

The rest of this paper is organized as follows. In \sec{ModelSec1}, we
present models for the various physical foreground components. 
%%% including a discussion of what effects should be 
%%% considered foregrounds in the first place.
In \sec{ModelSec2}, we present our mathematical formalism for 
foreground removal.
In \sec{ResidualSec}, we apply this to the \boom, \map\ and \planck\
missions and compute the level of foreground residuals in the
cleaned map for various scenarios.
In \sec{JointSec}, we compute the extent to which this residual 
contamination degrades the measurement of cosmological parameters,
both when the foreground power spectra are known and when they
must be computed from the CMB data itself.
In both cases, we study how robust our results are 
to variations in the foreground model.
We summarize our conclusions in 
\Sec{ConclusionsSec}.

\section{Foreground models 1: the physics}
\label{ModelSec1}

The foreground model described in this section is summarized in Table \ref{ForegModelTab}.
We make three models: one pessimistic (PESS), one middle-of-the-road (MID) 
and one optimistic (OPT). Since we want to span the entire range of
uncertainties, we have made both the PESS and OPT models rather extreme
in the (lamentably many) cases where 
observational constraints are weak or absent.
The MID model is intended to be fairly realistic, but somewhat
on the conservative (pessimistic) side.
% In \sec{JointSec}, we will parameterize $\Theta_k(\nu)$ and
% $C\kk_\l$ by a smaller set of numbers that can be measured 
% directly from the data.
A FORTRAN code evaluating these models has been made available at
{\it www.sns.ias.edu}/$\sim${\it max}/{\it foregrounds.html},
and we will continually update this as our foreground knowledge improves.

\subsection{Notation}

Our foreground model involves specifying the following
quantities for each physical component $k$ and each of the four types 
of polarization power ($P=T$, $E$, $B$ and $X$):
\begin{enumerate}
\item Its average frequency dependence $\Theta_{\kk}^P(\nu)$.
%,
%in units of thermodynamic temperature, i.e. $\Delta T$ around
%the CMB blackbody.
\item Its frequency coherence $\coh_\kk^P$.
\item Its spatial power spectrum $C_{\l\kk}^P$.
\end{enumerate}

Although this notation will be described in great detail 
in \sec{ModelSec2}, some clarifications are already in order at this point.
$\Theta_{\kk}^P(\nu)$ gives 
the frequency dependence of the {\rms} fluctuations in thermodynamic temperature
referenced to the CMB blackbody.  
For reference, antenna temperature is converted to thermodynamic temperature 
by multiplying by
\beq{cEq}
\conv = \left({2\sinh{x\over 2}\over x}\right)^2,
\eeq
where $x\equiv h\nu/k\Tcmb\approx \nu/56.8\,\GHz$.
Specific intensity or surface brightness is converted to antenna
temperature by
\beq{c2Eq}
\convv \equiv
{1\over x^2}\>{1\over 2k} \left({hc\over k\Tcmb}\right)^2
\approx
{1\over x^2}\>{10\,{\rm mK}\over\MJysr}.
\eeq 

We assume that the frequency dependence is
independent of polarization type and angular scale.
Note that the latter is not the same as assuming that the frequency 
dependence of the sky brightness does not vary with position on the sky.
%in some given part of the sky, 
%since the latter typically varies slightly with position.
The frequency coherence $\coh_{(k)}^P$ quantifies this spectral variation
as described
in \sec{ModelSec2}. For the purpose of this section, 
it is sufficient to know that $\coh\approx {1/\sqrt{2}\Da}$,
where $\Da$ is the dispersion in the foreground 
spectral index across the sky.
If we write the foreground specific intensity in the form
$\specint=f(\nu)\nu^\alpha$ 
for some shape function $f$,
then $\Delta\alpha$ is simply the {\rms} fluctuation in $\alpha$.
Because our foreground models choose $\Theta$ and $\coh$ to be independent
of the polarization type, we will suppress the $P$ superscript in
this section.  We consider the general case in \sec{JointSec}.

We define $C_\l$ in the usual manner, namely as the
variance of the amplitude of fluctuations in the $\ell^{\rm th}$ multipole.
We then model the power spectra of all components 
except the CMB anisotropies and the thermal Sunyaev-Zel'dovich (SZ)
effect as power laws 
\beq{PowerLawPowerEq}
C_{\l\kk}^P =(\p A)^2\l^{-\beta},
\eeq
where $\beta$ and the normalization $pA$ depend 
on the type of foreground ($k$) and polarization ($P$) as shown in Table \ref{ForegModelTab}. 
For convenience, we factor the normalization into two terms:
$A$ gives the normalization of the unpolarized component
and $p$ gives the relative normalization 
of the polarized components. 
We will explore more general power spectrum models in $\sec{JointSec}$.

\subsection{What is a foreground and what is a signal?}

\begin{figure*}[tb] 
\centerline{\epsfxsize=18cm\epsffile{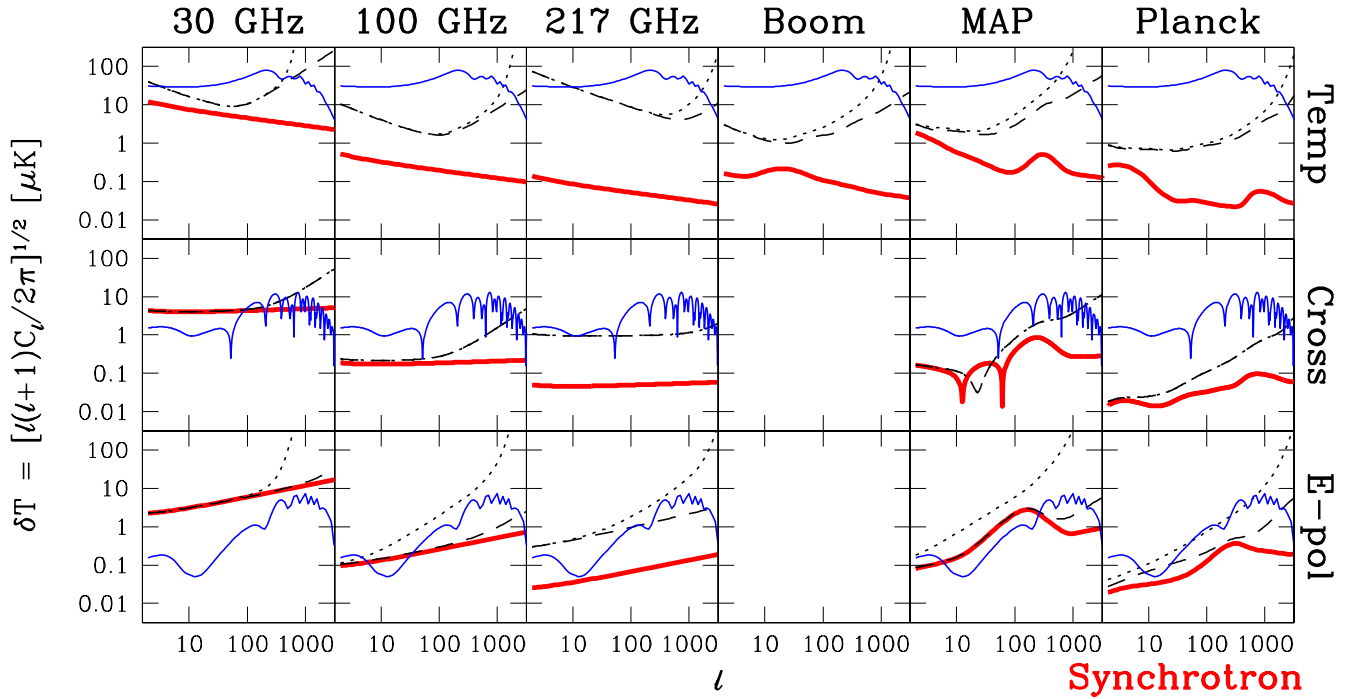}}
\caption{\label{SynchFig}\footnotesize%
The MID model for synchrotron radiation (heavy line).
The first three columns show the uncleaned amplitude as a function of
scale at 30, 100 and 217 GHz.  The rows show the temperature ($T$),
cross-correlation ($X$) and $E$-channel polarization, respectively.
For reference, the CMB power spectrum of our
our fiducial $\Lambda$CDM cosmology (\S~\protect\ref{CosmologyDef})
is also shown (thin solid line) together with the 
total foreground power including (dotted) and excluding
(dashed) Planck detector noise.
The second three columns show the foregrounds amplitude when the maps are
cleaned according to the optimal procedure in \S~\protect\ref{ResidualSec};
this method assumes that the foreground properties are well-known.
The cleaning depends on the experimental specifications; we show
results for \boom, \map\ and \planck. 
There is no polarization data in the 
{\boom} column, since this in an unpolarized experiment.
}
\end{figure*}

Of the multitude of physical mechanisms that create microwave
fluctuations in the sky, where should the line be drawn between 
what constitutes a cosmic signal and what is to be considered
foreground contamination?
All workers in the field agree that effects occurring 
around or before recombination at $z\sim 10^3$ constitute
signal, whereas 
dust, free-free and synchrotron radiation are foregrounds,
regardless of whether the origin is in the Milky Way or in extragalactic
objects. For the remaining effects, the distinction is less clear 
and somewhat arbitrary.
It has been common to label all effects occurring long after recombination
(see Refregier 1999 for a recent review)
as foregrounds, which would then include, \eg, the late integrated
Sachs-Wolfe (ISW) effect (Sachs \& Wolfe 1967;
Boughn \& Crittenden 1999) and gravitational lensing
of the CMB. 
We will take a different and more goal-oriented approach.
When the goal is to measure cosmological parameters, the crucial
issue is not when or how the signal was created, but how reliably it
can be calculated. 
We therefore make the following operational definition of 
what constitutes a foreground:
\begin{itemize}
{\item\it
A foreground is an effect whose dependence on cosmological parameters
we cannot compute accurately from first principles at the present time.
}
\end{itemize}
With this definition, gravitational lensing of the CMB, 
the late ISW effect, and the Ostriker-Vishniac (OV) effect
(Ostriker \& Vishniac 1986; Vishniac 1987) are {\it not} foregrounds, 
even though the latter is $\second$ order and non-Gaussian
(Hu {\etal} 1994; Dodelson \& Jubas 1995) and the two former jointly create
a non-Gaussian bispectrum (Zaldarriaga \& Seljak 1999; Goldberg \& Spergel 1999).
On the other hand, patchy reionization
and the thermal SZ effect
are foregrounds, since their calculation requires 
hydrodynamics simulations of reionization 
(reviewed in Haiman \& Knox 1999) and galaxy formation.

\subsection{Diffuse galactic foregrounds: synchrotron, free-free \& dust emission}

Our knowledge of Galactic foregrounds improved substantially
during 1998. Whereas older models (\eg, TE96) were mainly based on
extrapolations from frequencies far outside the CMB range, 
a number of statistically significant detections of cross-correlation between new 
CMB maps and various foreground templates 
now allow us to normalize many foreground signals directly at the
frequencies of interest.

\subsubsection{Synchrotron radiation}

\begin{figure*}[tb] 
\centerline{\epsfxsize=18cm\epsffile{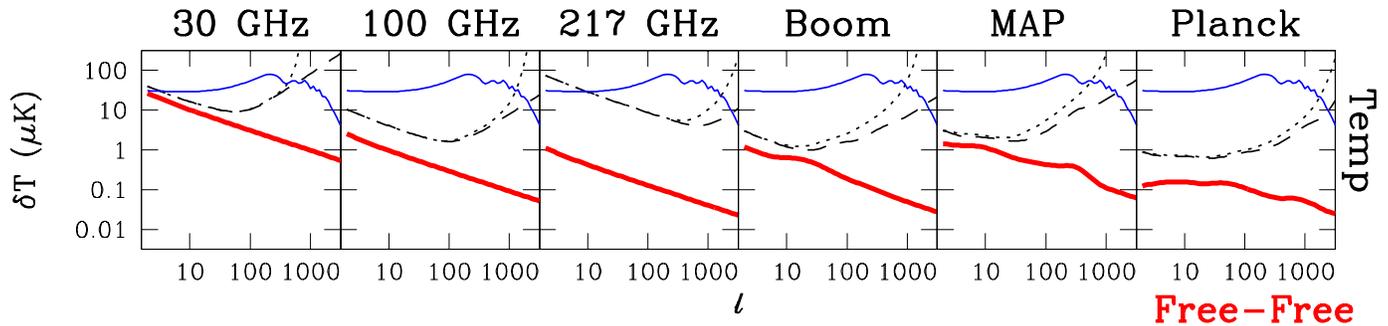}}
\caption{\label{BremsFig}\footnotesize%
Same as Fig. \protect\ref{SynchFig}, 
but for free-free emission. 
}
\end{figure*}

For synchrotron emission in our Galaxy
(see Smoot 1999 for a recent review), we 
model the frequency dependence as $\Theta_{(\rm synch)}(\nu)\propto \conv(\nu)\nu^{-\alpha}$.
Because the spectral index $\alpha$ depends on the energy distribution
of relativistic electrons (Rybicki \& Lightman 1979), 
it may vary somewhat across the sky.
One also expects a spectral steepening towards higher frequencies,
corresponding to a softer electron spectrum 
(Banday \& Wolfendale 1991; Fig 5.3 in Jonas 1999).
Based on the data described in Platania {\etal} (1998),
we take $\alpha=2.8$ for our MID estimate for the unpolarized
intensity, with a spectral uncertainty $\Delta\alpha=0.15$.
As to the power spectrum $\l^{-\beta}$, the 408 MHz Haslam map suggests 
$\beta$ of order 2.5 to 3.0 down to its resolution limit $\sim 1^\circ$
(TE96, Bouchet {\etal} 1996), although the interpretation is complicated
by striping problems (Finkbeiner {\etal} 1999). The Parkes survey 
(Duncan 1997; Duncan 1998, hereafter D98)
enables an extension of this down to $4'$, \ie, $\l\sim 900$, 
and gives $\beta\approx 2.4$ (de Oliveira-Costa {\etal} 1999b)
-- we adopt this value to be conservative, since we will 
normalize on large angular scales.
This agrees qualitatively with theoretical power spectrum estimates 
assuming isotropic turbulence
with a $k^{-11/3}$ Kolmogorov spectrum for the Galactic magnetic field
(Tchepurnov 1997).  

For the polarized synchrotron component, our observational 
knowledge is unfortunately very incomplete. The only available 
measurement of the polarized synchrotron power spectrum is
from the 2.4 GHz D98 maps, which 
exhibit a much bluer power spectrum in polarization than in intensity,
with $\beta\sim 1.0$ instead of $2.5$ (de Oliveira-Costa {\etal} 1999b).
However, at least part of this patchiness is due to modulations in
Faraday rotation 
%along the line of sight 
by small-scale variations 
in the Galactic magnetic field. 
These results therefore cannot be
readily extrapolated to higher frequencies such as 50 GHz, where
Faraday rotation (which scales as $\nu^{-2}$) becomes irrelevant.
A second difficulty lies in extrapolating from the D98 observing region
around the Galactic plane to higher latitudes, where the smaller
mean distance to visible emission sources may well result in 
less small-scale power in the angular distribution.
The polarization maps of Brouw \& Spoelstra (1976) extend to high Galactic
latitudes and up to 1.4 GHz but unfortunately are undersampled, making
it difficult to draw inferences 
about the polarized power spectrum from them.
To bracket the uncertainty, we take $\beta=1.0$ for PESS, 
$\beta=1.4$ for MID and $\beta=3$ 
(the same power spectrum slope as for the unpolarized intensity)
for OPT.

Although Faraday rotation softens the frequency dependence to $\alpha\sim 1.6$
for $\nu\simlt 5\>\GHz$ (de Oliveira-Costa {\etal} 1999b),
we assume that the polarization fraction saturates to a constant value 
for $\nu\gg 10\>\GHz$, as Faraday rotation becomes irrelevant. 
We therefore use the same $\alpha$ and $\Delta\alpha$ 
for polarized and unpolarized synchrotron
radiation.

For the MID scenario, we normalize the unpolarized synchrotron component to 
the cross-correlation with the 
19 GHz map found by de Oliveira-Costa {\etal} (1998).
This gives $\sigma=52\pm 17\mK$ on the 
$3^\circ$ scale\footnote{
For a Gaussian beam with {\rms} width $\theta$,
the {\rms} fluctuations $\sigma$ are given by 
\beq{rmsEq}
\sigma^2 = \sum_{\l=2}^\infty e^{-\theta^2\l(\l+1)} C_\l.
\eeq
The angular ``scale'' mentioned here and elsewhere generally refers
to the full-width-half-maximum (FWHM) beamwidth, given by
FWHM$=\sqrt{8\ln 2}\theta$.
}
for a $20^\circ$ galactic cut, retaining roughly the cleanest 
65\% of the sky.
% and is slightly conservative 
% since a $30^\circ$ galactic cut and should be more representative
% of the cleanest 50\% of the sky. 
This agrees well with the 
synchrotron amplitude obtained in the cross-correlation analyses using
the Tenerife 10 and 15 GHz maps (de Oliveira-Costa {\etal} 1999a; Jones 1999).
For the PESS model, we use the $7.1\mK$
upper limit from COBE DMR found by K96)
at 31.5 GHz on the $7^\circ$ scale.

The degree of synchrotron polarization typically varies between 10\% and
75\% on large scales (Brouw \& Spoelstra 1976), so 
we normalize our models to give
10\% (OPT), 30\% (MID) and 75\% (PESS) rms polarization on 
COBE scales.  Because the polarization power spectra in the 
MID and PESS models are blue-tilted relative to the intensity 
power spectra, the rms polarization exceeds 100\% in these models
on sub-degree scales.  This is physically possible because 
the $\l=0$ contribution to the intensity map has been ignored;
in an extreme case, it is possible to have polarization fluctuations
even with a perfectly smooth intensity map.   

%\beq{SynchroSpectrumEq}
%\Theta(\nu)\propto\nu^{-\alpha}.
%\eeq

\subsubsection{Free-free emission}
\label{BremsSec}

\begin{figure*}[tb] 
\centerline{\epsfxsize=18cm\epsffile{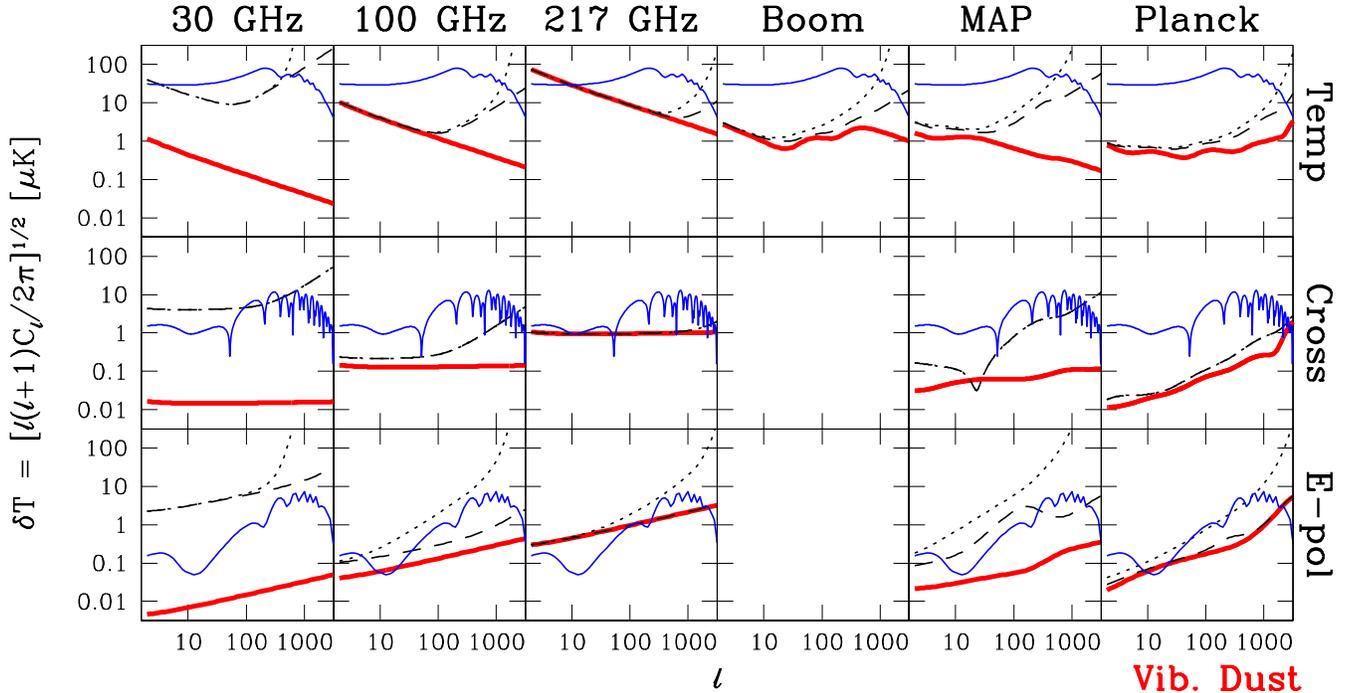}}
\caption{\label{vDustFig}\footnotesize%
Same as Fig. \protect\ref{SynchFig}, 
but for 
thermal (vibrational) dust emission. 
}
\end{figure*}

Of all diffuse Galactic foregrounds, free-free emission is the one whose
frequency dependence is best known.
We model it as a power law
$\Theta_{\rm (ff)} (\nu)\propto\conv(\nu)\nu^{-\alpha}$,
where $\alpha=2.15$ and $\Delta\alpha=0.02$.
In our OPT and MID scenarios, we assume that this emission
is completely unpolarized (Rybicki \& Lightman 1979). 
However, free-free emission can become polarized by Thomson scattering off of
free electrons within the H$_{\rm II}$ region itself 
(Keating {\etal} 1998; Davies \& Wilkinson 1999).
We therefore assume a 10\% polarization level in the PESS model, 
which corresponds to the most extreme case of an optically 
thick cloud and no line-of-sight superpositions 
of interloper H$_{\rm II}$-regions.

Although the spectrum of free-free emission 
is well-known, the amplitude and power spectrum are not.
Since dust dominates at high frequencies, synchrotron at low frequencies and
CMB in the intermediate range, it is difficult to obtain a 
spatial template of free-free emission. 
H$\alpha$ maps should be able to play this
role shortly (see McCullough {\etal} 1999 for a review), but in the interim, we must make 
do with more indirect estimates. K96 obtained a 2-sigma upper limit 
of $14.2\>\mK$ for the rms free-free 
fluctuations at 53 GHz by taking a linear combination
of the three COBE DMR maps that projected out the CMB---we use 
this normalization for our PESS model, and it is consistent with the upper limit
of Coble {\etal} (1999).
K96 also found a highly significant detection of a component
correlated with the DIRBE dust maps whose frequency dependence
was consistent with $\alpha=2.15$. Similar correlations 
have been detected for the Saskatoon data (de Oliveira-Costa {\etal} 1997), 
the 19~GHz map (de Oliveira-Costa {\etal} 1998) and the OVRO Ring 
experiment (Leitch {\etal} 1997)---see Kogut (1999) for a
review of this puzzle.
For our MID model,
we will follow K96 in assuming that this component is in fact free-free
emission, which gives an rms of $7.6\>\mK$ at 53 GHz on DMR scales for a 
$30^\circ$ galaxy cut. 
For the power spectrum shape, 
we assume $\beta=3$ for OPT and MID (as for dust) and
$\beta=2.2$ (as for synchrotron radiation) for PESS.
Again this agrees qualitatively with theoretical estimates 
%%% of the power spectrum
assuming a isotropic turbulence
with a Kolmogorov spectrum for electron density fluctuations 
in the interstellar medium (Tchepurnov 1997).  

\subsubsection{Dust}

\begin{figure*}[tb] 
\centerline{\epsfxsize=18cm\epsffile{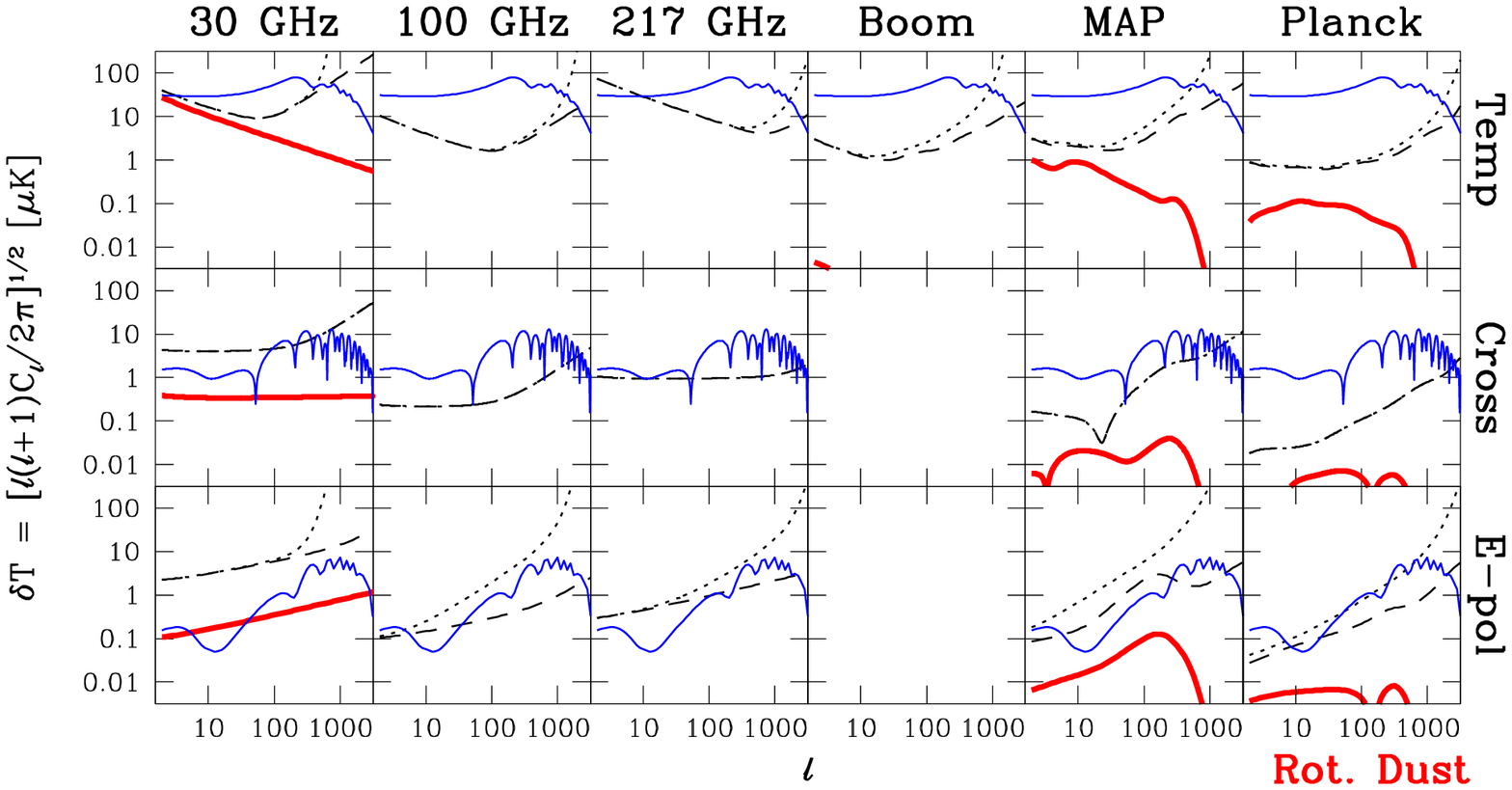}}
\caption{\label{rDustFig}\footnotesize%
Same as Fig. \protect\ref{SynchFig}, 
but for 
thermal spinning dust emission. 
}
\end{figure*}

For vibrational emission from dust grains in the interstellar medium,
we model the frequency dependence as 
\beq{DustSpectrumEq}
\Theta_{\rm(dust)}(\nu) \propto
\conv(\nu)\convv(\nu){\nu^{3+\alpha}\over e^{h\nu/kT_{\rm(dust)}}-1}.
\eeq
We assume a dust temperature $T_{\rm(dust)}=18\K$ (MID) 
and an emissivity $\alpha=1.7$
(K96). The effective emissivity could vary 
across the sky if the relative proportions of different types of
dust grains shift, and modulations in the dust temperature
with, {\eg}, galactic latitude, would further increase the 
dispersion in the frequency dependence.
Estimates of $\alpha$ have ranged between 1.4 and 2.0 across the sky and in
multi-component models (\eg, Reach {\etal} 1995).
Although recent work has weakened the evidence for 
multiple dust temperatures, at least in the cleanest parts of the sky
(see the discussion in BG99),  
joint analysis of the DIRBE and FIRAS data sets has given strong 
indications that two components with different emissivities
are present even at high Galactic latitudes 
(Schlegel {\etal} 1998; Finkbeiner \& Schlegel 1999).
We therefore we take $\Delta\alpha=0.3$ (MID).

As to the power spectrum $\l^{-\beta}$, the combined DIRBE and IRAS dust maps
suggest a slightly shallower slope $\beta=2.5$ (Schlegel {\etal} 1998)
than earlier work finding $\beta\approx 3.0$ 
(Gautier {\etal} 1992; Low \& Cutri 1994; Guarini {\etal} 1995; TE96).
However, a recent analysis of the DIRBE maps has shown no evidence
of a departure from an $\l^{-3}$ power law for 
$\l\simlt 300$ (Wright 1998); we will use this value for the MID model
because only the behavior 
at low $\l$ is important for the present analysis. 

Dust emission may be highly polarized if the grains align
in the local magnetic field (Wright 1987).
For the polarization power spectra, we use the models of 
Prunet {\etal} (1998b) and Prunet \& Lazarian (1999), which give
$\beta=1.3$ for $E$, $\beta=1.4$ for $B$, and $\beta=1.95$ for $X$.
This corresponds to about 1\% polarization in $E$ on the 
$7^\circ$ scale and greater polarization on smaller scales.
 
We normalize the (MID) unpolarized dust power spectrum using the 
DIRBE-DMR cross-correlation analysis of K96, which gives
{\rms} fluctuations of $2.9\muK$ at 53 GHz on the COBE angular 
scale.
This is is a factor 2.3 higher than the Prunet {\etal} model at 
100 GHz, and we boost their polarization normalization by the 
same factor to be conservative. The OPT and PESS normalizations are a factor
of 3 lower and higher, respectively, for $T$ on the $7^\circ$ scale.
The $E$ and $B$ normalization is a factor of three lower for OPT
but a factor 10 higher for PESS, the latter corresponding to 
about 15\% polarization on the 5' scale.

\subsubsection{``Anomalous'' dust emission}

\begin{figure*}[tb] 
\centerline{\epsfxsize=18cm\epsffile{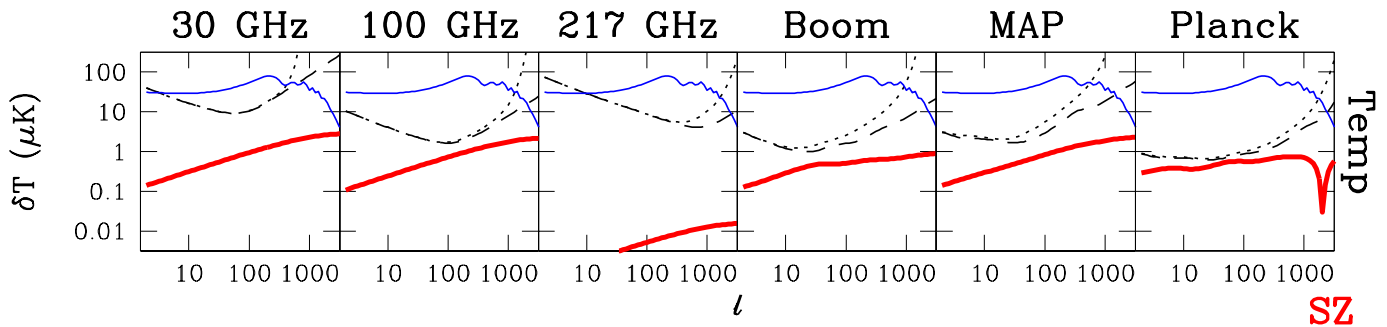}}
\caption{\label{rSZfig}\footnotesize%
Same as Fig. \protect\ref{SynchFig}, 
but for 
the thermal SZ effect from filaments. 
}
\end{figure*}

An alternative interpretation of the dust-correlated foreground
component described in \sec{BremsSec} has been proposed by 
Draine \& Lazarian (1998, hereafter DL98). They identify it as 
dust emission after all but radiating via rotational rather than
vibrational excitations. The latest Tenerife measurements
strongly support this idea
(de Oliveira-Costa {\etal} 1999) since the observed turnover 
in the spectrum with a decrease from 15 to 10 GHz is incompatible
with free-free emission alone.
This emission will be dominated by
the very smallest dust grains (more appropriately called
clusters, since they may consist of only $\sim 10^2$ atoms).
Many DL98 models are well fit by spectra of the form
of \eq{DustSpectrumEq}, but with rather unusual parameters.
For our MID model, we take the rather typical 
DL98 model that is fit by $T_{\rm(dust)}=0.25\K$, $\alpha=2.4$. However, the
range of theoretically and observationally allowed spectra 
is very large, and magnetic-dipole dust emission could have
yet another spectral signature (Draine \& Lazarian 1999).
We adopt a very large spectral uncertainty $\Delta\alpha=0.5$ to reflect this.
For our pessimistic model, we adopt an extremely blue ($\beta=1.2$) power
spectrum for this component, since the work of 
Leitch {\etal} (1997) indicates that 
this component may be very inhomogeneous on small scales.

We normalize our MID model
so that spinning dust accounts for the entire dust-correlated signal
at 31.5 GHz.
This double-counting is of course mildly conservative, 
since we normalized free-free emission in the same way. 
Given the complete absence of power spectrum measurements for this component, 
the MID model simply assumes the same power spectra 
as for regular dust emission,
both in intensity and polarization, as well as the same polarization
fractions. The PESS scenario gives 10\% polarization 
(Prunet \& Lazarian 1999).
In the OPT scenario, we assume no spinning dust component at all.

Throughout this paper, we are assuming that the different foreground
components are uncorrelated. 
%%% This simplifying assumption is likely to
%%% be too conservative, since we expect that vibrating dust is correlated 
%%% with spinning dust and perhaps with free-free emission as well.
This is probably not the case for, \eg, spinning and vibrating dust.
Once these correlations are better measured, one can take advantage
of this information to improve the foreground removal, as well as 
define linear combinations of the foregrounds that are uncorrelated.

\subsection{Thermal and kinematic SZ effect}
\label{SZsec}

The thermal SZ effect (Sunyaev \& Zel'dovich 1970)
is the characteristic distortion of the CMB spectrum
caused by hot ionized gas in galaxy clusters and filaments,
whereas the kinematic SZ effect is the temperature fluctuation
occurring when motion of such gas Doppler shifts the CMB spectrum.
The dominant part of the kinematic SZ effect caused by matter fluctuations
in the linear regime is known as the Ostriker-Vishniac (OV) effect
(Vishniac 1987),
and can be accurately computed using perturbation theory 
(Hu \& White 1996).
According to the definition we gave in \sec{ModelSec1}, a process
is a foreground only if it cannot be accurately computed at the 
present time, so only part of the kinetic SZ effect qualifies as
a foreground: the small correction to the OV effect caused by
nonlinear structures, whose computations would require
accurate hydrodynamics simulations. Since this correction is likely to be 
small, we will not attempt to model it in the present paper.

The thermal SZ effect, on the other hand, does qualify as a foreground
(Holder \& Carlstrom 1999).
Just as we assumed removal of bright radio and IR point sources, 
we will assume that cores of known clusters
have been discarded from the CMB maps.
In addition to removing known clusters, it has been estimated
that of order $10^4$ additional clusters can be detected 
(and removed) using the \planck\ data 
(de Luca {\etal} 1995; Aghanim {\etal} 1997; Refregier {\etal} 1998;
Refregier 1999), 
reducing both
the kinematic and thermal SZ effect from clusters to negligible levels. 
The SZ foreground will therefore be dominated by the thermal 
effect from filaments and other large-scale structures outside of clusters.
As our MID estimate of this effect, we 
use the semianalytic results of Persi {\etal} (1995),
whose $\Lambda$CDM model is well fit by the broken power law power
spectrum
\beq{PersiEq}
% {\l(\l+1)\over 4\pi}C_\l =
\l^2 C_{\l\rm(SZ)} =
%%% (0.086\mK\>A)^2
(0.26\mK\>A)^2
\left[\l^{n_1\gamma}+
\left({\ell\over\ell_*}\right)^{n_2\gamma}\right]^{1/\gamma},
\eeq
% set a         = 0.89e-14 = 0.046 uK^2 # Normalization
Here $n_1=1$ and $n_2=-2$ are the asymptotic slopes at
low and high $\ell$ respectively, while $\gamma=-0.25$ gives the
sharpness of the peak, which is located at $\ell_{peak}=4000$ using
$\ell_*\equiv (-n_1/n_2)^{-1/\gamma n_2}\ell_{peak}^{1-n_1/n_2}$.
\Eq{PersiEq} is normalized in the Rayleigh-Jeans limit $\nu\ll$ 56 GHz
for $A=1$.
% $A=0.00735\mK^2$.
% $A=-0.086\mK$. 
Our PESS model is normalized an order of magnitude higher, 
roughly in line with current observational upper limits.
Relativistic corrections to the frequency dependence 
are important for hot clusters (Wright 1979; 
Rephaeli 1995; Stebbins 1997).
Since we are throwing out the known clusters and the 
filaments that dominate the remaining effect are much 
cooler, the nonrelativistic SZ-spectrum
should be quite a good approximation. In thermodynamic temperature,
this is given by (Sunyaev \& Zel'dovich 1970)
\beq{SZspectrumEq}
\Theta_{\rm(SZ)}(\nu) \propto
2-{x\over 2}\coth{x\over 2}\to 1\quad\hbox{as}\quad x\to 0,
\eeq
where $x\equiv h\nu/k\Tcmb\approx \nu/56.8\,\GHz$.

\subsection{Detector noise}

\begin{table}[tb]\footnotesize
\caption{\label{tab:specs}}
\begin{center}
{\sc CMB Experimental Specifications}
\begin{tabular}{rcccc}
\tableskip\hline\hline\tableskip
Experiment & $\nu$ & FWHM & $10^6 \Delta T/T$ & $10^6 \Delta T/T$  \\
& & & (unpol) & (pol) \\
\tableskip\hline\tableskip
\boom 
& 90  & 20 & 7.4 & \nodata \\
& 150 & 12 & 5.7 & \nodata \\
& 240 & 12 & 10 & \nodata \\
& 400 & 12 & 80 & \nodata \\
\tableskip\hline\tableskip
\map
& 22 & 56 & 4.1 & 5.9 \\
& 30 & 41 & 5.7 & 8.0 \\
& 40 & 28 & 8.2 & 11.6 \\
& 60 & 21 & 11.0 & 15.6 \\
& 90 & 13 & 18.3 & 25.9 \\
\tableskip\hline\tableskip
\planck
& 30  & 33 & 1.6 & 2.3 \\
& 44  & 23 & 2.4 & 3.4 \\
& 70  & 14 & 3.6 & 5.1 \\
& 100 & 10 & 4.3 & 6.1 \\
& 100 & 10.7 & 1.7 & \nodata \\
%%% & 100 & 10.7 & 1.57 & 5.68 \\
& 143 & 8.0 & 2.0 & 3.7 \\
& 217 & 5.5 & 4.3 & 8.9 \\
& 353 & 5.0 & 14.4 & \nodata \\
& 545 & 5.0 & 147 & 208 \\
& 857 & 5.0 & 6670 & \nodata \\
\tableskip\hline
\end{tabular}
\end{center}
NOTES.---%
Specifications used for \boom, \map\ and \planck.
Frequencies $\nu$ are in GHz.  Full width at half maxima
(FWHM) of the beams are in arcminutes.
\boom\ covers a fraction $\fsky\approx 2.6\%$ of the sky, while we
assume a useful sky fraction of 65\% for 
\map\ and \planck.
$(w^P)^{-1/2} = \Delta T \times {\rm FWHM} \times \pi/10800$.  
In practice, we combine the two \planck\ 100 GHz channels
into one channel with FWHM of 10'.7 and $\Delta T/T$ of 1.57 and 
$5.68\times10^{-6}$ for unpolarized and polarized channels, respectively.
\end{table}

As first pointed out by Knox (1995), detector noise 
can be conveniently treated as an additional sky signal 
with power spectrum
\beq{NoiseClEq}
C_{\l\rm(noise)}^{P} = (w^{P})^{-1} e^{\theta^2\l(\l+1)}
\eeq
if the experimental beam is Gaussian with width $\theta$ in radians 
(the full-width-half-maximum is given by FWHM$=\sqrt{8\ln 2}\>\theta$).
Here the sensitivity measure $1/w^{P}$ is defined as the
noise variance per pixel times the pixel area in steradians for $P=T,E,B$.
As shown in Appendix A of Tegmark (1997b), \eq{NoiseClEq} remains valid
even for incomplete sky coverage --- the corresponding information loss 
causes correlations between the different noise multipoles, 
but not an increase in their variance.
The noise variance $(\Delta T/T)^2$ per pixel of
area FWHM$^2$ is given in Table \ref{tab:specs}.
We assume that that this pixel noise
is equal and uncorrelated for the two measured Stokes parameters 
$Q$ and $U$, which means that the same noise value applies to
$E$ and $B$ ($w^E=w^B$). We also assume that the noise 
is uncorrelated between intensity and polarization, 
so that $1/w^{X}=0$.
For an experiment like MAP where 
intensity/polarization is measured by adding/subtracting
pairs of linearly polarized receivers,
$w^E=w^B=w^T/2$ (one pair measures $Q$ and $T$,
another does $U$ and $T$, and all four measurements are independent
with identical variance).

%%% \begin{tabular}{||l||c|c|c|c||}
%%% \hline
%%% Channel spec.           &BOOM	&MAP    &LFI	&HFI\\
%%% \hline
%%% $\nu$ [GHz]             &90	&22     &       &100    \\
%%% FWHM [arcmin]           &20	&57     &       &10.7   \\
%%% $10^6 \Delta T/T$ (unpol)&7.4	&4.1     &       &1.7    \\
%%% $10^6 \Delta T/T$ (pol)&	&5.9     &       &       \\
%%% \hline
%%% $\nu$ [GHz]             &150	&30     &30  	&143    \\
%%% FWHM [arcmin]           &12	&41     &33	&8.0    \\
%%% $10^6 \Delta T/T$ (unpol)&5.7	&5.7	&1.6    &2.0    \\
%%% $10^6 \Delta T/T$ (pol)&	&8.0     &2.3    &3.7    \\
%%% \hline
%%% $\nu$ [GHz]             &240	&40     &44	&217    \\
%%% FWHM [arcmin]           &10	&28     &23  	&5.5    \\
%%% $10^6 \Delta T/T$ (unpol)&9.5	&8.2     &2.4 	&4.3    \\
%%% $10^6 \Delta T/T$ (pol)&	&11.6     &3.4    &8.9    \\
%%% \hline
%%% $\nu$ [GHz]             &400	&60     &70  	&353    \\
%%% FWHM [arcmin]           &12	&21     &14  	&5.0    \\
%%% $10^6 \Delta T/T$ (unpol)&80	&11.0     &3.6 	&14.4   \\
%%% $10^6 \Delta T/T$ (pol)&	&16.6     &5.1    &    \\
%%% \hline
%%% $\nu$ [GHz]             &       &90     &100 	&545    \\
%%% FWHM [arcmin]           &   	&13     &10  	&5.0    \\
%%% $10^6 \Delta T/T$ (unpol)&  	&18.3     &4.3 	&147    \\
%%% $10^6 \Delta T/T$ (pol)&	&25.9     &7.5    &208       \\
%%% \hline
%%% $\nu$ [GHz]             &       &       &       &857    \\
%%% FWHM [arcmin]           &       &       &       &5.0    \\
%%% $10^6 \Delta T/T$ (unpol)&      &       &       &6670   \\
%%% $10^6 \Delta T/T$ (pol)&	&       &       &       \\
%%% \hline
%%% \end{tabular}

\subsection{Point sources}
\label{PSsec}

\begin{figure*}[tb] 
\centerline{\epsfxsize=18cm\epsffile{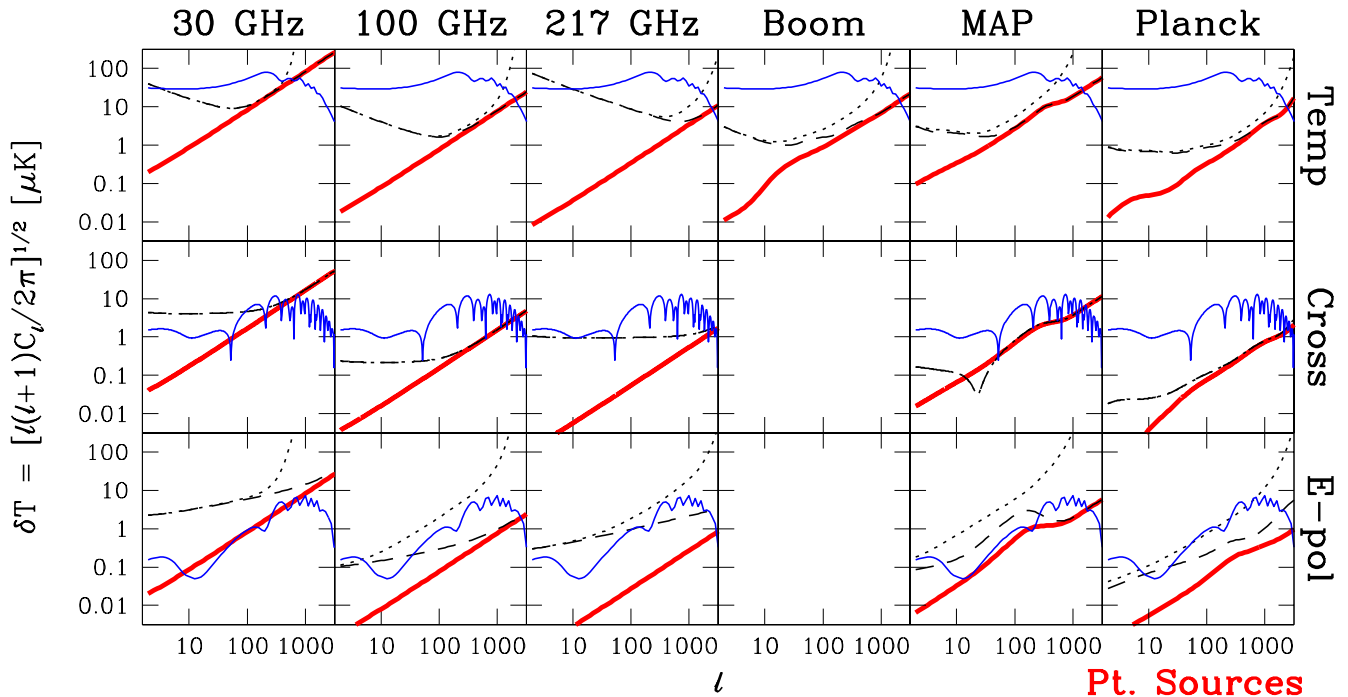}}
\caption{\label{psFig}\footnotesize%
Same as Fig. \protect\ref{SynchFig}, 
but for 
radio and far infrared point sources. 
}
\end{figure*}

The TE96 point-source model assumed that all sources above some
flux cut $S_c$ could be removed from the map
(by discarding the contaminated pixels, say) and gave the 
power spectrum due to Poisson fluctuations in the unresolved remainder.
Here we will make the conservative assumption that no external source
templates will be available at these frequencies, so that
point sources must be detected internally from the CMB maps themselves,
say as $5-\sigma$ outlyers.
Especially for high sensitivity experiments such as \planck, the main 
sources of confusion noise are the CMB fluctuations themselves
(and dust at very high frequencies).
It is therefore desirable to spatially band-pass filter the maps to suppress
CMB and detector noise fluctuations before performing the point-source search.
Tegmark \& de Oliveira-Costa (1998) derive such a procedure 
and find that the resulting minimal rms confusion noise $\sigma$
for point-source detection (in MJy) is given by 
\beq{MinimalSigmaEq}
\sigma(\nu) = [\conv(\nu)\convv(\nu)]^{-1} \left[
\sum_\l\lfac/\Cltot(\nu)\right]^{-1/2},
\eeq
where $\Cltot$ is the sum of the power spectra of other
foregrounds, noise and CMB.
Tegmark \& de Oliveira-Costa (1998) find that this filtering lowers the 
point-source detection threshold $\sigma$ 
by a factor between 2.5 and 18 for \planck.
Refregier {\etal} (1998) present such an analysis for the \map\ satellite.

Once the flux cut $S_c = 5\sigma$
%%% %\beq{FluxCutEq}
%%% %S_c={5\sigma\over[\conv(\nu)\convv(\nu)]^2}
%%% %\eeq
has been computed using our foreground and CMB model
(the latter is described in \sec{JointSec}),
we calculate the point-source 
power spectrum using the expression (TE96)
\beqa{psPowerEq}
C_{\l\rm(ps)}^T(\nu) &\equiv &
\left[\Theta_{(\rm ps)}(\nu)\right]^2 C_{\ell(\rm ps)}^T 
\nonumber\\ 
&= &[\conv(\nu)\convv(\nu)]^2\int_0^{S_c} 
\left[-{\partial n\over\partial S}\right](S,\nu)S^2dS.
\eeqa
Here $n(S,\nu)$ gives the source counts, \ie, the number of point 
sources per steradian whose flux exceeds $S$ at the frequency $\nu$.
We evaluate this integral, which is independent of $\l$, 
separately for each frequency channel
using the source count model of Toffolatti {\etal} (1998, 1999;
see also Guiderdoni {\etal} 1998, Guiderdoni 1999).
We then multiply the resulting power spectrum
by the normalization fudge factors $(\p A)^2$ 
given in Table \ref{ForegModelTab}.
These source count models are consistent with the upper limits from
the SCUBA experiment (Scott \& White 1999; Mann {\etal} 1999)
and other observations (Gawiser {\etal} 1998).
As stressed by, {\eg}, Franceschini {\etal} (1989),
point-source clustering can create additional large-scale power.
However, calculations of this effect 
(TE96; Toffolatti {\etal} 1998; Cress {\etal} 1996)
suggest that it is diluted by angular projection 
down to levels that are negligible
compared with the Poisson term of \eq{psPowerEq} 
({\it c.f.}, Scott \& White 1999).
The same holds for the effect of weak lensing 
modulation the flux cut (Tegmark \& Villumsen 1997).

This treatment is rather conservative in 
that it makes no assumptions about our ability to model 
the frequency dependence of point sources. In other words,
it assumes that one can remove a source
from a map only if one actually detects it at that
particular frequency. In practice, 
one might opt to discard pixels as contaminated if they
contain a detected point source at other nearby frequencies
as well, further reducing the residual $\sigmaps$.
Since most point sources have a spectrum substantially
different from CMB, the detection threshold can also be
pushed below that of \eq{MinimalSigmaEq} by taking linear 
combinations of band-pass filtered versions of different channels,
tailored to subtract out say the CMB and/or dust signals.

The frequency dependence of the residual point sources
has a distinctly bimodal distribution, corresponding to
radio sources (blazars, \etc)
and far infrared sources (early dusty galaxies, \etc).
Since these are modeled separately in Toffolatti {\etal} (1998), 
we treat them 
as two independent components, greatly reducing the 
effective spectral index uncertainty. We take $\Delta\alpha=0.5$ 
for the radio sources in the MID model. If measurements at different frequencies are
not taken simultaneously, time-variability of the sources
will increase this number (Gutierrez {\etal} 1999).
A more detailed model of the frequency coherence of IR point sources
is given by in Fig E.5 in the HFI report (AAO 1998), 
reprinted as Fig. 5b in BG99), 
suggesting that $\Delta\alpha$ may be smaller for this population.
We therefore assume $\Delta\alpha=0.3$ for the IR point sources (MID).

For the polarization power spectra, we conservatively assume that 
the radio sources are 10\% polarized and the IR sources are 5\% polarized. 
Point sources are one of the few foregrounds whose polarization
is {\it not} likely to be important.
This is because the amplitude relative to noise is always lower in
polarization: detector noise is typically ``141\% polarized'' in the sense
that it is at least as high in the polarization maps as in the intensity maps, 
usually by a factor $\sqrt{2}$.

\begin{figure*}[p] 
\centerline{\epsfxsize=18cm\epsffile{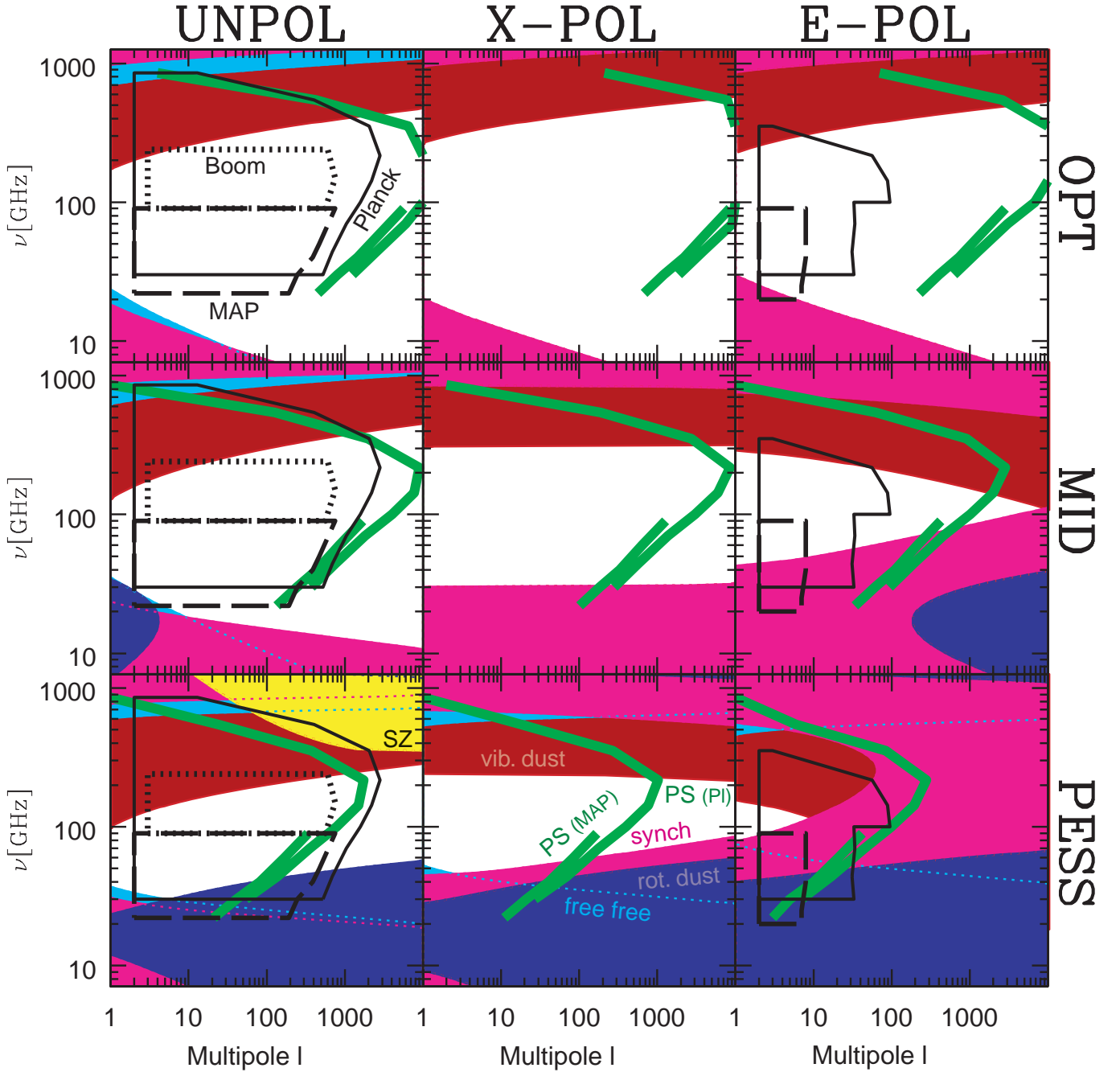}}
\caption{\label{lnuFig}\footnotesize%
This figure summarizes the frequency and scale dependence of our
foreground models for the optimistic (OPT), middle-of-the-road (MID) and
pessimistic (PESS) scenarios described in the text.
The colored regions show the parts of parameter space 
where the foreground fluctuations exceed a level 
$\delta T_*$ characteristic of the CMB, and correspond to
synchrotron (magenta), free-free (cyan) and vibrational 
dust emission (red),  
rotational dust emission (blue) and the thermal SZ effect (yellow).
For point sources, the residual is experiment-specific since it 
depends on the flux cut down to which point sources can be 
detected and excised---it is shown separately for \map\ and \planck\
as thick green lines.
The black boxes show where detector noise is less than $\delta T_*$
for \map\ and \planck. The thresholds in
$Q_{flat}\equiv (5/12)^{1/2}\delta T_*$ are $20\mK$, 
$3\mK$ and $0.5\mK$ for unpolarized, 
cross-polarized and $E$-polarized fluctuations, respectively. 
The $B$-spectra are similar to those shown for $E$-polarization.
}
\end{figure*}

\begin{figure*}[p] 
\centerline{\epsfxsize=18cm\epsffile{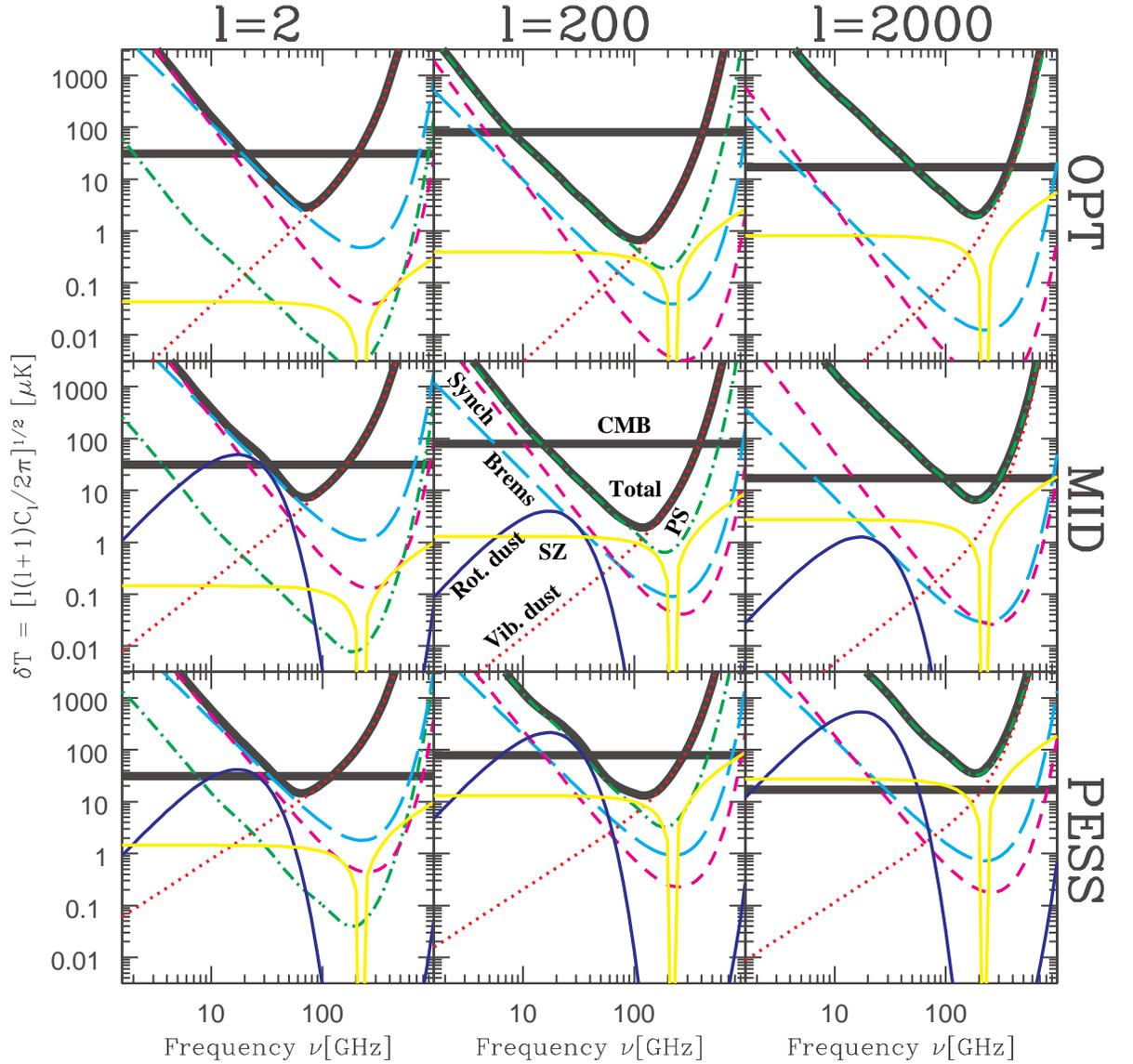}}
\caption{\label{MultipolesFig}\footnotesize%
The frequency dependence of our unpolarized foregrounds
is shown for three angular scales ($\l=2$, $\l=200$ and $\l=2000$)
for the optimistic (OPT), middle-of-the-road (MID) and
pessimistic (PESS) scenarios described in the text.
The thin curves correspond to synchrotron radiation (short-dashed), 
free-free emission (long-dashed), spinning dust
(solid), vibrational dust (dotted), point sources
(dot-dashed) and SZ (solid grey/yellow).
The thick curves show the CMB (horizontal line) and the total 
for all foregrounds.
}
\end{figure*}

\newcommand{\text}{{\it text}}
\begin{table*}[p]\footnotesize
\caption{\label{ForegModelTab}}
\begin{center}
{\sc Foreground Model Parameters}
\begin{tabular}{lc\colskipp cccc\colskipp cccc\colskipp cccc}
%\begin{tabular}{|lc|cccc|cccc|cccc|cccc|}

\tableskip\hline\hline\tableskip
&&
\multicolumn{4}{c\colskipp}{OPTIMISTIC}&
\multicolumn{4}{c\colskipp}{MIDDLE-OF-ROAD}&
\multicolumn{4}{c}{PESSIMISTIC}\\
&
&$\alpha$&$\Da$	&$\beta$	&$\p$	
&$\alpha$&$\Da$	&$\beta$	&$\p$	
&$\alpha$&$\Da$	&$\beta$	&$\p$\\	
\tableskip\hline\tableskip
Free-free&T	&2.15	&.01	&3	&1
		&2.15	&.02	&3	&1
		&2.10	&.04	&2.2	&1\\
emission&E	&\nodata&\nodata&\nodata&0
		&\nodata&\nodata&\nodata&0
		&2.10	&.04	&2.2	&0.1\\
	&B	&\nodata&\nodata&\nodata&0
		&\nodata&\nodata&\nodata&0
		&2.10	&.04	&2.2	&0.1\\
	&X	&\nodata&\nodata&\nodata&0
		&\nodata&\nodata&\nodata&0
		&2.10	&.04	&2.2	&0.3\\
$\nus=31.5$ GHz&&\multicolumn{4}{c\colskipp}{$A=30\mK$}
		&\multicolumn{4}{c\colskipp}{$A=70\mK$}
		&\multicolumn{4}{c}{$A=77\mK$}\\
\tableskip\hline\tableskip
Synchrotron&T	&2.9	&.1	&3	&1
		&2.8	&.15	&2.4	&1
		&2.6	&.3	&2.2	&1\\
radiation&E	&2.9	&.1	&3	&.1
		&2.8	&.15	&1.4	&.13
		&2.6	&.3	&1.0	&.25\\
	&B	&2.9	&.1	&3	&.1
		&2.8	&.15	&1.4	&.13
		&2.6	&.3	&1.0	&.25\\
	&X	&2.9	&.1	&3	&.2
		&2.8	&.15	&1.9	&.3
		&2.6	&.3	&1.6	&.4\\
$\nus=19$ GHz&	&\multicolumn{4}{c\colskipp}{$A=50\mK$}
		&\multicolumn{4}{c\colskipp}{$A=101\mK$}
		&\multicolumn{4}{c}{$A=192\mK$}\\
\tableskip\hline\tableskip
Vibrating&T	&2.0	&.1	&3	&1
		&1.7	&.3	&3	&1
		&1.4	&.5	&2.5	&1\\
dust	&E	&2.0	&.1	&3	&.01
		&1.7	&.3	&1.3	&.0022
		&1.4	&.5	&1.2	&.011\\
	&B	&2.0	&.1	&3	&.01
		&1.7	&.3	&1.4	&.0024
		&1.4	&.5	&1.2	&.011\\
	&X	&2.0	&.1	&3	&0.03
		&1.7	&.3	&1.95	&.0098
		&1.4	&.5	&1.85	&.02\\
$\nus=90$ GHz&	&\multicolumn{4}{c\colskipp}{$T=20$K, $A=9.5\mK$}
		&\multicolumn{4}{c\colskipp}{$T=18$K, $A=24\mK$}
		&\multicolumn{4}{c}{$T=16$K, $A=45\mK$}\\
\tableskip\hline\tableskip
Rotating&T	&\nodata&\nodata&\nodata&0
		&2.4	&.5	&3	&1
		&2.4	&1	&1.2	&1\\
dust	&E	&\nodata&\nodata&\nodata&0
		&2.4	&.5	&1.3	&.0022
		&2.4	&1	&1.2	&.1\\
	&B	&\nodata&\nodata&\nodata&0
		&2.4	&.5	&1.4	&.0024
		&2.4	&1	&1.2	&.1\\
	&X	&\nodata&\nodata&\nodata&0
		&2.4	&.5	&1.95	&.0098
		&2.4	&1	&1.2	&.2\\
$\nus=31.5$ GHz&&\multicolumn{4}{c\colskipp}{}
		&\multicolumn{4}{c\colskipp}{$T=0.25$K, $A=70\mK$}
		&\multicolumn{4}{c}{$T=0.25$K, $A=32\mK$}\\
\tableskip\hline\tableskip
Thermal	&T	&\text	&.01	&\text	&1
		&\text	&.02	&\text	&1
		&\text	&.05	&\text	&1\\
SZ	&E	&\nodata&\nodata&\nodata&0
		&\nodata&\nodata&\nodata&0
		&\nodata&\nodata&\nodata&0\\
	&B	&\nodata&\nodata&\nodata&0
		&\nodata&\nodata&\nodata&0
		&\nodata&\nodata&\nodata&0\\
	&X	&\nodata&\nodata&\nodata&0
		&\nodata&\nodata&\nodata&0
		&\nodata&\nodata&\nodata&0\\
$\nus=10$ GHz&	&\multicolumn{4}{c\colskipp}{eq.~(\ref{PersiEq}), (\ref{SZspectrumEq}), $A=.3$}
		&\multicolumn{4}{c\colskipp}{eq.~(\ref{PersiEq}), (\ref{SZspectrumEq}), $A=1$}
		&\multicolumn{4}{c}{eq.~(\ref{PersiEq}), (\ref{SZspectrumEq}), $A=10$}\\
\tableskip\hline\tableskip
Radio	&T	&\text	&.3	&0	&1
		&\text	&.5	&0	&1
		&\text	&1	&0	&1\\
point	&E	&\text	&.3	&0	&.05
		&\text	&.5	&0	&.1
		&\text	&1 	&0	&.2\\
sources	&B	&\text	&.3	&0	&.05
		&\text	&.5	&0	&.1
		&\text	&1 	&0	&.2\\
	&X	&\text	&.3	&0	&.1
		&\text	&.5	&0	&.2
		&\text	&1 	&0	&.3\\
&		&\multicolumn{4}{c\colskipp}{eq.~(\ref{psPowerEq}), $A=.3$}
		&\multicolumn{4}{c\colskipp}{eq.~(\ref{psPowerEq}), $A=1$}
		&\multicolumn{4}{c}{eq.~(\ref{psPowerEq}), $A=5$}\\
\tableskip\hline\tableskip
IR	&T	&\text	&.1	&0	&1
		&\text	&.3	&0	&1
		&\text	&.5	&0	&1\\
point	&E	&\text	&\nodata&0	&0
		&\text	&.3	&0	&.05
		&\text	&.5	&0	&.1\\
sources	&B	&\text	&\nodata&0	&0
		&\text	&.3	&0	&.05
		&\text	&.5	&0	&.1\\
	&X	&\text	&\nodata&0	&0
		&\text	&.3	&0	&.1
		&\text	&.5	&0	&.2\\
&		&\multicolumn{4}{c\colskipp}{eq.~(\ref{psPowerEq}), $A=.3$}
		&\multicolumn{4}{c\colskipp}{eq.~(\ref{psPowerEq}), $A=1$}
		&\multicolumn{4}{c}{eq.~(\ref{psPowerEq}), $A=5$}\\
\tableskip\hline
\end{tabular}
\end{center}
NOTES.---Our optimistic (OPT), middle-of-the-road (MID) and pessimistic (PESS)
foreground models. 
The frequency dependence is normalized so that $\Theta(\nus)=1$.
The power spectrum normalization is given by 
$(\p A)^2$, as specified by \eq{PowerLawPowerEq} (for free-free, synchrotron and 
dust emission), \eq{PersiEq} (for the thermal SZ effect) and
\eq{psPowerEq} (for point sources).
To avoid a profusion of large numbers in the table, we have factored 
the total normalization amplitude $\p A$ into an overall constant $A$
and a small dimensionless correction factor $\p$ 
that can be interpret polarization percentage (unless the polarized and 
unpolarized power spectra have different slopes).
The label ``{\it text}'' indicates that the parameterization is 
to be found in the text using the given equations.
\end{table*}

We conclude this section with some estimates of when 
point sources are important.
As shown in Tegmark \& Villumsen (1997), 
the rms fluctuation (in $\mu$K) due to residual point sources is  
\beq{sigpsApproxEq}
\sigmaps\approx\sqrt{\gamma-1\over 3-\gamma}N^{1/2}5\sigmaconf,
\eeq
where $N\equiv\pi\theta^2 n(S_c)$ is the number of sources
removed per beam area,
$\sigmaconf\equiv\sigma c c_*/2\pi\theta^2$ is the
confusion noise of \eq{MinimalSigmaEq}
converted from Jy into $\mK$, 
and the source counts have been approximated
by a power law $n'(S)\propto S^{-\gamma}$ near the flux cut.
Since relevant values for $\gamma$ are typically in the range 1.5--2.5
(see references in Tegmark \& Villumsen 1997), 
the first term is of order unity.
The best attainable $\sigmaconf$ is typically 3-5 times 
$\sigman$, the {\rms}
detector noise per pixel (Tegmark \& de Oliveira-Costa 1998).
Point sources have only a minor impact on a CMB 
experiment if $\sigmaps\ll\sigman$ because their power spectra
have the same shape as that for detector 
noise (apart from the noise increase below the beam scale).
\Eq{sigpsApproxEq} therefore tells us that 
using the CMB map itself for point-source removal is quite adequate
as long as $N\ll (4\times 5)^{-2}=0.002$. 
Conversely, if there are more sources per beam than this rule of 
thumb indicates, then an external point-source template will
be needed to reduce the point-source contribution to a
subdominant level.
% Figure 666 shows how 
% this criterion
% partitions CMB experiments into two classes: those for which 
% internal cleaning suffices and those which need external point 
% source data to reach their full potential.

\subsection{Foreground model summary}

The specifications of our foreground models are given in 
Table \ref{ForegModelTab}.
The power spectrum and frequency dependence is summarized in 
\fig{lnuFig}, which follows TE96 in showing where the various foregrounds
dominate over a typical CMB signal. 
%More details are given in 
%figures~\ref{CTfig}, \ref{CXfig} and~\ref{CEfig}, which show the power 
%foreground power spectra at three characteristic frequencies.
More details about each foreground are given in 
figures~\ref{SynchFig}--\ref{psFig}, which show the 
power spectra at three characteristic frequencies.
Figure~\ref{MultipolesFig} shows the frequency dependence of
the foregrounds on three different angular scales.

\section{Foreground models 2: the math}
\label{ModelSec2}

\subsection{Notation}

As described in T98 and further elaborated by White (1998),
foregrounds can be treated as simply an additional source of noise that is 
correlated between frequency channels. This leads to a natural
way of parameterizing them as well as to a useful way of removing them.
Let us first express this in its most general mathematical form, 
and then specialize to a case appropriate for our present
application of accuracy forecasting.

Consider a pixelized CMB sky map (the ``true sky'')
at some angular resolution $\theta_0$
consisting of $M$ numbers $x_1,...,x_M$, where $x_i$ is the 
temperature in the $\ith$ pixel.
Suppose that we have single-frequency data sets at our disposal at $\nfreq$
different frequencies $\nu_\freq$ 
($\freq=1,2,...,\nfreq$) consisting of $N_\freq$ numbers
$y^\freq_{1}$, $y^\freq_{2},...,y^\freq_{N_\freq}$, 
each probing some linear combination of the
sky temperatures $x_i$. 
Grouping these numbers into vectors $\x$, $\y^1$, $\y^2,...,\y^\nfreq$
of length
$M$, $N_1$, $N_2,...,N_\nfreq$ 
(these lengths are all generally all different), 
we can generally write 
\beq{ModelEq1}
\y^\freq = \A^\freq\x+\n^\freq
\eeq
for some known $N_\freq \times M$ scan strategy matrices $\A^\freq$ incorporating the beam shapes and
some random vectors $\n^\freq$ 
incorporating instrumental noise and foreground contamination.
The special case where the $\nfreq$ data sets are simply 
sky maps with resolution $\theta_0$ corresponds to $\A^\freq=\I$.
If the data sets are maps with different angular resolutions
$\theta_\freq\ge\theta_0$, 
then
\beq{SmoothingEq}
%\A^\freq_{ij} = \exp[-\theta_{ij}^2/2\Delta\theta_\freq^2]/2\pi\Delta\theta^2
\A^\freq_{ij} = {1\over 2\pi(\Delta\theta_\freq)^2} 
e^{-{1\over 2}{\left(\theta_{ij}\over\Delta\theta_\freq\right)^2}}
\eeq
if the beams are Gaussian, 
where $\theta_{ij}=\cos^{-1}(\rh_i\cdot\rh_j)$ is the 
angular separation between pixels in directions $\rh_i$ and $\rh_j$ and 
$\Delta\theta_\freq=(\theta_\freq^2-\theta_0^2)^{1/2}$ 
is the extra smoothing in map $\freq$.
\Eq{ModelEq1} is completely general, however, since the scan strategy matrix
$\A^\freq$ can also incorporate complications such as elliptical and 
non-Gaussian beams, triple beams, interferometer beams, or
oblong synthesized beams (e.g.\ Saskatoon).
Of course, the data sets need not be different channels observed by 
the same experiment---for instance, one might wish to 
use the 408 MHz Haslam survey as an additional ``channel''.

It is useful to define the larger $(\sum_f N_f)\times M$ 
matrix and the $(\sum_f N_f)$-dimensional vectors
\beq{GroupingEq}
\A\equiv\left(\bs\begin{tabular}{c}
$\A^1$\\
$\vdots$\\
$\A^\nfreq$
\end{tabular}\bs\right),\quad
\y\equiv\left(\bs\begin{tabular}{c}
$\y^1$\\
$\vdots$\\
$\y^\nfreq$
\end{tabular}\bs\right),\quad
\n\equiv\left(\bs\begin{tabular}{c}
$\n^1$\\
$\vdots$\\
$\n^\nfreq$
\end{tabular}\bs\right).\quad
\eeq
This allows us to rewrite \eq{ModelEq1} as 
\beq{ModelEq2}
\y=\A\x+\n,
\eeq
a set of linear equations that would be highly over-determined
if it were not for the presence of unknown noise $n$.

It is straightforward to include polarization information in 
our formalism. In this case, we wish to measure not one sky map 
but three: the unpolarized temperature map
$\xT$ and the ``electric'' and ``magnetic'' polarization maps
$\xE$ and $\xB$ 
(Kamionkowski {\etal} 1997; Zaldarriaga \& Seljak 1997). 
The latter are linearly related to the 
Stokes Q and U maps and have the advantage of being independent
of the choice of coordinate system and more directly linked to the physical
processes that make the CMB polarized.
Grouping them into a single vector
\beq{PolarizedxEq}
\x\equiv\left(\bs\begin{tabular}{c}
$\xT$\\
$\xE$\\
$\xB$
\end{tabular}\bs\right)
\eeq
and enlarging $\y$, $\n$ and $\A$ to include polarized measurements, 
we once again recover the form of \eq{ModelEq2}.
 
\subsection{Parameter estimation}

The general goal is to use the data set $\y$ to measure a set of physical 
parameters. These parameters, which we will denote $p_i$ 
($i=1,...,N$) 
and group together in a vector $\vp$, can be either cosmological parameters,
such as the true CMB sky temperatures $\x$ or 
model inputs like the baryon density $\Omega_b$,
or constants that parameterize the foreground model, such as the emissivity
$\alpha$ of thermal dust emission or the scale dependence $\beta$ of 
synchrotron radiation.
How accurately can this be done?
If the likelihood of observing $\y$ given these parameters is written
$\calL(\y;\vp)$, then the answer is contained in the {\it Fisher information matrix}
(Kendall \& Stuart 1969)
\beq{FdefEq}
\F_{ij}\equiv\bexpec{\partial^2\ln \calL\over\partial p_i\partial p_j}_\y,
\eeq
where the partial derivatives and the
averaging are evaluated using the true values of the parameters $\vp$.
The Cram\'er-Rao inequality shows that $(\F^{-1})_{ii}$ is the smallest
variance that any unbiased estimator of the parameter $p_i$ can have,
and we can generally think of $\F^{-1}$ as the best possible
covariance matrix for estimates of the vector $\vp$
(see Tegmark {\etal} 1997 for a review).

In \sec{ResidualSec}, we will present a foreground removal method
that recovers the CMB map $\x$ with these minimal error bars if the foreground
model is known. In \sec{JointSec}, we assess the accuracy to which 
cosmological parameters and foreground parameters can be measured jointly.

For the important case when all fluctuations are Gaussian
with mean\footnote{\label{MeanFootnote}Foregrounds 
typically do not have an expectation 
value of zero -- in fact, most of them are always positive.
This is one of the reasons why it can be advantageous 
to expand the maps in some set of basis functions and remove them
expansion coefficient by expansion coefficient instead of pixel by
pixel. For instance, in a Fourier or spherical harmonic decomposition,
it is typically only a single coefficient (the monopole) that will 
have a non-zero mean.
Alternatively, one can explicitly deal with the case of a non-zero mean 
including a constraint term (Bond {\etal} 1999).
}
$\expec{\y}=\bzero$,
\ie, when the vector
$\y$ has a multivariate Gaussian probability distribution of the form
\beq{GaussianEq}
\calL(\y;\vp) = (2\pi)^{-n/2} |\C|^{-1/2} e^{-{1\over 2}\y^t\C^{-1}\y},
\eeq 
the model is entirely specified by the covariance matrix 
$\C = \C(\vp)\equiv \expec{\y\y^t}$.
The Fisher matrix then becomes
\beq{GaussFisherEq}
\F_{ij} = {1\over 2}\tr
\left[\C^{-1}{\partial\C\over\partial p_i}\C^{-1}{\partial\C\over\partial p_j}\right].
\eeq
The covariance matrix $\C$, with contributions from 
CMB, foregrounds and detector noise, is therefore the
key quantity that our model must provide.
Modeling $\C$ is the topic of the next section.
% \S\ref{ModelSubsec}.

\subsection{Modeling the foreground covariance matrix}
\label{ModelSubsec}

\hyphenation{pre-ci-si-on} % But this doesn't help!

When removing foregrounds from upcoming high-precision experiments,
%%% to measure cosmological parameters to percent accuracies,
it may be desirable to  
work with $\C$ in its full generality, 
explicitly modeling correlations between different foregrounds, 
correlations between polarized and unpolarized foregrounds, 
correlations between foreground fluctuations levels and
galactic latitude, {\etc}
Indeed, the foreground removal method that will be given below 
in \eq{ComboEq1} requires no simplifications.
However, since the goal of the present paper is considerably 
more modest, we will make several simplifying approximations below.

\subsubsection{Transforming to spherical harmonics}
\label{SphHarmSec}

Let us first assume that all of the data sets are maps and that the
statistical properties of CMB, noise and foregrounds are 
isotropic\footnote{Galactic foregrounds such as dust, synchrotron and 
free-free emission are of course not statistically isotropic, since 
they are more prevalent close to the galactic plane. 
To be conservative, we will therefore assume that only the cleanest 
65\% of the sky is used (for a straight latitude cut, this
would correspond to discarding all pixels less than $20^\circ$ 
from the Galactic plane), and assume that the statistical 
properties of the remainder are isotropic
with a foreground amplitude corresponding to the dirtiest remaining 
region.

To take advantage of the fact that the contamination level depends
both on angular scale and Galactic latitude, it has been suggested
(Tegmark 1998) that the foreground removal be done not multipole
by multipole, but wavelet by wavelet, since $\C$ will become 
approximately block-diagonal in a suitable spherical wavelet
basis even when the foreground power depends on latitude.
Such wavelet bases are described by, \eg,
Cay\'on {\etal} (1999) and Tenorio {\etal} (1999).
}.
This allows us to make the matrix $\C$ block-diagonal by expanding 
the data sets $\y^\freq$ in spherical harmonics.
For notational convenience, we 
renormalize the expansion coefficients $\alm^{P\freq}$ of $\y^\freq$
(the polarization index $P=T$, $E$, and $B$)
by dividing out the effect of the beam $e^{-\theta_\freq^2 \ell (\ell+1)/2}$.
Then the covariance matrix 
%In this new basis, the covariance matrix 
%of the expansion coefficients $\alm^{P\freq}$ 
%($P=T$, $E$, and $B$)
takes the block-diagonal form\footnote{
\label{fskyFootnote}
When the sky coverage $\fsky<1$, certain multipoles become correlated
(Tegmark 1996). 
This reduces the effective number of uncorrelated modes
by a factor $\fsky^{-1}$, thereby increasing the sample 
variance on power measurements by the same factor
(Scott {\etal} 1994; Knox 1995). It also smears out sharp 
features in the power spectrum, but this effect is 
negligible as long the sky map is more than a few degrees wide
in its narrowest direction (Tegmark 1997b).
}
\beq{DiagonalNeq}
C_{\l m\l'm'}^{P\freq P'\freqq}\equiv
\expec{\alm^{P\freq*} a_{\l'm'}^{P'\freqq}} 
= \delta_{\l\l'}\delta_{mm'}\calC_\l^{P\freq P'\freqq}
\eeq
for some size $3F\times 3F$ power spectrum matrix\footnote{We use
script letters to indicate matrices of size $3F$ and bold letters
to indicate matrices and vectors of other sizes, in particular size $F$.}
$\calC_\l$ of
the true sky (as opposed to the beam smoothed sky).
This of course also involves dividing the noise $\n$ 
in equation (\ref{ModelEq1}) by 
the same factors, which allows us to recover the foregrounds on the 
true sky while altering the detector noise to the 
form in equation (\ref{NoiseClEq}).

The $\calC_\l$ matrix can be broken into a block-matrix form
\beq{BlockMatrix}
\calC_\l = \left(\begin{array}{ccc}%
\C^T_\l & \C^X_\l & \bfzero \\[2pt]
\C^X_\l & \C^E_\l & \bfzero \\[2pt]
\bfzero & \bfzero & \C^B_\l \end{array}\right),
\eeq
where the $\C^P_\l$ ($P=T,E,B,X$) are $F\times F$ matrices to specify
the correlation between different frequency channels for the intensity,
$E$-channel polarization, $B$-channel polarization, and
intensity-polarization cross-correlation, respectively.
Note that for the CMB and most foregrounds
cross-correlations between $B$ and either $T$ or $E$
vanish for symmetry reasons, $B$ has odd parity whereas 
$T$ and $E$ have even parity 
(Kamionkowski {\etal} 1997; Zaldarriaga \& Seljak 1997). 
This is not necessarily true for all foregrounds, so 
the $T-B$ and $B-E$ correlations may potentially contain
additional useful information about contamination.
For instance, the effective birefringence caused by
Faraday rotation though a uniform magnetic field 
is not invariant under parity and gives such ``forbidden''
cross-correlations (Lue {\etal} 1999).

%%% If we include polarization data indexed by $p=T,E,B$, then
%%% $\C_\l$ becomes a $3f\times 3f$ matrix $\C^{\freq p\freqq p'}_\l$. 

In terms of these power spectrum matrices, 
the Fisher matrix of \eq{GaussFisherEq}
reduces to
\beq{GaussFisherEq2}
\F_{ij} = {1\over 2}\sum_\l (2\l+1)\fsky\> 
\tr\left[\calC_\l^{-1}{\partial\calC_\l\over\partial p_i}\calC_\l^{-1}
{\partial\calC_\l\over\partial p_j}\right],
\eeq
where the matrix multiplications involve both polarization-type
and frequency.
Here the factor $(2\l+1)\fsky$ gives the effective number of uncorrelated
modes per multipole$^{\ref{fskyFootnote}}$, and the other factor gives
the information per mode.

\subsubsection{Separation into physical components}

We write $\n$ as a sum of detector noise and $K$ physically 
distinct foreground components (synchrotron emission, point sources, {\etc})
% $\n_k$ ($k=1,...,K$) 
and assume that 
these 
% components 
are all uncorrelated, both with each other and with the $\x$, the CMB.
% This means that $\N=\sum_{k=0}^\c\expec{\n_k\n_k^t}$,
% so the power spectrum matrix is simply given by
This means that the power spectrum matrix is given by a sum
\beq{CdefEq}
\calC_\l = \sum_{k=0}^{K+1} \calC_{\l\kk},
\eeq
where $\calC_{\l\kk}$ is the power spectrum matrix
of the $\kth$ component, the covariance matrix of its $\alm^P$ 
at different frequencies.
$\C_{\l(0)}$ denotes the CMB contribution,
$\C_{\l(1)}$ the detector noise.

\subsubsection{Frequency coherence}

It is convenient to factor these matrices into a spatial
term, a frequency dependence term, and a frequency correlation
term.  We therefore write
\beq{ClFactor}
\C_{\l\kk}^{P\freq\freqq}=C_{\l\kk}^P 
\Theta_{\kk}^{P\freq} \Theta_{\kk}^{P\freqq}
\R_{\kk}^{P\freq\freqq}.
\eeq 
We normalize the frequency spectrum 
$\Theta_{\kk}^{P\freq}\equiv\Theta_{\kk}^P(\nu_\freq)$ 
so that $\Theta_{\kk}^P(\nus)=1$, thereby absorbing 
the physical units into $C_{\l\kk}$, the angular
power spectrum of the $\kth$ component at the reference frequency $\nus$.
The frequencies $\nus$ are given in Table~\ref{ForegModelTab}
and are chosen to be where the constraints are strongest or
most relevant.  The correlation between different frequency channels
is then encoded in the matrix $\R$.

We will assume that the mean frequency dependence $\Theta_{\kk}^{P\freq}$
and the frequency correlations $\R_{\kk}^{P\freq\freqq}$ are 
independent of $\l$ for all foregrounds.  
We take this frequency-scale separability as the operational 
definition of a distinct component; however,
this is not necessarily
true for the physical components of \sec{ModelSec1}.
One could imagine decomposing these emission mechanisms into 
multiple components to take account of changes in frequency
dependence as a function of scale.

Let us label the detector noise as $k=1$.
Then $\Theta_{(1)}^{P\freq}$ is simply the {\rms} detector noise level in the
channel $\freq$ for polarization-type $P$. If this noise is 
uncorrelated between channels, 
we have $\R_{(1)}^P=\I$, the identity matrix. 
On the other hand, if the $\kth$ foreground component has 
the same spectrum $f(\nu)$
everywhere in the sky, it will have
$\alm^{P\freq}\propto f(\nu)$ and hence 
$\Theta_\kk^{P\freq}\propto f(\nu)$ and  
$\R_\kk^P=\E$, where $\E^{\freq\freqq}=1$ is the rank 1
matrix containing only ones.
Note that the CMB fluctuations fall into this category, 
\ie, $\R_{(0)}^P=\E$, since
their temperature is the same in all channels.
Real-world foregrounds will typically have
correlation matrices $\R_\kk$ that are intermediate between 
these two extreme cases of perfect correlation ($\R=\E$)
and no correlation ($\R=\I$).

Since we presently lack detailed measurements of the foreground 
correlation matrices $\R$, we will use the simple one-parameter model 
\beq{RmodelEq}
\R_\kk^{P\freq\freqq} \approx \exp\left\{
-{1\over2}\left[\ln(\nu_\freq/\nu_\freqq)\over\coh_\kk^P\right]^2\right\},
\eeq
that was derived in T98. We also explore some alternative models in 
\sec{JointSec}. 
The model parameter $\coh$, the
{\it frequency coherence}, determines by how many powers of 
$e$ we can change the frequency before the correlation between the
channels starts
to break down. The two limits $\coh\to 0$ and 
$\coh\to\infty$ correspond to the two extreme
cases $\R=\I$ and $\R=\E$ that we encountered above.
The T98 derivation of \eq{RmodelEq} shows that for a spectrum of the type 
\beq{SpectrumEq}
\specint=f(\nu)(\nu/\nus)^\alpha,
\eeq
one has as a rule of thumb that
\beq{DispersionEq}
\coh\approx {1\over\sqrt{2}\Da}.
\eeq
Here $\Da$ is the {\rms} dispersion across the sky of the
spectral index $\alpha$, and $f$ is some arbitrary 
function. 

The factorization into $\Theta$ and $\R$ in equation (\ref{ClFactor}) 
is appropriate for
the $T$, $E$, and $B$ block-elements, because these elements, like
the $\calC$ matrix itself, must be symmetric.  The block-elements $X$ 
are off-diagonal and therefore need not be symmetric.  Asymmetries
indicate that the correlation of the intensity at frequency $\nu_\freq$
and the $E$-polarization at frequency $\nu_\freqq$ differs from
that of the intensity at $\nu_\freqq$ and polarization at $\nu_\freq$.
We have no data to inform any specification of such asymmetries;
therefore, we adopt the same symmetric form for the $X$ elements as for the
diagonal elements.  In \sec{JointSec}, we do consider asymmetric
parameterizations of these off-diagonal elements.

In conclusion, our foreground model involves specifying the three quantities
given in \sec{ModelSec1} for 
each physical component $k$ and each of the four types of polarization
power ($P=T$, $E$, $B$ and $X$):
its average frequency dependence $\Theta_\kk^P(\nu)$,
its power spectrum $C_{\l\kk}^P$ and
its frequency coherence $\coh_\kk^P$.

\section{How accurately can foregrounds with known 
statistical properties be removed?}
\label{ResidualSec}

In this section, we use our foreground models
to compute the level to which foregrounds can be removed.
This is important for identifying which foregrounds are most
damaging and therefore most in need of further study. It 
is also useful for optimizing future missions
and for assessing the science impact of design changes 
to, \eg, \planck. 

The treatment in this section assumes that the statistical properties 
of the foregrounds
(power spectrum, frequency dependence and
frequency coherence) are known. In practice, these too must of course 
be measured using the data at hand, and we will treat this issue in 
\sec{JointSec}.

\subsection{Foreground removal}

Foreground removal involves 
inverting the (usually over-determined) system of noisy linear
equations\eqn{ModelEq2}.
Which unbiased estimate $\xt$ of the CMB map $\x$ has the smallest
{\rms} errors from foregrounds and detector noise combined?
Physically different but mathematically identical problems 
were solved in a CMB context by Wright (1996) and Tegmark (1997a), 
showing that if $\expec{\n}=\bzero$ (see footnote \ref{MeanFootnote}), then
the minimum-variance choice is
\beq{ComboEq1}
\xt = [\A^t\N^{-1}\A]^{-1}\A^t\N^{-1}\y,
\eeq
where $\N\equiv\expec{\n\n^t}$. Tegmark (1997a) also showed that this 
retains all the cosmological information of the original data sets
if the random vector $\n$ has a Gaussian probability distribution,
regardless of whether the CMB signal $\x$ is Gaussian 
or not.\footnote{
In other words, this foreground removal method 
is information-theoretically ``best'' (lossless) only 
if the foregrounds have a multivariate Gaussian probability distribution.
Generally they do not, in 
which case the advantage of this scheme is merely that it is 
the linear method that minimizes the total rms of foregrounds and noise.
Simulations by Bouchet {\etal} (1995) have show that linear removal schemes
are quite effective even when faced with non-Gaussian foreground
templates. 
However, non-linear techniques taking advantage
of the specific form of foreground non-Gaussianity can 
under some circumstances perform even better: \eg, the
maximum entropy method (Hobson {\etal} 1998),
the filtered threshold clipping for point sources as in \sec{PSsec}
(Tegmark \& de Oliveira-Costa 1998; Refregier {\etal} 1998),
or other techniques (Ferreira \& Magueijo 1997; Jewell 1999).
An additional advantage of linear methods is that their simplicity
allows the properties of the cleaned map to be 
computed exactly, which facilitates its interpretation and use for
measuring cosmological parameters.
For the linear method we describe, the cleaned map is
simply the sum of the true map and various residual  
contaminants whose power spectra can be computed analytically.
} 

Substituting \eq{ComboEq1} into \eq{ModelEq2} shows that
the recovered map is unbiased ($\expec{\xt}=\x$) 
and that the pixel noise $\err\equiv\xt-\x$ has the covariance matrix
\beq{ComboCovarEq}
\Sig\equiv\expec{\err\err^t}=[\A^t\N^{-1}\A]^{-1}.
\eeq
As long as $\expec{\nt}=\bzero$, the map $\xt$ remains unbiased
even if the model for the noise covariance $\N$ is incorrect.
As described in T98, this method 
generalizes and supersedes the multi-frequency
Wiener filtering technique for foreground subtraction of 
TE96 and Bouchet {\etal} (1996)\footnote{
TE96 proposed modeling spectral uncertainties in a given foreground by 
treating it as more than one component. For example, dust
emission could be modeled as 
\beq{MultiCompEq}
\specint = 
\sum_{i=1}^c a_i B(\nu)\left({\nu\over\nus}\right)^{\alphabar+\varepsilon_i}
\eeq
for a set of small emissivity variations $\varepsilon_i$.
It is easy to show that the T98 method is recovered in the limit
$c\to\infty$. The simplest case with 
$c=2$ and $\varepsilon_1=-\varepsilon_2=\varepsilon$ gives the special case
explored in the \planck\ HFI proposal (AAO 1998; BG99) and also tested for MAP 
(Spergel 1998, private communication): 
% \beq{BouchetEquivEq}
% B(\nu) = 
% a_1 B_0(\nu)\left({\nu\over\nus}\right)^{\alphabar+\varepsilon} +
% a_2 B_0(\nu)\left({\nu\over\nus}\right)^{\alphabar-\varepsilon}
% = 
% b_1 B_0(\nu)\left({\nu\over\nus}\right)^\alphabar +
% b_2 B_0(\nu)\left({\nu\over\nus}\right)^\alphabar \ln{\nu\over\nus}
% \eeq
\beqa{BouchetEquivEq}
\specint&=& 
a_1 B(\nu)\left({\nu\over\nus}\right)^{\alphabar+\varepsilon} +
a_2 B(\nu)\left({\nu\over\nus}\right)^{\alphabar-\varepsilon}\\
&=&
b_1 B(\nu)\left({\nu\over\nus}\right)^\alphabar +
b_2 B(\nu)\left({\nu\over\nus}\right)^\alphabar \ln{\nu\over\nus}
\eeqa
if $|\varepsilon\ln\nu/\nus|\ll 1$, where 
$b_1\equiv a_1+a_2$ and $b_2\equiv(a_1-a_2)\varepsilon$.
In our formalism, this TE96 two-component model
simply corresponds to the approximation that
the matrix $\R$ has rank 2.
}, and reduces to the special case of Dodelson (1996) for $\N=\I$. 
Note that whereas the full covariance matrix $\C$ was needed
to compute the general Fisher matrix in \sec{ModelSubsec}, only 
the covariance $\N$ of the noise and foreground components 
is needed here. This is because we do not care about sample variance 
when the parameters to be estimated are the CMB sky 
temperatures ($\vp=\x$).
In short, the foreground removal method described here 
requires no assumptions whatsoever about the CMB sky---we are not even
assuming that the CMB fluctuations are isotropic or Gaussian.

\subsection{How the different frequencies get weighted}

Expanding our data in spherical harmonics as above, 
we subtract the foregrounds separately for each
multipole $a\lm$ using \eq{ComboEq1}. The 
relevant vectors and matrices reduce to 
\beq{GroupingEq2}
\x\lm=\left(\bs\begin{tabular}{c}
$a\lm^T$\\[2pt]
$a\lm^E$\\[2pt]
$a\lm^B$
\end{tabular}\bs\right),\quad
\y\lm=\left(\bs\begin{tabular}{c}
$\a\lm^T$\\[2pt]
$\a\lm^E$\\[2pt]
$\a\lm^B$
\end{tabular}\bs\right),\quad
\eeq
\beq{Aeq}
\calA = 
\left(\bs\begin{tabular}{ccc}
$\e$	&$\bfzero$	&$\bfzero$\\[2pt]
$\bfzero$	&$\e$	&$\bfzero$\\[2pt]
$\bfzero$	&$\bfzero$	&$\e$
\end{tabular}\bs\right),
\quad
\calN_\l = 
\left(\bs\begin{tabular}{ccc}
$\N_\l^T$	&$\N_\l^X$ &$\bfzero$\\[2pt]
$\N_\l^X$	&$\N_\l^E$ &$\bfzero$\\[2pt]
$\bfzero$	&$\bfzero$    &$\N_\l^B$
\end{tabular}\bs\right).
\eeq
The $F$-dimensional vectors $\a\lm^T$, $\a\lm^E$ and $\a\lm^B$ give the 
measured multipoles at the $f$ different frequencies, i.e., the data
we wish to use to estimate the CMB multipoles in $\x$.
$\e$ is the $F$-dimensional row vector consisting entirely of ones,
$\calA$ is the $3F \times 3$ 
scan strategy matrix for a given 
$(\ell,m)$\footnote{$\calA$ takes on this trivial form due to
the renormalization of ${\bf y}_{\ell m}$ 
and ${\bf x}_{\ell m}$ to the true sky in \sec{SphHarmSec},
\ie, since beam effects have been eliminated.} 
and $\N^T_\l$, $\N^E_\l$, $\N^B_\l$ and $\N^X_\l$ are the 
$F\times F$ power spectrum matrices of the non-cosmic signal,
built by summing the covariance matrices of \eq{ClFactor}, e.g., 
\beq{Ntdef}
\N^T_\l = \sum_{k=1}^{K+1} \C_{\l\kk}^T.
\eeq
\Eq{ComboEq1} thus gives the solution $\xt_{\l m}=\calW_\l^t\y_{\l m}$, 
where we can write
\beq{WdefEq}
\calW_\l\equiv\calN_\l^{-1}\calA[\calA^t\calN_\l^{-1}\calA]^{-1}=
\left(\bs\begin{tabular}{ccc}
$\w_\l^T$	&$\w_\l^{T'}$	&$\bfzero$\\
$\w_\l^{E'}$	&$\w_\l^E$ 	&$\bfzero$\\
$\bfzero$	&$\bfzero$    	&$\w_\l^B$
\end{tabular}\bs\right)
\eeq
for some $F$-dimensional weight vectors $\w$, so 
\beqa{ExplicitSolutionEq}
\at\lm^T&=&\w_\l^T\cdot\a\lm^T+\w_\l^{T'}\cdot\a\lm^E,\\
\at\lm^E&=&\w_\l^E\cdot\a\lm^E+\w_\l^{E'}\cdot\a\lm^T,\\
\at\lm^B&=&\w_\l^B\cdot\a\lm^B.
\eeqa
% \beq{ExplicitSolutionEq}
% \left\{\bs
% \begin{tabular}{c}
% $a\lm^T=\w_\l^T\cdot\a\lm^T+\w_\l^{T'}\cdot\a\lm^E$\\
% $a\lm^E=\w_\l^E\cdot\a\lm^E+\w_\l^{E'}\cdot\a\lm^T$\\
% $a\lm^B=\w_\l^B\cdot\a\lm^B$
% \end{tabular}
% \right.
% \eeq
Since these are by construction unbiased estimators of the 
true multipoles, the weight vectors clearly satisfy 
$\e\cdot\w_\l^T=\e\cdot\w_\l^E=\e\cdot\w_\l^B=1$
(the estimates are weighted averages of the different measurements)
and $\e\cdot\w_\l^{T'}=\e\cdot\w_\l^{E'}=0$ (there is no 
mixing of polarizations). If the foregrounds and the detector noise 
lack correlations between $T$ and $E$, \ie, if $\N_\l^X=0$,
then $\calN$ becomes block-diagonal and
the solution simplifies to $\w_\l^{T'}=\w_\l^{E'}=0$.

\begin{figure}[tb] 
\centerline{\epsfxsize=\idlwidth\epsffile{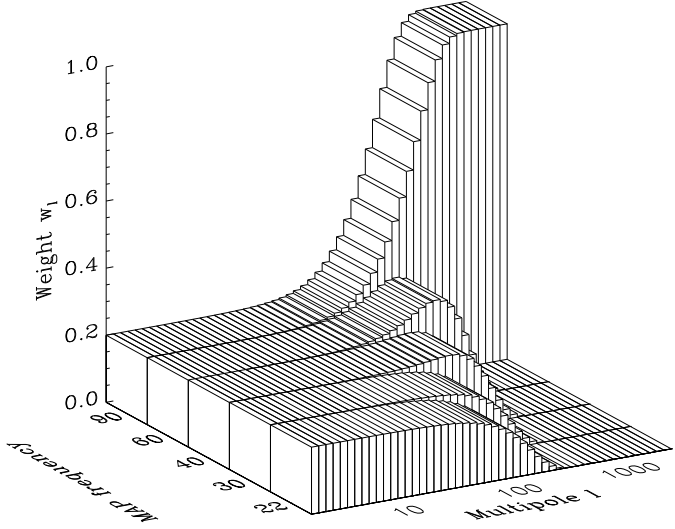}}
\caption{\label{wTfigMAPno}\footnotesize%
The weights $\w_\l^T$ with which the 5 unpolarized MAP 
channels are combined into a single map are plotted as
a function of angular scale $\l$ for the case of no foregrounds. 
Similar plots for the Wiener filtering method can be found in AAO (1998)
and BG99.
}
\end{figure}

\begin{figure}[tb] 
\centerline{\epsfxsize=\idlwidth\epsffile{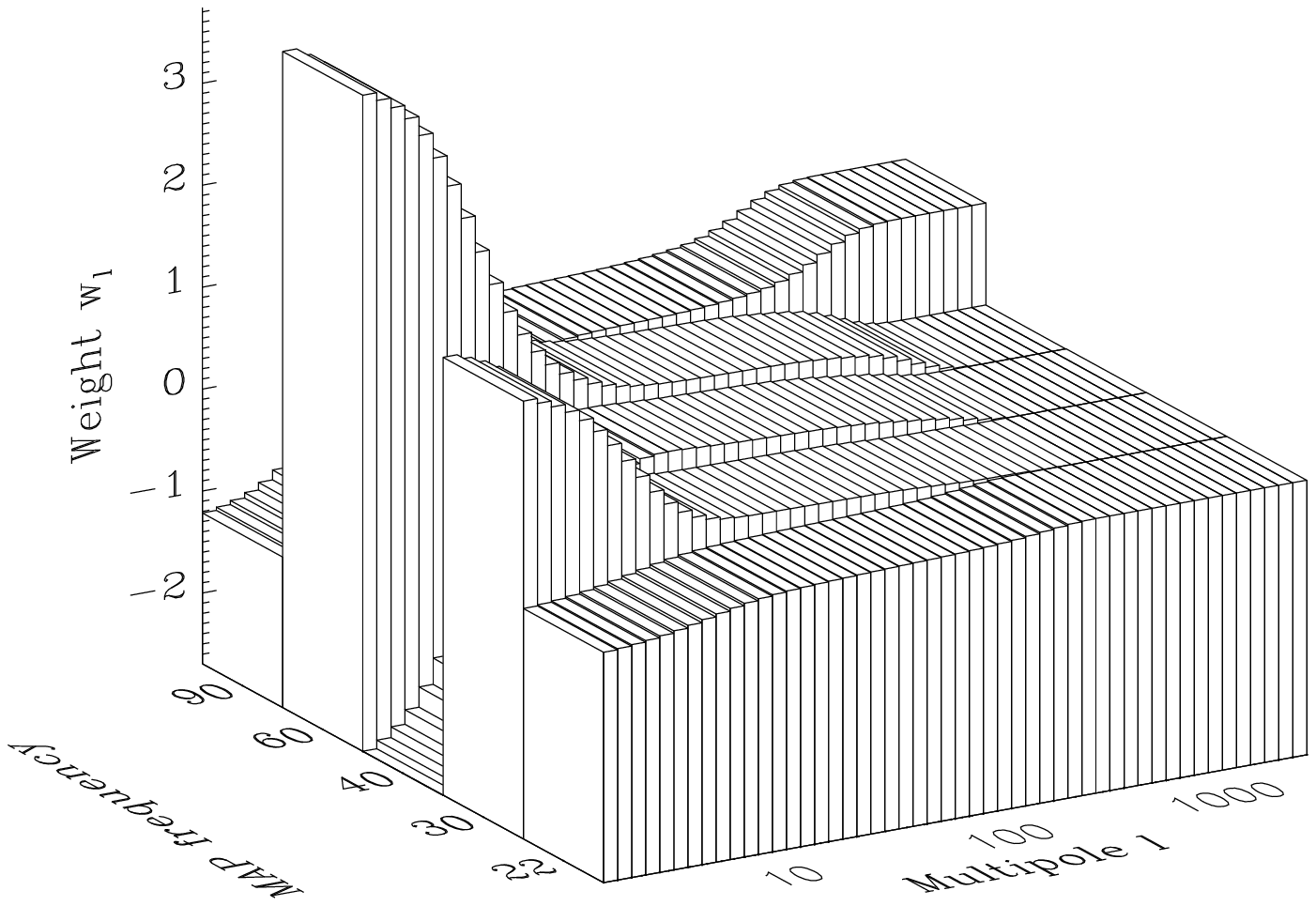}}
\caption{\label{wTfigMAPmid}\footnotesize%
Same as \fig{wTfigMAPno}, but for the MID foreground scenario. 
}
\end{figure}

These weight vectors are plotted for \map\ in 
figures~\ref{wTfigMAPno} and~\ref{wTfigMAPmid},
and some \planck\ examples will be shown in \sec{CoherenceDepSec}.
We have simplified these figures by
using the approximation of
ignoring foreground correlations between the $T$ and $E$ maps.
In other words, we plot the best choice of weighting 
satisfying $\w_\l^{T'}=\w_\l^{E'}=0$.  It is generally possible to do
slightly better.

A number of features of these figures are easy to interpret.
The foreground-free case of \fig{wTfigMAPno} corresponds to a 
standard minimum-variance weighting of the channels. Although the 
\map\ specifications are such that all five channels are equally
sensitive on large scales, the higher channels get more weight on small
scales because of their superior angular resolution.
Although \fig{wTfigMAPmid} shows that things get more complicated in
the presence of foregrounds, we recover this familiar inverse-variance 
weighting in the limit where foregrounds are less of a headache 
than detector noise, here for $\l\simgt 300$.
On angular scales where foregrounds constitute a major problem,
the weighting scheme works harder to subtract them out: the weights must
still add up to unity, but now some of them go negative and others
become as large as 3.
For instance, large positive weight is given to the  
60 GHz channel on large scales, balanced against a negative weight 
at 90 GHz (to subtract out vibrating dust) and
40 and 22 GHz (to remove synchrotron, free-free and spinning dust emission).
The greater the assumed amplitudes are for the foregrounds,
the more aggressively the cleaning method tries to subtract them
out with large positive and negative weights.
The price for this is of course that the residual detector 
noise becomes larger than for the minimum-variance weighting of
\fig{wTfigMAPno}.

\subsection{The three cleaned maps and their power spectra}

Transforming the cleaned multipoles $\at\lm$ back into real space, 
the final result of this 
foreground subtraction procedure is three cleaned maps of the CMB:
one intensity map $T$ and two polarization maps $E$ and $B$.
\Eq{ComboCovarEq} gives a $3\times 3$ covariance matrix of the form
\beq{ComboCovarEq2}
\Sig_\l=[\calA^t\calN_\l^{-1}\calA]^{-1}=\calW_\l^t \calN_\l \calW_\l = 
\left(\bs\begin{tabular}{ccc}
$\Nt^T$	&$\Nt^X$ &$0$\\[2pt]
$\Nt^X$	&$\Nt^E$ &$0$\\[2pt]
$0$	&$0$     &$\Nt^B$
\end{tabular}\bs\right),
\eeq
where 
$\Nt^T$, $\Nt^E$ and $\Nt^B$ are the cleaned power spectra
of the non-cosmic signals in the $T$, $E$ and $B$ maps,
and $\Nt^X$ is the cross-correlation between $T$ and $E$. 
These four power spectra are plotted in the rightmost panels of
figures~\ref{SynchFig}--\ref{psFig} for the cleaned \boom,
\map\ and \planck\ maps.

Note that although the CMB power spectrum emerges unscathed from
the map merging process (since the weights were always 
normalized to add up to unity), the input power spectra of the various
foregrounds generally get their shape distorted ($N_\l^P \ne \Nt^P$).
This is because the weighting is different for each $\l$-value, typically
suppressing foregrounds by a greater factor on those angular scales where
they are large and damaging than on scales where they are fairly negligible.
Indeed, the
rightmost 3 panels of Figures \ref{SynchFig}--\ref{psFig} show 
that rather complex power 
spectrum features can become imprinted on the least important 
foregrounds, as the need to subtract out more important foregrounds
shifts the relative channel weights around.

\subsection{Power spectrum error bars}

How accurately can we measure the four CMB power spectra
from these three cleaned maps? 
If we parameterize our cosmological model directly in terms
of the CMB power spectrum coefficients, \ie,
% $\vp_\l=(C_{\l{\rm(CMB)}}^T$, $C_{\l{\rm(CMB)}}^E$,
% $C_{\l{\rm(CMB)}}^B$, $C_{\l{\rm(CMB)}}^X)$,
\beq{plDefEq}
\vp_\l\equiv (C_{\l{\rm(CMB)}}^T, C_{\l{\rm(CMB)}}^E,
C_{\l{\rm(CMB)}}^B, C_{\l{\rm(CMB)}}^X),
\eeq
we can answer this question by computing the corresponding 
$4\times 4$ Fisher matrix $\F_\l$. 
Our measurement
$\xt_{\l m}$ of the 3-dimensional multipole vector $\x_{\l m}$
from \eq{GroupingEq2} has a covariance matrix
\beq{ComboCovarEq4}
\bCt_\l=\expec{\xt_{\l m}^*\xt_{\l m}^t} = 
\left(\bs\begin{tabular}{ccc}
$\Ct^T$	&$\Ct^X$ &$0$\\[2pt]
$\Ct^X$	&$\Ct^E$ &$0$\\[2pt]
$0$	&$0$     &$\Ct^B$
\end{tabular}\bs\right).
\eeq
Here $\Ct^T$, $\Ct^E$, $\Ct^B$ and $\Ct^X$ are the total
power spectra in the cleaned maps,
combining the contributions from CMB, detector noise and foregrounds,
\eg, $\Ct^P = C_{\l{\rm(CMB)}}^P + \Nt^P$.
Since $\xt_{\l m}$ is by assumption Gaussian-distributed, 
our sought-for $4\times 4$ Fisher matrix $\F_\l$ is
given by
\beq{GaussFisherEq3}
\F_{\l PP'} = {1\over 2}\>\tr
\left[\bCt^{-1}{\partial\bCt\over\partial C_{\l{\rm(CMB)}}^P}\bCt^{-1}{\partial\bCt\over\partial C_{\l{\rm(CMB)}}^{P'}}\right],
\eeq
which after some algebra reduces to 
% \beq{PowerFisherEq}
% \F_\l={1\over D_\l^2}
% \left(\bs\begin{tabular}{cccc}
% ${1\over 2}(\Ct^E)^2$	&${1\over 2}(\Ct^X)^2$ 	&0		&$-\Ct^E\Ct^X$\\[2pt]
% ${1\over 2}(\Ct^X)^2$	&${1\over 2}(\Ct^T)^2$ 	&0		&$-\Ct^T\Ct^X$\\[2pt]
% $0$		&$0$		&${D_\l^2\over(\Ct^B)^2}$	&$0$\\[2pt]
% $-\Ct^E\Ct^X$	&$-\Ct^T\Ct^X$	&0 		&$(\Ct^X)^2+\Ct^T \Ct^E$
% \end{tabular}\bs\right),
% \eeq
\beq{PowerFisherEq}
\F_\l={1\over D_\l^2}
\left(\bs\begin{tabular}{cccc}
${1\over 2}E_\l^2$	&${1\over 2}X_\l^2$ 	&0		&$-E_\l X_\l$ \\[2pt]
${1\over 2}X_\l^2$	&${1\over 2}T_\l^2$ 	&0		&$-T_\l X_\l$\\[2pt]
$0$		&$0$		&$\displaystyle{D_\l^2 \over 2 B_\l^{2}}$	&$0$\\[2pt]
$-E_\l X_\l$	&$-T_\l X_\l$	&0 		&$T_\l E_\l + X_\l^2$
\end{tabular}\bs\right),
\eeq
where $D_\l\equiv T_\l E_\l - X_\l^2$.
We have used the shorthand convention
$P_\ell = \tilde C_\ell^P$ here (and only here) for space reasons.
This is the information content in a single multipole $\x_{\l m}$.
Since we have effectively have $(2\l+1)\fsky$ independent 
modes$^{\ref{fskyFootnote}}$
that each measure the four power spectrum
coefficients in $\vp_\ell$, the 
full $\F_\l$ is $(2\l+1)\fsky$ times that 
given by \eq{PowerFisherEq}.
Inverting this matrix gives the best attainable covariance matrix 
$\M$ for our
4-vector $\vp_\ell$ of measured power spectra:
%\fsky^{-1}(2\l+1)^{-1} \F_\l^{-1}$
\beq{PowerCovEq3} 
\M = \fsky^{-1}(2\l+1)^{-1} \F_\l^{-1},
\eeq
where  
% \beq{PowerCovEq}
% \M_\l={2\fsky^{-1}\over 2\l+1}
% \left(\bs\begin{tabular}{cccc}
% $(\Ct^T)^2$	&$(\Ct^X)^2$ & 0	&$\Ct^T\Ct^X$\\[2pt]
% $(\Ct^X)^2$	&$(\Ct^E)^2$ & 0	&$\Ct^E\Ct^X$\\[2pt]
% $0$		&$0$		&$(\Ct^B)^2$	&$0$\\[2pt]
% $\Ct^T \Ct^X$	&$\Ct^E \Ct^X$	& 0 	&$[(\Ct^X)^2+\Ct^T \Ct^E]/2$
% \end{tabular}\bs\right)
% \eeq
\beq{PowerCovEq}
% \M_\l={2\fsky^{-1}\over 2\l+1}
\F_\l^{-1} = 2 
\left(\bs\begin{tabular}{cccc}
$T_\l^2$	&$X_\l^2$ & 0	&$T_\l X_\l$\\[2pt]
$X_\l^2$	&$E_\l^2$ & 0	&$E_\l X_\l$\\[2pt]
$0$		&$0$		&$B_\l^2$	&$0$\\[2pt]
$T_\l X_\l$	&$E_\l X_\l$	& 0 	&${1 \over 2}[T_\l E_\l+X_\l^2]$
\end{tabular}\bs\right)
\eeq
Zaldarriaga {\etal} (1997) showed that this same covariance 
matrix
%%% \footnote{
%%% { Useless factoid off the day, pending deletion:}
%%% In terms of the dimensionless correlation coefficient between 
%%% intensity and $E$-polarization,
%%% $r\equiv C_X/\sqrt{C_T C_E}$, the correlation
%%% matrix corresponding to $\M$ is simply
%%% \beq{PowerCovEq3}
%%% {\M_{ij}\over\sqrt{\M_{ii}\M_{jj}}}= 
%%% \left(\bs\begin{tabular}{cccc}
%%% $1$	&$r^2$	&$0$	&$\rho$\\
%%% $r^2$	&$1$	&$0$	&$\rho$\\
%%% $0$	&$0$	&$1$	&$0$\\
%%% $\rho$	&$\rho$	&$0$	&$1$\\
%%% \end{tabular}\bs\right),
%%% \eeq
%%% where $\rho\equiv[(1+r^{-2})/2]^{-1/2}$.
%%% The determinant is $(1+r^4)(1-r^2)/(1+r^2)$, so $\M$ is nonsingular
%%% as long as $|r|<1$.
%%% }
was actually obtained when measuring the power
spectrum in the maps in the usual way, with estimators
$(2\l+1)^{-1}\sum_{m=-\l}^\l|a_{\l m}|^2$, which demonstrates that 
this method retains all the 
power spectrum information available.

The analogous derivation of the Fisher matrix for other parameters 
upon which the CMB power spectrum depends,
say a parameter vector $\vp'$ of cosmological 
parameters ($h$, $\Omega_b$, \etc), shows that it can
be expressed in terms of this matrix:
\beq{SimpleFisherEq}
\F_{ij} = \sum_\l (2\l+1)  \fsky
\left({\partial\vp\over\partial p'_i}\right)^t 
\F_\l
\left({\partial\vp\over\partial p'_j}\right).
\eeq

\begin{figure}[tb] 
\vskip-0.5cm
\centerline{\epsfxsize=15cm\epsffile{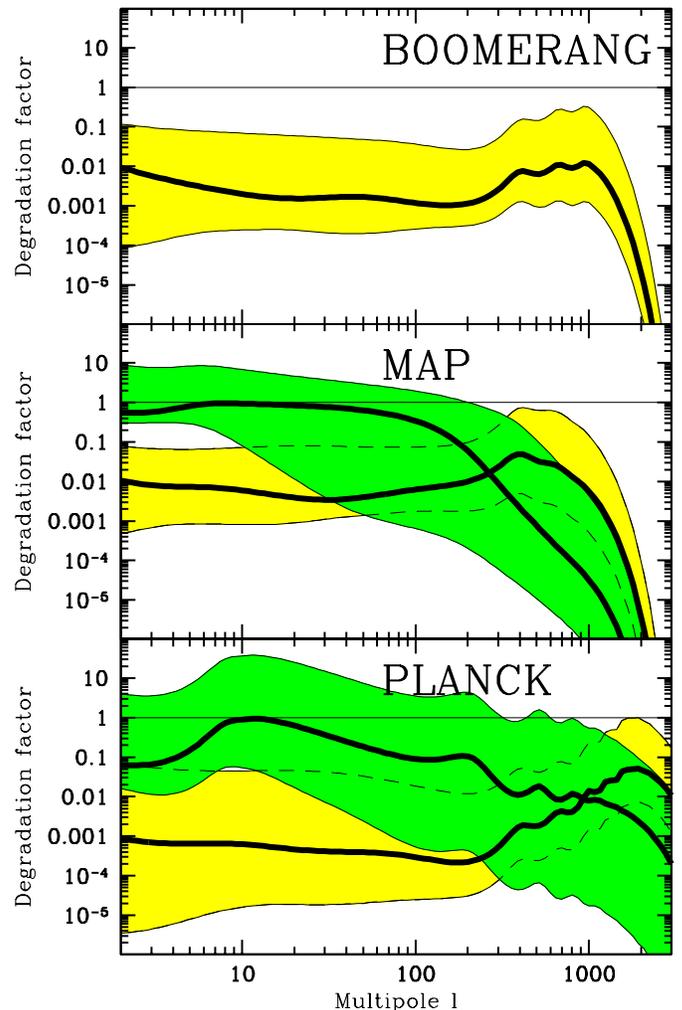}}
\vskip-1cm
\caption{\label{DegradationFig}\footnotesize%
Degradation factors.
The fraction $\DF-1$ by which foregrounds increase the power
spectrum error bars is shown for \boom\ (top), 
\map\ (middle) and \planck\ (bottom).
Each shaded band shows the range of uncertainty between 
the PESS and OPT models, with the MID case 
indicated by a heavy curve. The lighter band is for
intensity $T$, the darker one for $E$ polarization.
}
\end{figure}

The variance of a measured power spectrum coefficient $C_\l^P$ is
given by the corresponding diagonal element of \Eq{PowerCovEq3}, 
so the error bars take the particularly simple form
$\Delta C_\l^P = [(2\l+1)\fsky/2]^{-1/2} {\tilde C}_\l^P$.
Let us define the {\it degradation factor} $\DF$ as 
the factor by which these error bars increase in the presence of foregrounds.
For the $T$, $E$ and $B$ cases, we have $\Delta C_\l^P\propto C_\l^P$, 
so this factor becomes simply
\beqa{FDFeq}
\DF^P 
&\equiv& {\Delta\Cl^P\over\Delta\Cl^P|_{\hbox{no foreg}}} 
= {\Ct^P\over\Ct^P|_{\hbox{no foreg}}}\nonumber\\
&=& 1 + {{\tilde C}_{\l({\rm foreg})}^P\over 
C_{\l({\rm cmb})}^P+{\tilde C}_{\l({\rm noise})}^P},
\eeqa
where 
${\tilde C}_{\l({\rm foreg})}^P$ is the sum of the powers
${\tilde C}_{\l\kk}^P$ of all foreground components.
The expression becomes more complicated for the $X$ case,
where $(\Delta C_\l^X)^2 = [(\Ct^X)^2+\Ct^T \Ct^E]/2$.
These degradation factors are plotted in \fig{DegradationFig}
for the $T$ and $E$ cases, for \boom, \map\ and \planck\ and our 
PESS, MID and OPT scenarios.
Here and throughout, we use the $\Lambda$CDM cosmology presented
in \sec{CosmologyDef}.

For the $T$ case, we see that foregrounds never increase error
bars by more than a factor of $10\%$ in the MID scenario and
2 in the PESS scenario.
For $E$, the MID foregrounds never cost more than a factor of two,
whereas the PESS case degrades \planck\ (which has the most to lose because
of its high sensitivity) about twenty-fold at $\l\sim 10$.
Since noise is negligible at low $\l$, the foregrounds are competing only
with sample variance here, and so the $E$ degradation is caused by 
the polarized
foreground power being substantial compared to the CMB power.
Since detector noise always dominates at high $\l$, the degradation asymptotically 
goes away as $\l\to\infty$. For the unpolarized case, the degradation
is seen to be worst between these two limits, around the beam scale of 
each experiment, where point-source power can become comparable
to both CMB and noise.

Two other foreground degradation measures have been previously 
used in the literature.
The most closely related one is the {\it foreground degradation factor}
of Dodelson (1996), which is the ratio of the {\rms} noise in the 
cleaned maps for the cases with and without foregrounds.
This assumed that foregrounds could be subtracted out completely,
\ie, that $\Delta\alpha=0$ and that there were more components
channels than foregrounds.
The ``quality factor'' (Bouchet {\etal} 1999; BG99)
is the amount by which multifrequency Wiener filtering suppresses
the power of the CMB in the cleaned map, assuming $\Delta\alpha=0$,
and was defined for each foreground component separately.
The most important difference is that our degradation factor is
relative
to the noise {\it plus} sample variance, 
since our focus is on power spectra and 
measurement of cosmological parameters.

\subsection{Dependence on assumptions about amplitude}

The MID model in Figure \ref{DegradationFig} gives our
estimate of how small the noise and foreground power spectra can be made
in the cleaned maps, while comparing the OPT and PESS models 
indicates the range of uncertainty.
Let us now discuss the effects of model assumptions in more detail.

The foreground behavior enters in two different ways:
\begin{enumerate}
\item The {\it assumed} foreground behavior determines 
the weights $\w$ that we use when cleaning the maps.
\item The {\it true} foreground behavior determines 
the actual foreground residual in the cleaned maps.
\end{enumerate}
Let us make this distinction explicit by using 
$\calNa$ to denote our assumed foreground matrix (our prior),
as distinguished from the true matrix $\calN$.
The resulting foreground contamination is then given by
(suppressing the $\l$ subscript)
\beq{ComboCovarEq3}
\Sig=[\calA^t\calNa^{-1}\calA]^{-1}[\calA^t\calNa^{-1}\calN\calNa^{-1}\calA]
[\calA^t\calNa^{-1}\calA]^{-1},
\eeq
which only reduces to \eq{ComboCovarEq2} if $\calNa=\calN$, \ie, if our
model is correct. We will still recover unbiased CMB maps
$T$, $E$ and $B$ even if our model is incorrect, but generally with
larger foreground contamination than would be attainable with
a correct model.

\Eq{ComboCovarEq3} shows that the contaminant power spectra in $\Sig$ depend
linearly on $\calN$.  Thus the complicated residual foreground power spectra 
depicted in
the 3 rightmost panels of Figures \ref{SynchFig}--\ref{psFig} 
(computed \eq{ComboCovarEq3} by taking $\calN$ to be the 
contribution from a single foreground component)
may be thought of as the result of multiplying the rather featureless 
foreground power spectra that are actually on the sky
by {\it known} transfer functions.

%This does of course not cause problems for spotting 
%residual foregrounds, since the ``transfer function'' by which 
%each foreground power spectrum gets modulated is determined 
%by the weights $\w_\l$ and is therefore fully known.
%Since $\calN$ can be written as a sum of contributions
%from noise and the individual foregrounds, these power spectra
%can be similarly decomposed, as shown in the
%rightmost panels of Figures \ref{SynchFig}--\ref{psFig}.
%If the amplitude of a given foreground exceeds that in the model by some 
%factor, the residual in the cleaned map will simply be larger by the same
%factor.
%
%Note that although the CMB power spectrum emerges unscathed from
%the map merging process (since the weights were always 
%normalized to add up to unity), the input power spectra of the various
%foregrounds generally get their shape distorted.
%This is because the weighting is different for each $\l$-value, typically
%suppressing foregrounds by a greater factor on those angular scales where
%they are large and damaging than on scales where they are fairly negligible.
%Indeed, the
%rightmost panels of Figures \ref{SynchFig}--\ref{psFig} show 
%that rather complex power 
%spectrum features can become imprinted on the least important 
%foregrounds, as the need to subtract out more important foregrounds
%shifts the relative channel weights around.
%This does of course not cause problems for spotting 
%residual foregrounds, since the ``transfer function'' by which 
%each foreground power spectrum is modulated is determined 
%by the weights $\w_\l$ and is therefore fully known.
%

\subsection{Dependence on assumptions about frequency coherence}

\subsubsection{Effect of changing $\Da$}
\label{CoherenceDepSec}

\begin{figure}[tb] 
\vskip-1cm
\centerline{\epsfxsize=\smwidth\epsffile{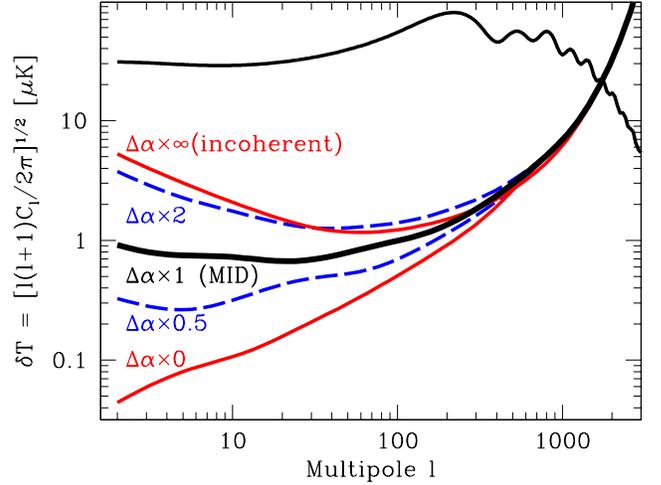}}
\vskip-1cm
\caption{\label{CoherenceFig1}\footnotesize%
The effect of frequency coherence. 
The total power spectrum from noise and foregrounds in the
cleaned \planck\ $T$ map is shown for five different assumptions about
frequency coherence, corresponding to multiplying
all values of $\Da$ from the MID model (heavy curve) by
$\infty$, 2, 1, 0.5 and 0, respectively (from top to bottom).
The fiducial CMB power spectrum is shown for comparison.
}
\end{figure}

\begin{figure}[tb] 
\centerline{\epsfxsize=\idlwidth\epsffile{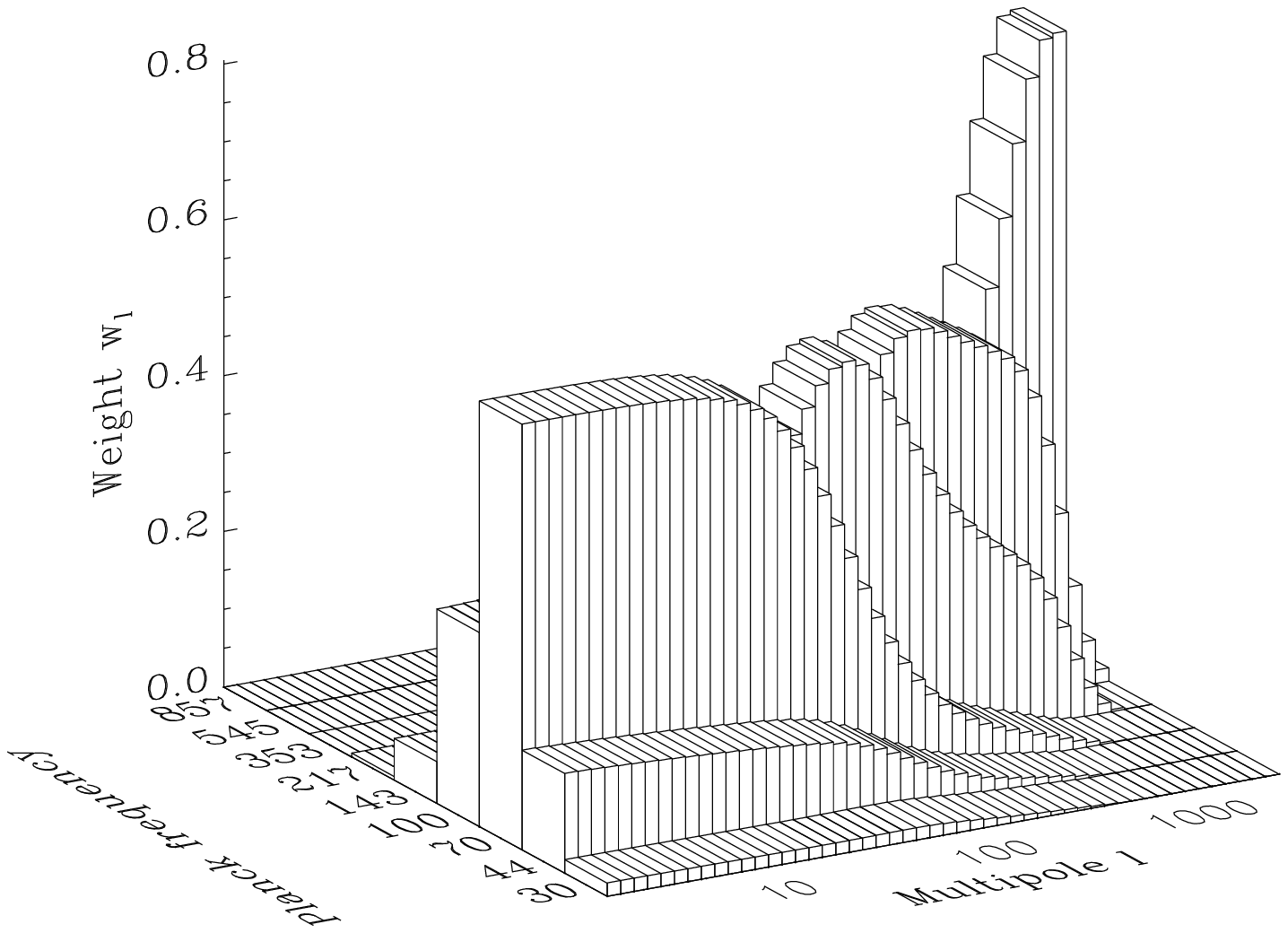}}
\caption{\label{wTfigPlanckInc}\footnotesize%
The weights $\w_\l^T$ with which the unpolarized maps at
the 9 \planck\   
frequencies are combined into a single map are plotted as
a function of angular scale $\l$ for the MID model,
but with completely incoherent foregrounds ($\Delta\alpha=\infty$). 
}
\end{figure}

\begin{figure}[tb] 
\centerline{\epsfxsize=\idlwidth\epsffile{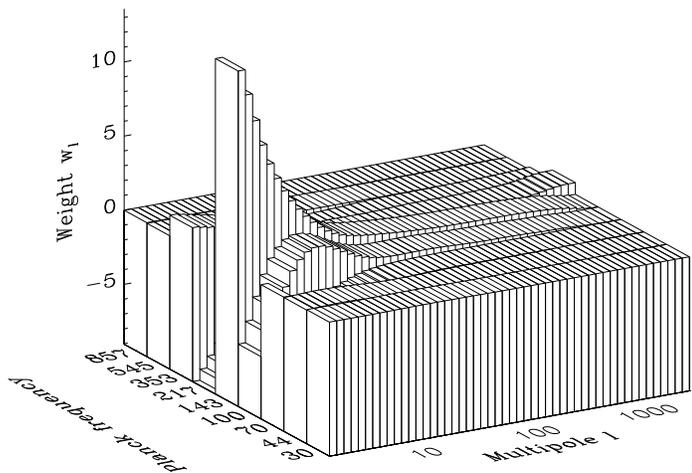}}
\caption{\label{wTfigPlanckMID}\footnotesize%
The weights $\w_\l^T$ with which the unpolarized maps at
the 9 \planck\   
frequencies are combined into a single map are plotted as
a function of angular scale $\l$ for the MID foreground model.
}
\end{figure}

\begin{figure}[tb] 
\centerline{\epsfxsize=\idlwidth\epsffile{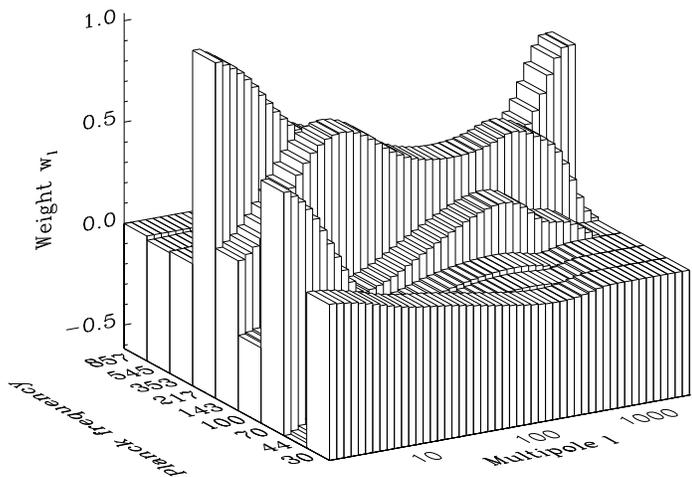}}
\caption{\label{wTfigPlanckCoh}\footnotesize%
The weights $\w_\l^T$ with which the unpolarized maps at
the 9 \planck\   
frequencies are combined into a single map are plotted as
a function of angular scale $\l$ for the MID model,
but with perfectly coherent foregrounds ($\Delta\alpha=0$). 
}
\end{figure}

In general, the less coherent a foreground is, the more difficult it is to 
remove.
\Fig{CoherenceFig1} shows this effect.
All panels use the scale and frequency dependence of the MID model,
but with the frequency coherence spanning the range between
the extreme cases $\coh=\infty$ and $\coh=0$.
As expected, the situation generally gets worse as we progress
from ideal perfectly coherent foregrounds (bottom curve) to 
realistic (middle three curves) and completely incoherent ones (top curve),
The incoherent case corresponds to no foreground subtraction whatsoever, 
simply averaging the \planck\ channels with inverse-variance weighting.

While less coherence is usually a bad thing, 
\fig{CoherenceFig1} shows a subtle exception to this rule at 
$\l\sim 100$. 
Here weak coherence is seen to be worse than no coherence at all.
Figures~\ref{wTfigPlanckInc}-\ref{wTfigPlanckCoh}
shed more light on this perhaps surprising behavior
by showing how the channel weighting changes as we 
increase the frequency coherence.
These figures correspond to three of the five curves
in \fig{CoherenceFig1} (the top, middle and bottom ones).
\fig{wTfigPlanckInc} shows the case of completely incoherent foregrounds
($\Delta\alpha=\infty$). It gives an inverse-variance weighting just
as in \fig{wTfigMAPno}, but with the variance receiving a contribution
from foregrounds as well as noise.
In \Fig{wTfigPlanckMID}, we see that the poor coherence between
widely separated channels is forcing the method to do much of the
foreground subtraction using neighboring channels, using
costly large-amplitude weights at 100, 143 and 217 GHz on large
scales. In contrast, the case of ideal foregrounds shown in
\fig{wTfigPlanckCoh} is seen to be much easier to deal with, requiring
no weights exceeding unity in amplitude.
For instance, the small dust contribution at low frequencies
can be subtracted out essentially for free, 
by a tiny negative weight for the dust-dominated 545 GHz channel.

The non-monotonic behavior (where things eventually start improving again 
when the coherence becomes sufficiently low)
does not occur if there
is merely a single foreground component present with no detector noise.
Instead, it results from an interplay between foregrounds and noise.
A perfectly incoherent foreground can be efficiently dealt with 
in the same way as noise:
by inverse-variance weighting the channels as in \fig{wTfigPlanckInc}, 
the incoherent foreground 
fluctuations will average down, whereas a coherent foreground would not.
Typically, the worst case is for $\coh$ of order unity, 
although the exact value depends on the
other foregrounds. 
\Fig{CoherenceFig1} thus shows that although we do not appear to live
in the worst of all possible worlds, we are only off by a
small factor!

\subsubsection{Effect of incorrect assumptions about $\Da$}

\begin{figure}[tb] 
\vskip-1cm
\centerline{\epsfxsize=\smwidth\epsffile{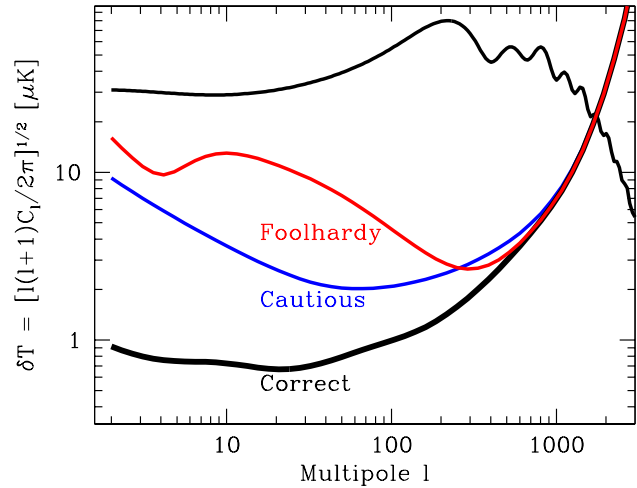}}
\vskip-1cm
\caption{\label{CoherenceFig2}\footnotesize%
The effect of faulty assumptions about frequency coherence.
The total power spectrum from noise and foregrounds in the
cleaned \planck\ $T$ map is shown for the MID model using 
three different assumptions for the cleaning process: 
the correct (MID) $\Da$-values (bottom), 
$\Da=\infty$ (middle, cautious) and
$\Da=0$ (top, foolhardy).
The fiducial CMB power spectrum is shown for comparison.
}
\end{figure}

What if our model is incorrect?
\Fig{CoherenceFig2} uses \eq{ComboCovarEq3} to show the effect of 
two kinds of errors: being too optimistic and being too pessimistic
about ones abilities to model the frequency dependence of foregrounds.
For all there curves, the MID model is used as the truth, but the 
weights $\w$ for the foregrounds subtraction are for different assumptions
about the frequency coherence.
Not surprisingly, correct assumptions give the best 
removal, showing the importance of accurately measuring the
actual frequency coherence of foregrounds.
The curve labeled ``cautious'' shows the
result of assuming incoherent foregrounds ($\Da=\infty$), 
corresponding 
to no foreground subtraction at all, merely inverse-variance averaging with
no negative weights as in \fig{wTfigPlanckInc})
whereas the one labeled ``foolhardy'' illustrates the effect
of assuming ideal foregrounds ($\Da=0$, using the weights of
\fig{wTfigPlanckCoh}).
The fact that the former generally lies beneath the latter shows that 
when faced with uncertainty about $\Da$, 
it is better to err on the side of caution: in our example,
foreground removal based on the overly optimistic model 
is performing worse than no foreground removal
at all.

\subsubsection{Effect of functional form of coherence}

\begin{figure}[tb] 
\centerline{\epsfxsize=\smwidth\epsffile{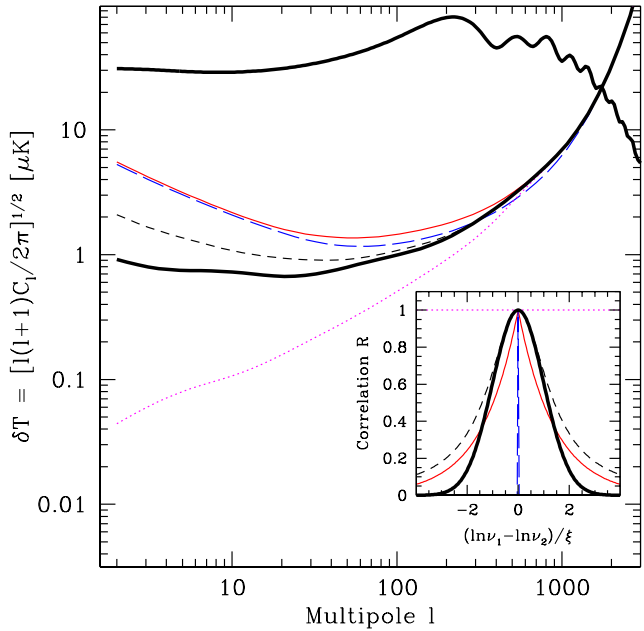}}
\caption{\label{CoherenceFig3}\footnotesize%
Changing the functional form of frequency coherence.
The total power spectrum from noise and foregrounds in the
cleaned \planck\ $T$ map is shown for the MID model using 
different shapes for the coherence function $f$, 
shown with the corresponding line type in the inset.
The shapes are Gaussian (heavy solid curve), 
Lorentzian (short-dashed), exponential (thin solid curve),
flat (dotted) and completely incoherent (long-dashed).
The fiducial CMB power spectrum is shown for comparison.
}
\end{figure}

We can rewrite our coherence model of \eq{RmodelEq} as 
\beq{RmodelEq2}
\R_\kk^{P\freq\freqq} 
= f\left(\ln\nu_\freq-\ln\nu_\freqq\over\coh_\kk^P\right),
\eeq
where $f(x)=e^{-x^2/2}$.
The derivation in T98 did not show that $f(x)$ was 
a Gaussian, merely that behaved like a parabola 
at the origin with $f(0)=1$, $f'(x)=0$ and $f''(0)=-1$.
Since $f$ (which we will 
term the coherence function)
has not yet had its shape accurately measured for any foreground,
we repeat the Planck analysis for a variety of such functions of 
the form 
\beq{RmodelEq3}
f(x) = \left(1+{x^2\over 2n}\right)^{-n}.
\eeq
The case $n=\infty$ recovers the Gaussian of \eq{RmodelEq}, 
$n=1$ gives a Lorentzian, {\etc}

Figure \ref{CoherenceFig3} shows
that the shape of the far wings of $f$ is only of 
secondary importance---the main question is
how correlated neighboring channels are, 
which for $\coh\gg 1$ depends mainly on
the curvature of $f$ near the origin. Narrowing 
the wings (increasing $n$) usually helps slightly,
once again demonstrating that more coherence is not necessarily 
a good thing.
For comparison, \Fig{CoherenceFig3} shows the 
case where $f(x)=1$ at the origin and vanishes elsewhere 
and the case $f(x)=1$, corresponding to the limits
$\Da=\infty$ and $\Da=0$, respectively.
Also shown is the 
exponential coherence function $f(x) = \exp[|x|/\sqrt(2)$.
This is seen to be quite a conservative choice, giving
even larger residuals than the $\Da=\infty$ case, 
since the correlations between neighboring channels are 
strong enough to be important but not good enough to 
be really useful for foreground subtraction.
A generalization of this exponential coherence function
will come in handy in \sec{JointSec}, where our goal is 
to be as pessimistic as possible with the intent
to destroy parameter estimation with foregrounds.

%%% %\subsection{Leftover snippet}
%%% %
%%% %In addition to using information that is literally prior 
%%% %(the DIRBE and HASLAM maps, results from earlier CMB experiments, etc)
%%% %to construct prior foreground model, a wealth of useful information
%%% %about foregrounds will of course be contained in the upcoming 
%%% %CMB data sets themselves. We will return to the issue of how this 
%%% %information can be used in \sec{JointSec}.

%%%%%%%%%%%%%%%%%%%%%%%%
%\input daniel.tex
%%%%%%%%%%%%%%%%%%%%%%%%

%%% Max -- The commented definitions are not presently in use, but
%%% the uncommented ones are.  You may need to check for conflicts.
%%% I suggest that you use onecolfloat.sty; it allows proper floats in
%%% emulateapj.  I think that you can include this document in your
%%% TeX file by simply doing \input{name.tex}. /Daniel

%%% Daniel - I've moved all the definitions to the
%%% beginning, removed some unused ones and 
%%% switched "o" to "O" for cosmological parameter
%%% macros involving \Omega rather than \omega, 
%%% to avoid colliding with the defs in the rest of
%%% the paper. /Max

\section{Simultaneous Estimation of Foregrounds and Cosmology}
\label{JointSec}

To this point, we have considered the case in which the statistical
properties of the foregrounds are known exactly. 
Moreover we have shown that incorrect assumptions about these 
properties can lead to foreground removal strategies that do more harm
than good. 
This begs the
question of whether we will in fact know these foregrounds to the
level needed for accurate subtraction.  The sky maps from CMB
satellite missions will provide some of the most relevant data
on this question.  Hence, we will next consider the case in which
cosmological parameters and the foreground model are simultaneously
estimated from the CMB data.

At this point, the calculation becomes less well-defined.  If we
are allowed to consider arbitrary excursions around the fiducial
model, then cosmological parameter estimation fails completely.
There is no mathematical way to exclude a foreground that matches
the CMB frequency dependence, is perfectly coherent, and has
an arbitrarily inconvenient power spectrum (say, one mimicking the
variation of a cosmological parameter).  Physically,
however, we believe this to be unreasonable.  We therefore must
construct a parameterized model of foregrounds that allows for
a reasonably, but not completely, general coverage of the 
possibilities.

\subsection{Cosmological Parameters}
\label{CosmologyDef}

We adopt a low-density, spatially flat adiabatic CDM model for our cosmology.
The model has a matter density of $\Omega_m=0.35$,
a baryon density of $\Ob=0.05$, a massive neutrino density
of $\On=0.0175$ (one massive species with a mass of $0.7\eV$), and
a cosmological constant $\Ol=0.65$.  The primordial
helium fraction is $Y_P=0.24$.
The Hubble constant is $H_0 = 100h\kmsmpc = 65\kmsmpc$.
The universe is reionized suddenly at low redshift with an optical depth 
of $\tau=0.05$.  The primordial power spectrum is scale-invariant,
so the scalar spectral index is $\ns=1$.  There are no
tensors ($\ts=0$).  The model is normalized to the COBE-DMR experiment.
This is the same model that was used in 
Eisenstein {\etal} (1999, hereafter ``E99''), 
and further details on the above choices can be found
there.

We will ask how well CMB data can constrain a 10-dimensional 
excursion around this parameter space.  The parameters are 
$\Omhh$, $\Obhh$, $\Onhh$, $\Ol$, $\tau$, $\yp$ (constrained to
vary by 0.02 at 1-$\sigma$), $\ns$, $\ns'$, $\ts$, and 
the normalization.  $ns'$ (denoted $\alpha$ in E99) 
is the running of the scalar tilt,
\beeq
\ns(k) = \ns(\kfid)+\ns'\ln(k/\kfid),
\eeq
where $\kfid=0.025\impc$.  Note that $\Om=1-\Ol$ and 
$h=\sqrt{(\Omhh)/(1-\Ol)}$ 
are defined implicitly and hence can vary.  
This parameter space is identical
to that of E99 except that we have not included spatial
curvature.  There is a severe degeneracy between curvature 
and the cosmological constant
(Bond {\etal} 1997; Zaldarriaga {\etal} 1997).
Since this degeneracy is 
best broken by using non-CMB data 
(\eg, galaxy redshift surveys or SN Ia), 
we choose to focus only on the well-constrained combination
of the two parameters here.
Details of how we perform the numerical derivatives of the power
spectra with respect to these cosmological parameters can be found in
E99.  Fortunately, all the derivatives with respect to 
the foregrounds (\sec{JointSec}) can be done analytically,
so that no new numerical problems are introduced.

\subsection{Foreground Parameters}

We allow for uncertainty in the foregrounds by adding a large 
number of parameters to the models specified in \sec{ModelSec1}.  
Recall that
each component was specified by a frequency dependence, a frequency
coherence, and a power spectrum for each polarization type 
($T$, $E$, $B$, and $X$).  
For each type and each component, we now include parameters to
allow excursions in frequency dependence, frequency coherence, 
and spatial power.
As described below, we use of order fifty additional parameters
denoted by vectors $\bfq$, $\bfr$ and $\bfs$
for each foreground component.  

\begin{figure}[tb] 
\vskip-1.1cm
\centerline{\epsfxsize=\smwidth\epsffile{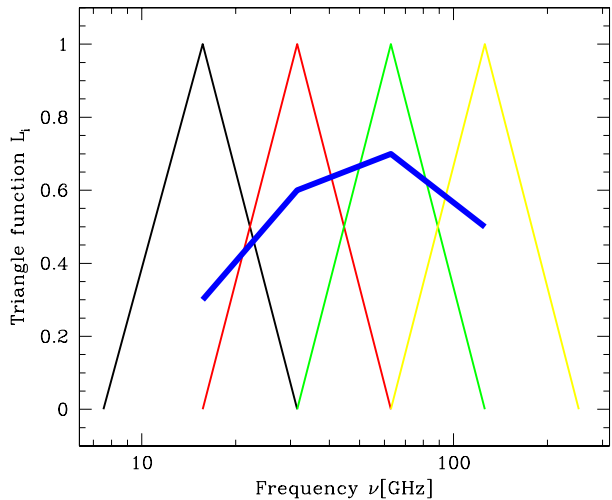}}
\vskip-1.0cm
\caption{\label{TriangleFig}\footnotesize%
The figure shows the four triangle functions $L_i$ that span
the excursions in frequency dependence for foregrounds seen
by MAP. Any function that is piecewise linear between 
the four break points (e.g., the heavy curve)
can be written as a linear combination of these functions
in this range.
}
\end{figure}

For the frequency dependence, we allow a piecewise power-law excursion
in thermodynamic temperature around the fiducial model
\beq{ThetaExcursion}
\ln \Theta_{(k)}^P (\nu) = \left. \ln \Theta_{(k)}^P(\nu) \right|_{\rm fid} 
	+ L\left(\ln\nu; \bfq_{(k)}^P\right).
\eeq
Here, the function $L$ is a linear interpolation between the values 
$(\bfq=q_1,\ldots,q_{n_\nu})_{(k)}^P$
at the breakpoints $\nu^1,\ldots,\nu^{n_\nu}$, so
$L(\ln\nu^i;\bfq_{(k)}^P)=q_{i(k)}^P$ and is a straight line between breakpoints
and the fiducial model (``fid'') has $\bfq_{(k)}^P=0$.
This means that we can rewrite it as 
\beeq\label{LdecompEq}
L(\ln\nu;\bfq)= \sum_{i=1}^{n_\nu} q_i L_i(\ln\nu),
\eeq
where the functions $L_i$ have triangular shape as
illustrated in \fig{TriangleFig}:
$L_i(\ln\nu^i)=1$, goes linearly to zero at the neighboring 
breakpoints, and vanishes everywhere else.

We allow a separate frequency excursion for $T$, $E$, and $B$
for each foreground component.
For the cross correlation we allow for the possibility that
the excursions in the correlation of the temperature at $\nu$ 
and $E$-polarization at $\nu'$ is not the same as the temperature
at $\nu'$ with the $E$-polarization at $\nu$.
We therefore define two excursions for the cross-correlation 
$\bfC_{\l(k)}^X(\nu,\nu') \propto \Theta_{(k)}^{X_T}(\nu)\Theta_{(k)}^{X_E}(\nu')$.  One of the
excursions affects the temperature index, the other the polarization
index.  $\left. \Theta_{(k)}^{X_T} \right|_{\rm fid} = \Theta_{(k)}^{X}$ and likewise
for ${X_E}$ since
we have assumed that the fiducial model is symmetric in this respect. 

Since the breakpoints $\nu^i$ specify the degrees of freedom in 
the foreground model, they are independent of the number and location
of observing frequencies $\nu_i$ for any given experiment.
We choose them to be evenly
spaced in $\ln\nu$, with a factor of 2 in frequency between each,
and are centered at the geometric mean of the highest and lowest 
frequencies of the experiment.
%%% Because the interpolation is linear, one need not specify any breakpoints
%%% beyond the first outside the frequency range of the experiment.
This means that we specify $n_\nu=4$ breakpoints for 
\boom\ (67.1, 134, 268, and 537 GHz) and 
\map\ (15.7, 31.5, 62.9, and 126 GHz) and 
$n_\nu=6$ for \planck\ (28.3, 56.7, 113, 227, 454, 907 GHz).

In the $n_\nu=2$ limit, the new parameters correspond to varying
the normalization and power-law exponent of the frequency dependence
of the foreground component.  We chose to extend this freedom by
piecewise-linear interpolation rather than smoother options for
technical reasons.  Splines have non-local behavior; frequencies
far outside the range of the CMB experiment would affect the results
inside the range.  Polynomials are scale-free, so that a polynomial
of a given order will adapt itself to the particular experimental
specifications so as to put the maximal number of wiggles inside
the frequency range.  This would mean that the effective 
number of degrees of freedom in the foreground model would
not be consistent from one experiment to the next.  Simple
interpolation avoids these problems: one need not specify any
breakpoints beyond the first outside the frequency range of the
experiment, and the scale for variations in the foreground is
independent of the experiment.

We don't allow frequency variations for the thermal SZ component,
because its spectrum is theoretically known. As discussed in
\sec{SZsec}, relativistic corrections are expected to be negligible
for filaments, so we ignore this complication.

To include these parameters in the Fisher matrix of \eq{GaussFisherEq2},
we must specify the derivatives of $\calC_\ell$ with respect
to the interpolation parameters $\bfq_{(k)}^P$.  Clearly only the
elements of the same component and polarization type are affected.
For $P=T$, $E$, and $B$, the derivatives of that submatrix are
\beeq
\left. {\partial C_{\ell(k)}^{P\freq\freqq}\over \partial q_{i(k)}^P} \right|_{\rm fid} 
%&&=C_{\ell(k)}^{P\freq\freqq} \left.\left[
%{\partial L(\ln\nu_\freq; \bfq_{(k)}^P)\over \partial q_{i(k)}^P} + 
%{\partial L(\ln\nu_\freqq;\bfq_{(k)}^P)\over \partial q_{i(k)}^P} 
%\right]\right|_{{\bfq=0}} \nonumber\\ \nonumber
%%% \left.{\partial L(\ln\freq; \bfq_{(k)}^P)\over \partial q_{i(k)}^P}\right|_{\bfq=0} + 
%%% \left.{\partial L(\ln\freqq; \bfq^{(k)P})\over \partial q_i^{(k)P}}\right|_{\bfq=0} 
%%% \right]_{|{\bfq=0}} \nonumber\\ \nonumber
%&&
= C_{\l(k)}^{P\freq\freqq} \left[L_i(\ln\nu_\freq) + L_i(\ln\nu_\freqq)\right],
\eeq
where we have used \eq{LdecompEq}.
For $P=X$, the derivatives are
\beeqa
{\partial C_{\ell(k)}^{X\freq\freqq}\over \partial q_{i(k)}^{X_T}} = 
C_{\ell(k)}^{X\freq\freqq} L_i(\ln\nu_\freq),\\
{\partial C_{\ell(k)}^{X\freq\freqq}\over \partial q_{i(k)}^{X_E}} = 
C_{\ell(k)}^{X\freq\freqq} L_i(\ln\nu_\freqq) .
\eeqa

For the frequency coherence, we adopt the exponential model for
the matrix $\bfR$ using
\beeq
\label{expcoh}
\bfR^{\freq\freqq} = \exp\left[-\left|\int^{\nu_\freqq}_{\nu_\freq}\,
\Delta(\nu)d(\ln\nu)\right|\right].
\eeq
In the fiducial model, $\Delta(\nu)=\Delta\alpha=1/\sqrt{2}\coh$ 
independent of frequency.
Now we allow excursions of the form
\beeq
\Delta(\nu) = \Delta\alpha + L(\ln\nu; \bfr),
\eeq
where $L$ is a linear interpolation function as before, parameterized 
by $\bfr=r_1,\ldots,r_{n_\nu}$. We use the same spacing
of the interpolation points $\nu^1,\ldots,\nu^{n_\nu}$
as in the frequency dependence case described above. 

As above, for each foreground component except SZ, we allow separate
excursions for each of $T$, $E$, and $B$ and two excursions for $X$.
The derivatives of $\calC$ are ($P=T$, $E$, $B$)
\beeq
{\partial C_{\l(k)}^{P\freq\freqq}\over \partial r_{i(k)}^{P}} = 
- C_{\l(k)}^{P\freq\freqq} \left|\int^{\nu_\freqq}_{\nu_\freq}\,d(\ln\nu)
L_i(\ln\nu)\right|.
\eeq
For the cross-correlation $X$, we allow for an asymmetric excursion
by invoking two excursions of the above form and setting either 
the upper ($\freq<\freqq$) or lower ($\freq>\freqq$) triangle 
elements to zero.

For the spatial power, we consider excursions of the form
\beeq
\ln C_{\ell(k)}^P = \left.\ln C_{\ell(k)}^{P}\right|_{\rm fid} +
L(\ln\ell; \bfs),
\eeq
where $L$ is a linear interpolation function.  The parameters
$\bfs=s_1,\ldots,s_{n_\ell}$ are the values of $L$ at a grid of $\ell$
that begins at $\ell=2$ and increased by factors of $e$.  As
we sum the Fisher contribution to $\ell_{\rm max}=2800$, this
gives $n_\ell=9$ grid points.  However, the overall normalization
(i.e. moving all the $s_i$ the same amount) is degenerate with
an overall shift in the frequency dependence (eq.\ [\ref{ThetaExcursion}]),
so in cases other than the thermal SZ, we hold the middle spatial
$s_j$ ($\l=109$) equal to zero.

Separate excursions are allowed in $P=T$, $E$, $B$, and $X$
for each foreground component, including the SZ effect.
The derivatives of $\calC$ with respect to these parameters are 
\beeq
{\partial \bfC_{\ell(k)}^{P} \over \partial s_{i(k)}^P} = 
\bfC_{\ell(k)}^P L_i(\ln\ell),
\eeq
where all terms involving different $k$ and $P$ are zero, as before.
The $L_i$ spatial functions are defined analogously to \eq{LdecompEq}.

In short, for most foreground components, we have $10n_\nu+4n_\ell-4$
free parameters in $(\bfq,\bfr,\bfs)$.  
For the thermal SZ, we have only $4n_\ell$.  However,
for unpolarized components, the parameters for polarization excursions
have zero derivatives, so we remove them.  This leaves 385 (489) 
parameters for \map\ (\planck) for the MID model, 257 (325) for the OPT
model, and 441 (561) for the PESS model.  For the MID model, 
105 (129) of the parameters are associated with the intensity
anisotropies, so there are 105 parameters for our \boom\ estimates.

We allow these parameters to vary without external priors in almost 
all cases.  As we will show below, the CMB experiments are able 
to constrain the foreground model well enough to extract the cosmic
signal.  The one exception is the thermal SZ effect with the 
\map\ experiment.  The frequency dependence of the SZ is similar
to the cosmic temperature variations for frequencies much below
200 GHz.  If we allow unfettered variations in the SZ spatial power
spectrum, there are significant degradations in the performance
of \map\ on cosmological parameters.  However, these degeneracies 
correspond to very large departures from the fiducial SZ level.
We therefore include a prior (for \map\ only) that the SZ power spectrum cannot
vary by more than a factor of 10 from the fiducial level
(i.e., the parameters $z_i$ cannot exceed 10 at 1-$\sigma$ confidence).
This is an extremely generous prior---numerical simulations 
are surely correct to within a factor of 10 at 68\% confidence---but
it substantially reduces the degradation of the \map\ performance
in the presence of SZ signals.
In detail, for the MID model, \map\ with $T$ alone could measure 
$\Obhh$ to 0.0036 with the SZ variations being omitted, 
0.0037 with the prior described above, and 0.0075 with a prior of $10^6$
on the SZ variations.  For this and other parameters, the prior of
10 removes variations in the SZ as a source of cosmological uncertainty
in the MID model.  The PESS model, with its 10-fold increase in the
fiducial SZ level, has 10-20\% differences between results with a prior of 10
and those with no SZ variations.
\planck\ and \boom\ can control the SZ to better than a factor of 10,
so no prior is applied.

\subsection{Cosmological Parameters in the Presence of Foregrounds}

Because the variations in the foreground model have effects at all $\ell$,
we cannot express the effects of the foregrounds as a simple degradation
of the error bars at each multipole (c.f.\ Fig.\ \ref{DegradationFig}).  
Excursions from
the fiducial model produce changes in frequency and scale dependence that 
can compensate both each other and cosmological signals in complicated ways.  
In other words, with this more complicated foreground model,
one does not recover a cleaned CMB power spectrum as an intermediate 
step of the analysis, but must proceed 
directly to the estimation of the parameters characterizing 
the models for foregrounds and cosmology.
To quantify the effects of the foregrounds, we will therefore simply
compare the final marginalized error bars on cosmological parameters.

\begin{table}[tb]\footnotesize
\caption{\label{tab:foregvar}}
\begin{center}
{\sc Marginalized Errors with Foreground Variations\\}
\begin{tabular}{l\colskip l\colskip ccc}
\tableskip\tableline\tableline\tableskip
& & \multicolumn{3}{c}{Foregrounds}\\
Experiment & Quantity & None & Known & Unknown \\
\tableskip\tableline\tableskip
\boom\ (T)
&  $\ln(\Om h^2)$		 & 0.45 & 1.007 & 1.282 \\
&  $\ln(\Ob h^2)$		 & 0.27 & 1.008 & 1.242 \\
&  $m_\nu$ (eV) $\propto\On h^2$	 & 3.5 & 1.006 &  1.51 \\[\tskip]
&  $\ns(\kfid)$ 		 & 0.30 & 1.007 & 1.203 \\
&  $\Ol$			 & 0.57 & 1.007 & 1.287 \\[\tskip]
&  $\tau$			 & 1.3 & 1.016 &  1.87 \\
&  $\ts$ 			 & 1.2 & 1.007 &  1.52 \\
\tableskip\tableline\tableskip\map\ (T)
&  $\ln(\Om h^2)$		 & 0.20 & 1.027 & 1.393 \\
&  $\ln(\Ob h^2)$		 & 0.12 & 1.027 & 1.453 \\
&  $m_\nu$ (eV) $\propto\On h^2$	 & 0.87 & 1.017 &  1.65 \\[\tskip]
&  $\ns(\kfid)$ 		 & 0.11 & 1.026 & 1.332 \\
&  $\Ol$			 & 0.23 & 1.026 & 1.324 \\[\tskip]
&  $\tau$			 & 0.31 & 1.014 &  1.69 \\
&  $\ts$ 			 & 0.42 & 1.023 & 1.240 \\
\tableskip\tableline\tableskip\map\ (TP)
&  $\ln(\Om h^2)$		 & 0.080 & 1.208 &  1.66 \\
&  $\ln(\Ob h^2)$		 & 0.051 & 1.201 &  2.01 \\
&  $m_\nu$ (eV) $\propto\On h^2$	 & 0.57 & 1.078 &  2.06 \\[\tskip]
&  $\ns(\kfid)$ 		 & 0.041 & 1.264 &  2.63 \\
&  $\Ol$			 & 0.091 & 1.230 &  1.74 \\[\tskip]
&  $\tau$			 & 0.018 &  1.90 &  3.33 \\
&  $\ts$ 			 & 0.16 & 1.309 &  1.86 \\
\tableskip\tableline\tableskip\planck\ (T)
&  $\ln(\Om h^2)$		 & 0.062 & 1.014 & 1.042 \\
&  $\ln(\Ob h^2)$		 & 0.035 & 1.013 & 1.040 \\
&  $m_\nu$ (eV) $\propto\On h^2$	 & 0.55 & 1.010 & 1.029 \\[\tskip]
&  $\ns(\kfid)$ 		 & 0.041 & 1.013 & 1.031 \\
&  $\Ol$			 & 0.080 & 1.013 & 1.039 \\[\tskip]
&  $\tau$			 & 0.23 & 1.015 & 1.074 \\
&  $\ts$ 			 & 0.18 & 1.011 & 1.035 \\
\tableskip\tableline\tableskip\planck\ (TP)
&  $\ln(\Om h^2)$		 & 0.016 & 1.056 & 1.160 \\
&  $\ln(\Ob h^2)$		 & 0.0094 & 1.028 & 1.165 \\
&  $m_\nu$ (eV) $\propto\On h^2$	 & 0.24 & 1.032 & 1.075 \\[\tskip]
&  $\ns(\kfid)$ 		 & 0.0076 & 1.109 & 1.303 \\
&  $\Ol$			 & 0.022 & 1.051 & 1.151 \\[\tskip]
&  $\tau$			 & 0.0036 &  1.69 &  1.96 \\
&  $\ts$ 			 & 0.0073 &  4.04 &  6.58 \\
\tableskip\tableline
\end{tabular}
\end{center}
NOTES.---%
Marginalized errors for some cosmological parameters within the 12-dimensional
adiabatic CDM family of cosmological models and the exponential
coherence function (eq.~[\protect\ref{expcoh}]) for the foregrounds.  ``None'' column lists 
$1-\sigma$ error in case where there are no foregrounds.  
``Known'' column lists the relative degradation when our foreground MID model
is added under the assumption that the statistical properties of the
foregrounds are known exactly.
``Unknown'' column lists the relative degradation when the statistical
properties of the foregrounds must be simultaneously estimated within
a generous parameterization of possible models.
Results for \map\ with intensity only (T) and with intensity and polarization
(TP) are shown; likewise for \planck.
$\ns(\kfid)$ is the logarithmic derivative of the scalar primordial
power spectrum at $\kfid=0.025\impc$; in the presence of $\ns'\ne0$,
$\ns$ is a function of scale.
Cosmological model is $\Om=0.35$, $\Ob=0.05$, $\On=0.0175$ ($m_\nu=0.7\eV$), 
$\Ol=0.65$, $h=0.65$, $\ns=1$, $\ns'=0$, $\tau=0.05$, $Y_p=0.24$, and $\ts=0$. 
$\nt=0$ and cannot vary.  
\end{table}

For display purposes, we will focus on the performance on four parameters,
chosen to illustrate important aspects of the interplay between 
foregrounds and cosmological signals:  
\begin{enumerate}
\item The baryon density $\Obhh$ as measured only from the temperature information.
This parameter is sensitive to the structure of the acoustic peaks and
to the diffusion scale (Hu \& Sugiyama 1995).  

\item $\Obhh$ as measured from both intensity and polarization.
The polarization of the acoustic peaks substantially improve the accuracy with which
this parameter can be measured (Zaldarriaga {\etal} 1997) 
--- mainly through the X power spectrum, as we will see in 
\sec{txebSec}.

\item The reionization optical depth $\tau$ as measured from intensity and
polarization.  This is dominated by the $E$-channel signal at large
angular scales (Hogan {\etal} 1982), 
thereby testing how well the diffuse polarized galactic signals can
be removed.

\item The tensor-to-scalar ratio $\ts$ as measured from intensity and
polarization.  This is the only cosmic signal in the B-channel polarization
(Kamionkowski {\etal} 1997; Zaldarriaga \& Seljak 1997).

\end{enumerate}

\begin{figure}[tb] 
\centerline{\epsfxsize=\colwidth\epsffile{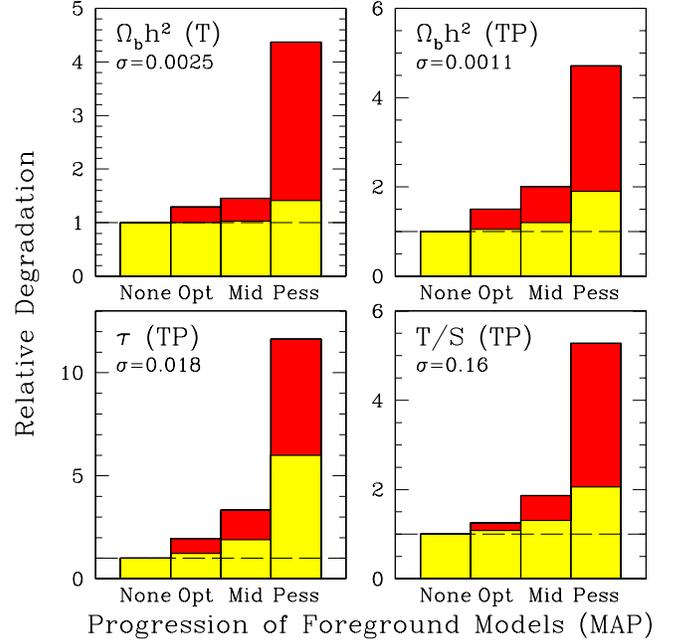}}
\caption{\label{fig:omp_map}\footnotesize%
Relative degradation in error bars from \map\ on four cosmological parameters
as the amplitude of foregrounds are increased.  
({\it top-left}) Behavior of $\Obhh$ with intensity information only (T). 
({\it proceeding clockwise}) $\Obhh$, $\ts$, and $\tau$
with intensity and polarization information (TP).
Bars show the error bar for each foreground case relative to the
no-foregrounds case; 
the $1-\sigma$ error of the latter is listed in each panel.
The histograms show results for a series of foreground models,
ranging from no foregrounds to our OPT, MID, and PESS models.
({\it lightly-shaded}) Results with foregrounds of known properties.
({\it heavily-shaded}) Results with foregrounds whose parameters
must be simultaneously estimated from the CMB data.
}
\end{figure}
%%% Relative degradation in error bars from \map\ on
%%% four cosmological parameters as we increase the severity
%%% of the foregrounds.  Results for the OPT, MID, and PESS
%%% foreground models are shown relative to the results with
%%% no foregrounds.

\begin{figure}[tb] 
\centerline{\epsfxsize=\colwidth\epsffile{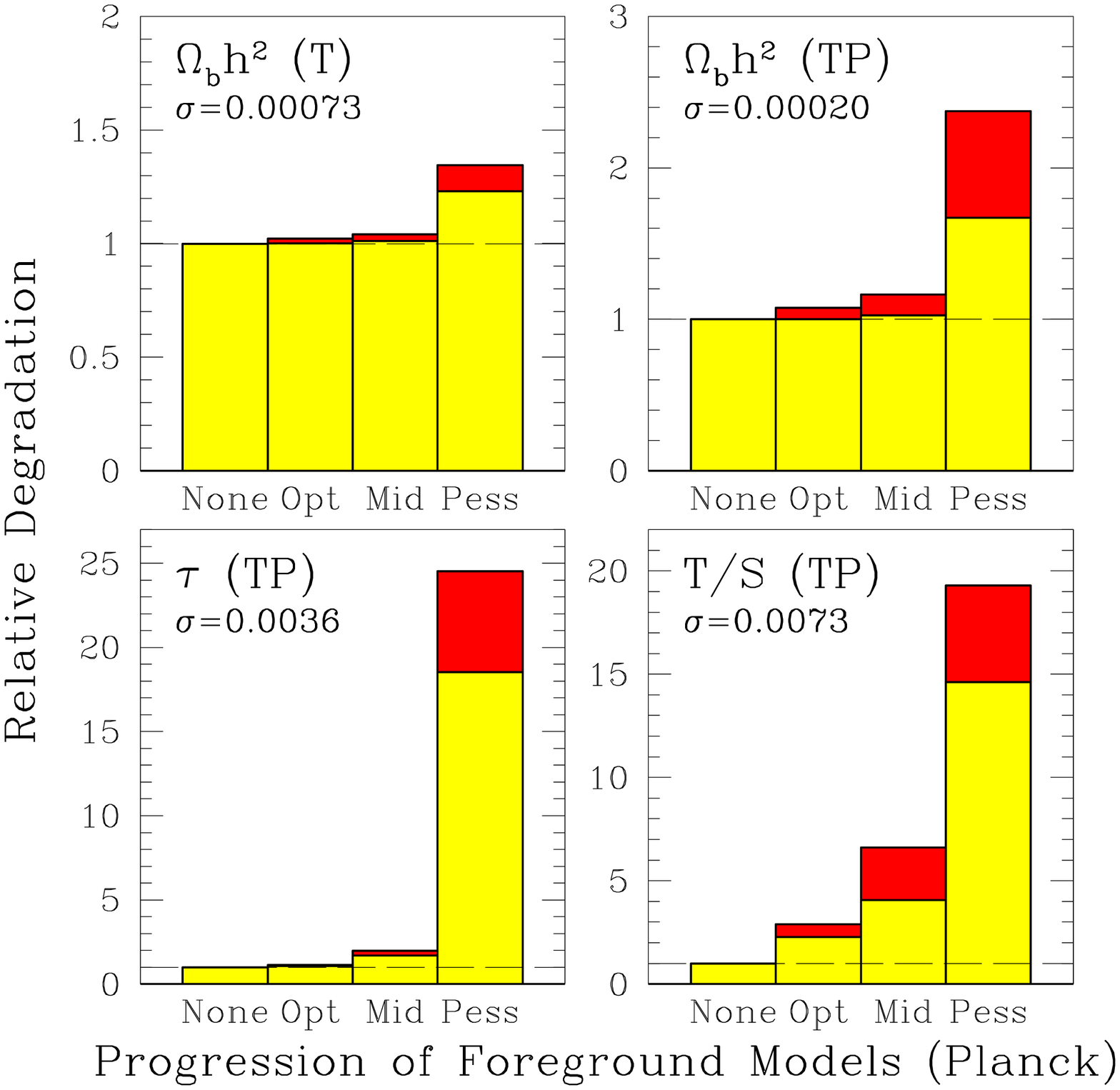}}
\caption{\label{fig:omp_planck}\footnotesize%
As Figure \protect\ref{fig:omp_map}, but for \planck.
}
\end{figure}

In Figures \ref{fig:omp_map} and \ref{fig:omp_planck}, we show the
degradation of the \map\ and \planck\ performance on these parameters
in the presence of our OPT, MID, and PESS foreground models.
In each case, the baseline is the performance in the absence of
any foregrounds at all.  The degradations in the presence of
foregrounds are shown for both the case of known properties and
the case of unknown properties.

The performance on $\Obhh$ with and without polarization is very
encouraging.  The degradations are less than a factor of 2 in
all cases but the PESS model in \map\ (where it reaches 4-5).  
\planck\ is able to survive even the PESS model with only a factor of 2
increase in the projected errors.  The reason for this strong 
performance is the detailed structure of the acoustic peaks.
Even if cleaning is imperfect, the foreground power spectra do
not have the oscillatory behavior of the cosmic derivatives and
can therefore be distinguished from variations in cosmological
parameters.  Note that most of the degradation can be attributed
to uncertainties in the foreground model; the performance in the
case of known foreground properties is nearly perfect.

The situation is somewhat less rosy for the large-angle polarization
signals.  $\tau$ and $\ts$ both have unique signatures in the
large-angle polarization, where signals from the acoustic peaks are quite
weak.  In the absence of foregrounds, even small cosmic signals can be
detected because their sample variance is equally low.  With foregrounds, the
signal-to-noise is considerably worse.  Performance is correspondingly
poorer, and the results do depend more sensitively on the severity of
the foregrounds.  However, one should note that for the OPT and MID
models, even the large-angle polarization signal can be cleaned to
reasonable accuracy, yielding excellent constraints on $\tau$ and
$\ts$.  For the PESS model, the degradation is generally more than a
factor of 10, although the errors for \planck\ would still be
interesting ($\sim\!5\%$ for $\tau$, $\sim\!10\%$ for $\ts$).

Note that the extra frequency coverage and sensitivity of \planck\ does
not imply that it will necessarily suffer less relative degradation
than \map\ from the presence of foregrounds; although \planck\ will
clean more effectively, its baseline projections were more ambitious.

In Table \ref{tab:foregvar}, we display the numerical results
for \boom, \map, and \planck\ with the MID foreground model.
The errors without foregrounds are shown, followed by the relative
degradations in the presence of foregrounds with and without 
knowledge of their properties.  For the satellites, we show 
results considering intensity information alone and then both
intensity and polarization information.  \boom\ is slightly more
robust against foregrounds than \map, reaching 3-fold degradations
in the PESS model.  Because our foreground model has more components
centered at lower frequencies, the higher frequency range of \boom\ may 
shift it away from the reach of the model's variations.

\subsection{Which details matter?}

\begin{figure}[tb] 
\centerline{\epsfxsize=\colwidth\epsffile{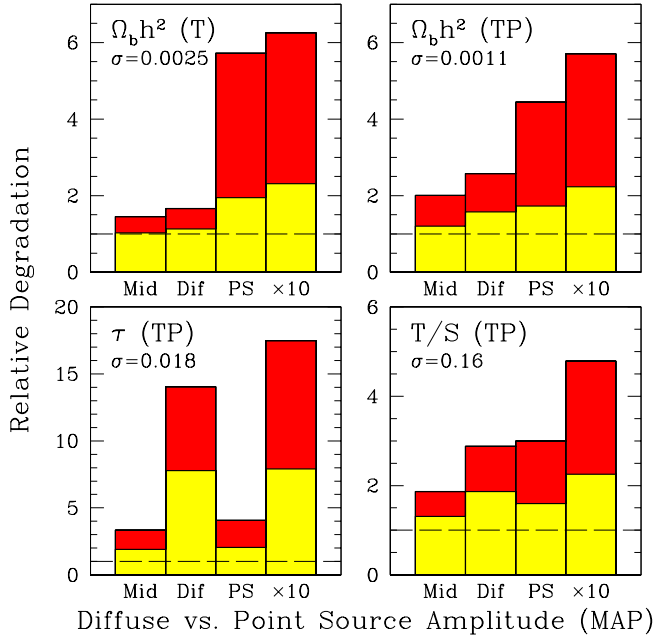}}
\caption{\label{fig:difps_map}\footnotesize%
As Figure \protect\ref{fig:omp_map}, but increasing the
diffuse foregrounds and point sources separately.
``Mid'' bars refer to the MID foreground model, 
simultaneously estimating the foreground parameters.
``Diff'' bars show the results when the diffuse components (i.e. all
those that are not point sources) are increased by 10 in amplitude.
``PS'' bars show the results when the point source and SZ components have
their amplitude (after PSF fitting) increased by a factor of 10.
``$\times10$'' bars show the combined result.
All values are shown as the fractional increase relative to the 
results with no foregrounds.  This figure shows the case for \map.
}
\end{figure}

\begin{figure}[tb] 
\centerline{\epsfxsize=\colwidth\epsffile{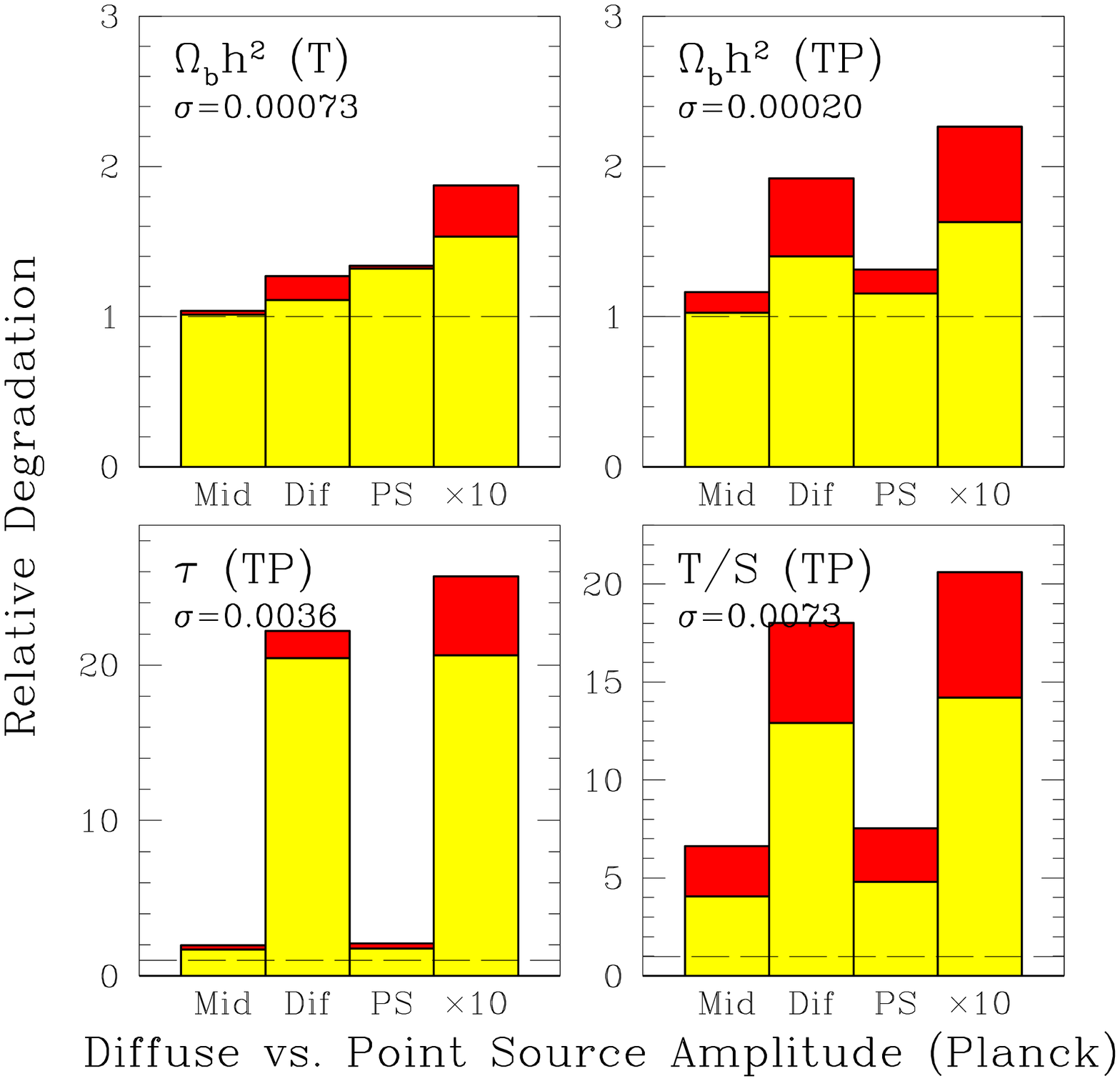}}
\caption{\label{fig:difps_pl}\footnotesize%
As Figure \protect\ref{fig:difps_map}, but for \planck.
}
\end{figure}

\subsubsection{Dependence on foreground amplitude}

As we increase the amplitudes of the foregrounds, which components most
affect which cosmological parameters?  We consider artificially boosting
the amplitudes of the MID foreground model by a factor of 10 
(i.e., a factor of 100 in power).  
In Figures \ref{fig:difps_map}
and \ref{fig:difps_pl}, we apply this factor of 10 increase separately
to the diffuse components and to the SZ and point source 
components.  Generally, the errors on $\tau$ and $\ts$ are primarily
affected by the diffuse components rather than the point sources, 
while the results for $\Obhh$ are the reverse.  However, there are
some mild exceptions.  For \planck\ with polarization, $\Obhh$ 
is sensitive to the amplitude of the diffuse components.  This is 
because we assumed the power spectrum of the polarization of dust
and synchrotron to be considerably bluer than that of the intensity.
Hence, when increased in amplitude, these foregrounds contaminate the
acoustic peak structure in the polarization and degrade the performance
somewhat.  Also, for \map\, the results on $\ts$ are sensitive to both
diffuse and point-source components.  \map\ does not make a strong 
detection of the tensor signal in the polarization and is therefore
reliant upon the large-angle signal in the intensity.  Boosting the point source
amplitude confuses the comparison between $\ell\approx50$ and
$\ell\approx500$ that would test for the presence of tensors.
The trends in Figures \ref{fig:difps_map} and \ref{fig:difps_pl}
are insensitive to whether foreground parameters are assumed to
be known or not.

\subsubsection{Dependence on frequency coherence}

As was discussed in \sec{CoherenceDepSec},
the cleaning of foregrounds is usually more effective when a map at one frequency
gives a good estimate of the foreground's presence at another frequency.
This is governed by the covariance matrix $R$ and the frequency coherence 
$\Delta\alpha$.  Because the parameters of this matrix are very poorly
known at the present time, it is important to check that our results 
are insensitive to our choices in this sector.

As described above, the parameter $\Delta\alpha$ sweeps between the
two extremes of perfect correlation between frequency channels and 
total independence.  As shown in T98, both of these cases have
desirable properties for removal of foregrounds.  In the former case,
there is a particular combination of the frequency maps that completely
removes the component in question.  In the latter case, one uses the
fact that any correlation between different maps must be cosmic signal.
Of course, neither perfect correlation nor total independence is 
correct, and the intermediate case admits a less complete cleaning 
of foregrounds.

\begin{figure}[tb] 
\centerline{\epsfxsize=\colwidth\epsffile{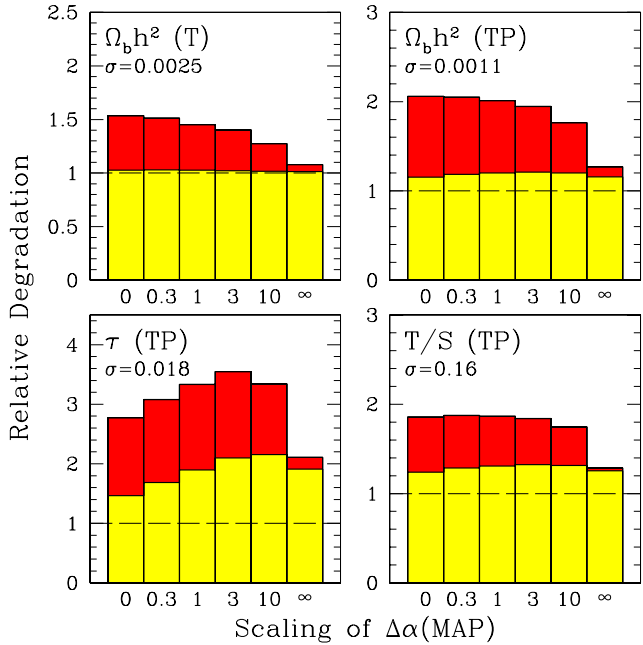}}
\caption{\label{fig:daseries_map}\footnotesize%
As Figure \protect\ref{fig:omp_map}, but altering the frequency
coherence $\Delta\alpha$ of the fiducial foreground model.  We scale
all $\Delta\alpha$ (see Table \protect\ref{ForegModelTab}) by a factor of
0, 0.3, 1, 3, 10, and $\infty$.  As shown in T98, perfect coherence
and perfect decoherence are the best-cases; intermediate values yield
worse performance.  All errors are shown relative to those in the
case with no foregrounds.
}
\end{figure}

\begin{figure}[tb] 
\centerline{\epsfxsize=\colwidth\epsffile{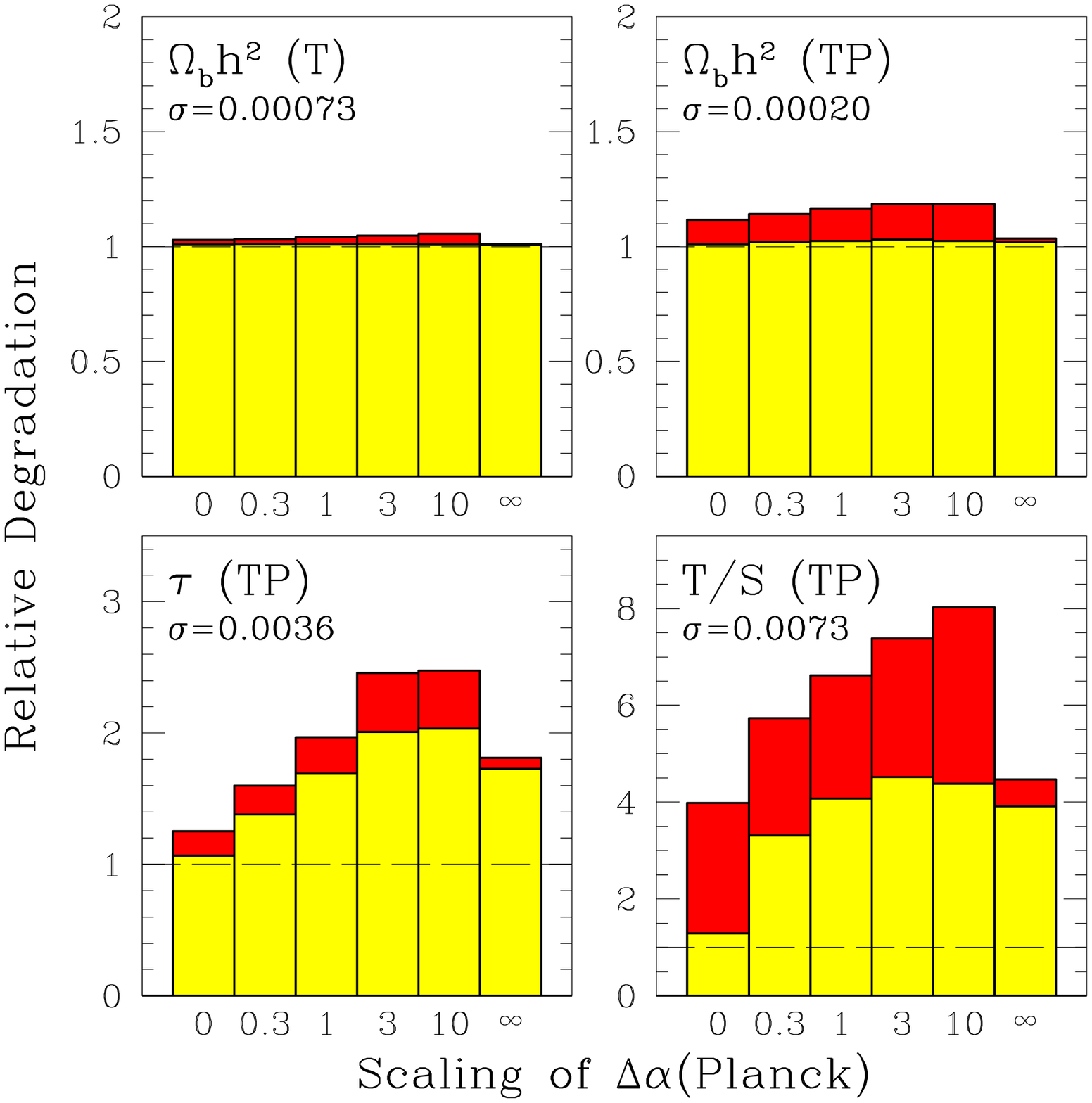}}
\caption{\label{fig:daseries_pl}\footnotesize%
As Figure \protect\ref{fig:daseries_map}, but for \planck.
}
\end{figure}

In Figures \ref{fig:daseries_map} and \ref{fig:daseries_pl}, we show a
sequence of models in which we scale all of the $\Delta\alpha$'s in the
MID model by a constant that ranges from zero (perfect correlation) to
$\infty$ (complete independence).  As expected, the errors on
cosmological parameters increase as one moves away from the extremes
and reaches a maximum in the middle.  This peak typically occurs when
the $\Delta\alpha$'s of the foregrounds are multiplied by $\sim\!3$,
but the actual location varies from case to case.  However, because the
peak is broad, the errors from our base model are actually rather close
to the maximum.  We therefore conclude that our treatment of the
covariance between the frequency channels has been sufficiently
conservative.

\subsubsection{Dependence on foreground model complexity}

We now consider turning off certain sets of variations to 
examine which variations are causing the most degradation.  The
results are shown in Table \ref{tab:foregpriors}.  We
separate the foreground parameters into three sets, namely
those involving the frequency dependence (the $\bfq_\kk^P$),
the frequency coherence (the $\bfr_\kk^P$),
and the spatial power spectrum (the $\bfs_\kk^P$).
By convention, the overall normalization of the foreground component is
carried by the frequency dependence, not the spatial power spectrum.
We include these sets one at a time and in pairs to investigate
which is most important.  Considered singly, uncertainties in the shape
of the power spectrum generally increase the error bars the most,
although uncertainties in frequency coherence are more important for
$\ts$ in \map.  Taken together, uncertainties in the frequency
dependence and coherence are important for $\ts$ for both \map\ 
and \planck.

\begin{table*}[tb]\footnotesize
\caption{\label{tab:foregpriors}}
\begin{center}
{\sc Adding and Removing Knowledge of Foregrounds\\}
%%% \begin{tabular}{rcccc\colskipp cccc}
%%% \tableskip\tableline\tableline\tableskip
%%% & \multicolumn{4}{c\colskipp}{\map} & \multicolumn{4}{c}{\planck} \\
%%% Foreground Knowledge & $\Obhh(T)$ & $\Obhh(TP)$ & $\tau(TP)$ & $\ts(TP)$
%%% & $\Obhh(T)$ & $\Obhh(TP)$ & $\tau(TP)$ & $\ts(TP)$ \\
%%% \tableskip\tableline\tableskip
\begin{tabular}{rc\colskippp cccc\colskippp cccc}
\tableskip\tableline\tableline\tableskip
& \boom & \multicolumn{4}{c\colskippp}{\map} & \multicolumn{4}{c}{\planck} \\
Foreground Knowledge & $\Obhh(T)$ 
& $\Obhh(T)$ & $\Obhh(TP)$ & $\tau(TP)$ & $\ts(TP)$
& $\Obhh(T)$ & $\Obhh(TP)$ & $\tau(TP)$ & $\ts(TP)$ \\
\tableskip\tableline\tableskip

Known Properties
& 1.000 % Output0512/boom.md.var.flnofg.j
& 1.000 & 1.000 & 1.000 & 1.000 % Output0512/map.md.var.flnofg.j
& 1.000 & 1.000 & 1.000 & 1.000 \\ % Output0512/planck.md.var.flnofg.j
\tableskip\tableline\tableskip
Unknown $C_\l$ 
& 1.133 % Output0512/boom.md.var.flcl9.j
& 1.162 & 1.202 & 1.320 & 1.085 % Output0512/map.md.var.flcl9.j
& 1.014 & 1.029 & 1.097 & 1.185 \\ % Output0512/planck.md.var.flcl9.j
Unknown $\Theta$ 
& 1.036 % Output0512/boom.md.var.flfr9.j
& 1.022 & 1.164 & 1.222 & 1.175 % Output0512/map.md.var.flfr9.j
& 1.012 & 1.073 & 1.010 & 1.121 \\ % Output0512/planck.md.var.flfr9.j
Unknown $\R$ 
& 1.000 % Output0512/boom.md.var.flcoh9.j
& 1.014 & 1.026 & 1.078 & 1.021 % Output0512/map.md.var.flcoh9.j
& 1.000 & 1.010 & 1.016 & 1.084 \\ % Output0512/planck.md.var.flcoh9.j
\tableskip\tableline\tableskip
Unknown $\Theta$ \& $\R$ 
& 1.106 % Output0512/boom.md.var.flcl0.j
& 1.089 & 1.292 & 1.374 & 1.288 % Output0512/map.md.var.flcl0.j
& 1.019 & 1.098 & 1.071 & 1.482 \\ % Output0512/planck.md.var.flcl0.j
Unknown $C_\l$ \& $\R$ 
& 1.151 % Output0512/boom.md.var.flfreq0.j
& 1.230 & 1.376 & 1.458 & 1.227 % Output0512/map.md.var.flfreq0.j
& 1.014 & 1.049 & 1.109 & 1.290 \\ % Output0512/planck.md.var.flfreq0.j
Unknown $C_\l$ \& $\Theta$ 
& 1.164 % Output0512/boom.md.var.flcoh0.j
& 1.194 & 1.390 & 1.545 & 1.281 % Output0512/map.md.var.flcoh0.j
& 1.018 & 1.093 & 1.107 & 1.296 \\ % Output0512/planck.md.var.flcoh0.j
\tableskip\tableline\tableskip
Unknown, except: \\
Known Free-free
& 1.098 % Output0512/boom.md.var.flff0.j
& 1.348 & 1.616 & 1.755 & 1.414 % Output0512/map.md.var.flff0.j
& 1.023 & 1.132 & 1.161 & 1.619 \\ % Output0512/planck.md.var.flff0.j
Known Synchrotron
& 1.140 % Output0512/boom.md.var.flsy0.j
& 1.387 & 1.622 & 1.622 & 1.378 % Output0512/map.md.var.flsy0.j
& 1.024 & 1.117 & 1.115 & 1.411 \\ % Output0512/planck.md.var.flsy0.j
Known Vibrating Dust
& 1.172 % Output0512/boom.md.var.flvd0.j
& 1.208 & 1.278 & 1.266 & 1.136 % Output0512/map.md.var.flvd0.j
& 1.024 & 1.122 & 1.133 & 1.492 \\ % Output0512/planck.md.var.flvd0.j
Known Rotating Dust
& 1.221 % Output0512/boom.md.var.flrd0.j
& 1.320 & 1.585 & 1.656 & 1.367 % Output0512/map.md.var.flrd0.j
& 1.024 & 1.127 & 1.140 & 1.515 \\ % Output0512/planck.md.var.flrd0.j
Known Thermal SZ
& 1.221 % Output0512/boom.md.var.flsz0.j
& 1.385 & 1.661 & 1.756 & 1.421 % Output0512/map.md.var.flsz0.j
& 1.022 & 1.137 & 1.163 & 1.626 \\ % Output0512/planck.md.var.flsz0.j
Known Radio PS
& 1.185 % Output0512/boom.md.var.flrps0.j
& 1.238 & 1.447 & 1.691 & 1.364 % Output0512/map.md.var.flrps0.j
& 1.007 & 1.063 & 1.143 & 1.589 \\ % Output0512/planck.md.var.flrps0.j
Known Infrared PS
& 1.209 % Output0512/boom.md.var.flips0.j
& 1.218 & 1.411 & 1.671 & 1.348 % Output0512/map.md.var.flips0.j
& 1.022 & 1.107 & 1.097 & 1.535 \\ % Output0512/planck.md.var.flips0.j
\tableskip\tableline\tableskip
All Unknown
& 1.221 % Output0512/boom.md.var.flsz10.j
& 1.415 & 1.674 & 1.756 & 1.424 % Output0512/map.md.var.flsz10.j
& 1.024 & 1.137 & 1.163 & 1.626 \\ % Output0512/planck.md.var.flsz10.j
%%% & 1.232 % Output0512/boom.md.var.flnone.j
%%% & 2.868 & 2.427 & 1.775 & 1.603 % Output0512/map.md.var.flnone.j
%%% & 1.027 & 1.137 & 1.163 & 1.626 \\ % Output0512/planck.md.var.flnone.j

\tableskip\tableline
\end{tabular}
\end{center}
NOTES.---%
Errors on cosmological parameters as we alter the knowledge of the 
foreground model.  All numbers are listed relative to the results
when the foreground properties are known; note that this differs
from the convention of Table \protect\ref{tab:foregvar}.  
All results use the MID foreground model.  A prior of 10 has been
used on the SZ component except where stated.
In the first half of the table, we progressively 
introduce each of the three different types of variations, applying
them to all of the foreground components.
The sets of foreground parameters for the 
frequency dependence ($\bfq_\kk^P$), the frequency coherence ($\bfr_\kk^P$), 
and the shape of the spatial power spectrum ($\bfs_\kk^P$)
are denoted by $\Theta$, $\R$, and $C_\l$, respectively.
Note that the normalization of the fluctuations is carried by the
$\Theta$ uncertainties.
Higher errors indicate that the experiment's performance is particularly
sensitive to those uncertainties of the foregrounds.
In the second half of the table, we consider all the foreground
properties to be unknown except for those of a given component.
Lower errors indicate that external information on that foreground
would be particularly valuable.
\end{table*}

\subsubsection{Dependence on foreground type}

We next consider the results when all foregrounds properties
are unknown save for those of a single component.  This can
identify the component about which external information would
have the most importance in improving cosmological inferences.
The results are again shown in Table \ref{tab:foregpriors}.
For \boom, we find that uncertainties in the foregrounds are 
not contributing much
additional degradation beyond the mere presence of the foregrounds;
the largest remaining concern is that free-free or synchrotron emission might
have a high-frequency contribution.
For \map, improving knowledge of the vibrating dust has the
most impact, on both the large-angle polarization signals
and the small-angle acoustic features.  Better control of 
point sources would help $\Obhh$ from temperature information,
but one should recall that our foreground model allows non-monotonic
excursions in the power spectrum of the point sources and so may
be overly pessimistic.  Further, \map\ suffers significant degradation
unless the thermal SZ is controlled by an external prior
of a factor of 10, so robust calculations of the power spectrum
of this effect as a function of cosmology will be required.
For \planck, no foreground makes an enormous
difference by itself, although radio point sources have the largest
(but still small) effect.

%%% \begin{figure}[tb] 
%%% \centerline{\epsfxsize=\colwidth\epsffile{fgseries_planck.ps}}
%%% \caption{\label{fig:fgseries_pl}\footnotesize%
%%% As Figure \protect\ref{fig:fgseries_map}, but for \planck.
%%% }
%%% \end{figure}

\subsubsection{Dependence on polarization type}
\label{txebSec}

\begin{figure}[tb]
\vskip-0.3cm
\centerline{\epsfxsize=5.6in\epsffile{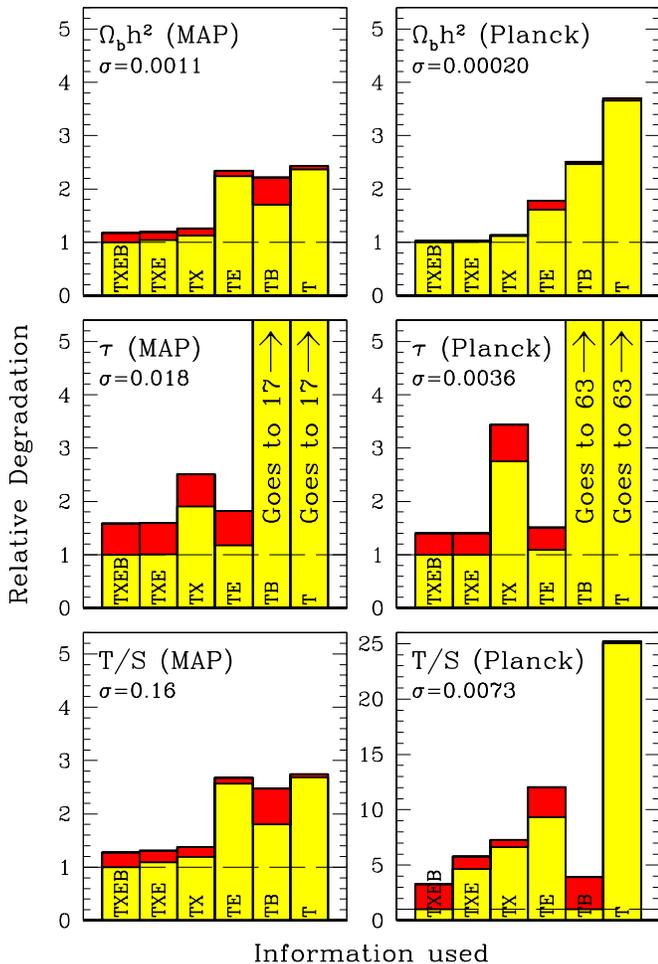}}
\vskip-0.9cm
\caption{\label{TXEBfig}\footnotesize%
Information about cosmological parameters from 
different polarization types.
In each panel, the bars show the relative
error bars using all information ($TXEB$), all
but $B$-polarization ($TXE$), $T$ and $X$ only ($TX$),
$T$ and $E$ only ($TE$) and unpolarized intensity alone ($T$).
Results are shown both without any foregrounds (yellow)
and for the MID foreground scenario with known properties (red). 
Note that these color conventions differ from those in 
Figure \protect\ref{fig:omp_map}--\protect\ref{fig:daseries_pl}
and \protect\ref{fig:fgseries_map} and also that this figure
uses a Gaussian rather than exponential coherence function.
The rightmost bars in the $\tau$ panels extend far off the scale.
}
\end{figure}

For which polarization type would prior knowledge of the foreground
properties most help cosmological parameter estimation?
The answer to this question
depends on where the cosmological parameter information 
is coming from in the first place, and this in turn depends on 
the parameter in question.
Limiting ourselves first to the case of known foreground properties, 
we can answer this question using \eq{SimpleFisherEq}.
If foregrounds and/or systematic errors made
the measured power spectrum $\Ct^P$ completely unusable, this would
correspond to adding in infinite amount of noise
to the element $\M_{PP}$ of the  
$4\times 4$ covariance matrix of \eq{PowerCovEq3}.
The Fisher matrix $\F_\l$ 
of \eq{SimpleFisherEq} therefore gets replaced by
\beq{InfoLossEq}
\F'_\l \equiv\lim_{t\to\infty} 
\left[\F^{-1}_\l + \J t\right]^{-1},
\eeq
where the $\J$ is a $4\times 4$ matrix with zeroes everywhere
except in element $(P,P)$; 
$\J_{P'P''}\equiv\delta_{PP'}\delta_{PP''}.$
This corresponds to simply crossing out 
row $P$ and column $P$ of $\F_\l$, inverting the remaining 
$3\times 3$ matrix, and 
padding with zeroes. For example,
if we drop the information from 
$X$-polarization, we obtain
\newcommand{\skippp}{@{\hspace{0.27in}}}
\beqa{SubFisherEq}
\F'_\l&=&2 
\left(\bs\begin{tabular}{cc}
$\>\left[\bs\begin{tabular}{ccc}
$T_\l^2$	&$X_\l^2$ 	&$0$\\[2pt]
$X_\l^2$	&$E_\l^2$ 	&$0$\\[2pt]
$0$		&$0$		&$B_\l^2$
\end{tabular}\bs\right]^{-1}$&
%%%
$\bs\begin{tabular}{ccc}
$0$\\[2pt]
$0$\\[2pt]
$0$
\end{tabular}\bs$\\
%%%
$\hskip-0.5truecm\begin{tabular}{c\skippp c\skippp c}
$0$		&$0$		&$0$
\end{tabular}\bs$&
%%%
$0$
\end{tabular}\bs\right)\\
&=&
{2\over D_\l^2}
\left(\bs\begin{tabular}{cccc}
$E_\l^2$	&$-X_\l^2$ 	&0		&$0$ \\[2pt]
$-X_\l^2$	&$T_\l^2$ 	&0		&$0$\\[2pt]
$0$		&$0$		&$\displaystyle{D_\l^2 \over B_\l^{2}}$	&$0$\\[2pt]
$0$		&$0$		&0 		&$0$
\end{tabular}\bs\right),
\eeqa
where the notation is the same as in \eq{PowerCovEq}.
Likewise, omitting two of the four power spectra corresponds to 
crossing out two rows and columns before inverting, \etc, so we can 
compute the attainable accuracy on cosmological parameters using 
any subset of the four power spectra $T$, $E$, $B$ and $X$.

\Fig{TXEBfig} shows the results for our
sample of three cosmological parameters using
five such subsets.
Comparing $T$ alone with the other cases illustrates the well-known fact
that polarization helps substantially, especially 
with $\tau$ and $T/S$ 
(see \eg, Hogan {\etal} 1982; Bond \& Efstathiou 1987;
Zaldarriaga {\etal} 1997). We find that 
all parameters that are sensitive to the acoustic peaks 
are like $\Obhh$ in that 
the bulk of the polarization gain is coming from $X$-polarization,
manifested by
$T+X$ giving smaller error bars than $T+E$ and
by the combination $T+X+E$ being only 
marginally better than $T+X$. On the other hand for 
$\tau$, $E$ is seen to be more important 
than $X$ for picking up the large-scale bump caused
by early reionization. 
The $B$ channel receives contributions from gravity waves
alone.  However it dominates the measurement of $T/S$
only for Planck because MAP does not have enough signal-to-noise 
to yield interesting constraints on the $B$-polarization. 

These results imply that a better understanding of foreground polarization
in $X$ would most improve errors for $\Obhh$, $E$ for 
$\tau$ and ultimately $B$ for $T/S$.  
We also test these conclusions in the case of
simultaneous estimation of foreground and cosmological parameters
by placing priors separately on each 
of the polarization types; the results confirm these tendencies.

\section{Conclusions} 
\label{ConclusionsSec}

We have presented a comprehensive treatment of microwave foregrounds
and the manner in which they 
degrade our ability to measure cosmological parameters with the CMB.
Having developed three quantitative models, we compute their effect
upon the \boom, \map\ and \planck\ missions, including
the level of foreground residuals in the
cleaned maps for various scenarios and 
the extent to which this residual 
contamination would degrade the measurement of cosmological parameters.
We consider both the case when the foreground power spectra are known and 
the case in which they must be computed from the CMB data itself.
Our foreground model can be found 
at {\it www.sns.ias.edu/$\sim$max/foregrounds.html} together with
software implementing our cleaning algorithm. 

Our results are generally encouraging, in that the experiments
perform well in the face of rather severe foreground models.
This success derives from the fact that the cosmic signals can be
distinguished from foregrounds by their
frequency dependence, their frequency coherence, {\it and} their
spatial power spectra.  With these handles on the cosmic signal,
we find that 
the error bars on most cosmological parameters are degraded by
less than a factor of two for our best-guess foreground model and
by less than a factor of five in our most pessimistic scenario.
Effects producing large-angle polarization signals, 
namely reionization and tensor perturbations, suffer more because
of their intrinsically small cosmic amplitude, but even these can 
be accurately extracted in most cases.

\subsection{The most damaging foregrounds}

One useful result of this work is that it highlights which foregrounds
are potentially most damaging for precision cosmology and therefore most in 
need of further study.
We find that allowing for uncertainties in the properties of the 
foregrounds does cause a substantial degradation in performance
relative to the case of known foreground properties.  In the study
of the acoustic peaks, these uncertainties were dominant; however,
in the study of large-angle polarization, the mere presence of sample
variance from the foregrounds was more important for \planck.

Taken alone, the uncertainties in the shape of the power spectra
were more important that the uncertainties in either the frequency
dependence or the frequency coherence.  In the case of tensors,
the combination of frequency dependence and frequency coherence
were particularly important.  Of course, combinations of excursions are
usually worse than the sum of individual excursions.

In the case of \map, adding external information about vibrating dust
made the most improvement in the results.  Point source information 
also helped in the temperature data on the acoustic peaks. 
Knowing the level of thermal SZ fluctuations from filaments 
to within a factor of 10
{\it a priori} noticeable improved the results, so order-of-magnitude
limits on this effect from simulations or observations will be valuable.
In the case of \boom, restricting the ability of free-free and
synchrotron emission to pollute the 90 GHz channel was most important.
Clearly, \map\ and \boom\ will complement each other in their constraints
on possible pathologies of the intensity of foregrounds, as the 
experiments cover relatively low and high frequencies, respectively.
In the case
of \planck, the foreground cleaning in the MID model was sufficiently
good that most foregrounds had only a minor impact.  
The largest degradations were due to radio point sources and synchrotron
radiation.

Since we found that temperature-polarization cross-correlation
carries much more information than
$E$-polarization on most cosmological parameters (the exceptions
being $\tau$ and $T/S$), it is clearly important to 
accurately model and measure the cross-correlation between polarized
and $E$-unpolarized foregrounds. 

\subsection{Robustness}

\begin{figure}[tb] 
\centerline{\epsfxsize=\colwidth\epsffile{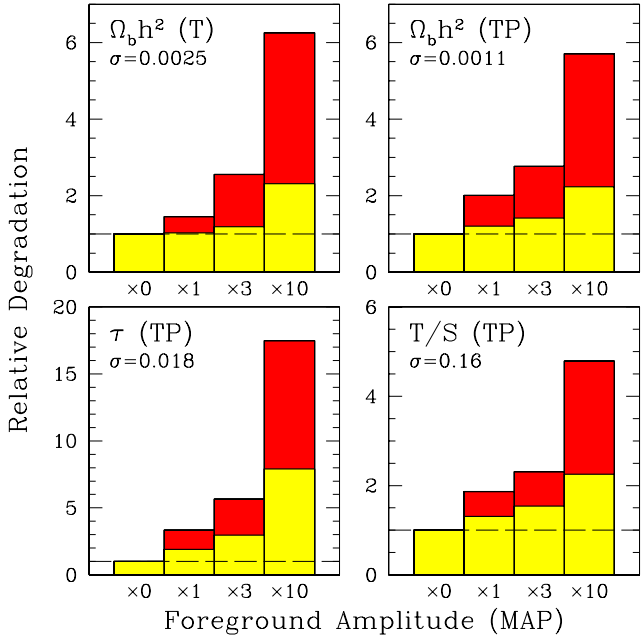}}
\caption{\label{fig:fgseries_map}\footnotesize%
As Figure \protect\ref{fig:omp_map}, but showing the relative degradation 
in error bars from \map\ on four cosmological parameters
as the amplitude of foregrounds are increased.  
%%% ({\it top-left}) Behavior of $\Obhh$ with intensity information only (T). 
%%% ({\it proceeding clockwise}) $\Obhh$, $\ts$, and $\tau$
%%% with intensity and polarization information (TP).
%%% Bars show the error bar for each foreground case relative to the
%%% no-foregrounds case; 
%%% the $1-\sigma$ error of the latter is listed in each panel.
The histograms show results for a series of foreground models
based on our MID model, with amplitudes of all components
multiplied by 0 (i.e. no foregrounds), 1, 3, and 10.
The results are scaled to the no-foreground case; the $1-\sigma$
errors in this case are listed in each panel.
({\it lightly-shaded}) Results with foregrounds of known properties.
({\it heavily-shaded}) Results with foregrounds with unknown parameters
that must be simultaneously estimated from the CMB data.
}
\end{figure}

How robust are these results? Have we been to conservative or too optimistic?
In general, we have tried to err on the side of caution, 
occasionally to the point of playing the role of the Devil's advocate.
We view the MID model as slightly cautious and the PESS model as quite
extreme, on the verge of being ruled out by current constraints.
We have also been conservative in not taking advantage of foreground
dependence on Galactic latitude except in the crudest way, 
with a Galactic cut.
In the same spirit, we have not included information from non-CMB templates
such as the DIRBE or Haslam maps. 
The formalism we have presented is general enough that both of these types of 
additional information can be included, the latter simply by including the
foreground templates as additional ``channels'' in the analysis.

However, there are also ways in which real-world foregrounds may be worse
than we have assumed. 
We have made the simplifying assumption that each physical component is
separable in $\l$ and $\nu$, \ie, that only the amplitude
(not the shape) of its power spectrum depends on frequency.
This needs to be tested empirically, and may reveal that certain 
foregrounds decompose into several separable subcomponents.
Perhaps most importantly, we have
modeled foregrounds as Gaussian, which is certainly incorrect at some level. 
Our removal method still
succeeds in minimizing the rms residual even if the foregrounds are non-Gaussian, 
and all our plots of residual power spectra remain correct (since they involve
second moments only), but the error bars on the measured power spectra 
(which involve fourth moments)
that propagate into the calculations of cosmological parameter accuracy will change 
in this case, probably for the worse.
As mentioned, the variance of a measurement of say $C_\l$ will 
be $2/N$ for the Gaussian case, where $N=(2\l+1)\fsky$ is the
effective number of independent modes that probe this quantity.
For a measurement of the band power between $\l_1$ and $\l_2$,
we have $N=[(\l_2+1)^2-\l_1^2]\fsky$ modes.
Foreground non-Gaussianity typically correlates these modes, reducing
the effective number of independent modes and thereby increasing the 
variance on the measured multipole or band power.
We explore the effect of such errors in a very crude way in 
\fig{fig:fgseries_map},
by simply increasing the foreground amplitudes by various factors $Q$.\footnote{
It is easy to show that asymptotically, 
as $\Q\to\infty$ and foregrounds dominate over sample variance and
detector noise, the parameter error bars will scale as as $Q^2$.
\Fig{fig:fgseries_map} shows that we are far from that limit,
with a foreground increase giving a much weaker response.
}
An amplitude increase $Q$ causes an increase in the power
spectrum of $Q^2$, corresponding to a variance increase of $Q^4$ and 
a reduction of $N$ by $Q^4$. For instance, increasing 
all foreground amplitudes by a factor $Q=10$ corresponds to reducing
the number of independent modes by 10,000. This is likely to
be more severe than the actual level of foreground non-Gaussianity, 
since it would imply, \eg, that all 10,000 multipole modes $a_{\l m}$ up to 
$\l=100$ would be almost perfectly correlated.
It should be noted that in the extreme case where non-Gaussianity gives 
perfect correlations between neighboring multipoles, foregrounds become 
trivial to remove by projecting out an overall offset. In other words, the
worst possible case lies somewhere in between the extremes of no 
mode correlation and complete mode correlation. A detailed study of 
the non-Gaussian properties of foregrounds would certainly be worthwhile,
using, \eg, the WOMBAT compilation of foreground data
(Gawiser {\etal} 1999).

\subsection{Comparison with other work}

A number of excellent treatments of foregrounds and their impact on CMB 
measurements have been published. Thorough and recent ones
that are particularly relevant to this paper are
those done
for the \planck\ proposal (TE96; Bouchet {\etal} 1996; 
Bersanelli {\etal} 1996; Bouchet {\etal} 1998; AAO 1998; BG99) 
and K99. 
Although these studies did not compute the accuracies with which 
cosmological parameter could be measured, they all calculated
residual power spectra in the cleaned maps and their associated error bars,
which can be compared with ours.
We typically find slightly higher levels of residual foreground.
Apart from minor differences in the assumed foreground power spectra
etc, this is because we do not assume that the foreground
covariance between different frequencies is a matrix of rank 1 or 2.
The former assumption (made in, \eg, TE96 and Bersanelli {\etal} 1996) corresponds
to assuming $\Delta\alpha=0$, \ie, perfect frequency coherence.
The latter, used in for instance BG99 and K99, is equivalent to assuming that
each foreground can be decomposed into two perfectly coherent components.

The elegant treatment of K99 gives foregrounds even more leeway than we have,
with thousands of free parameters, allowing their power spectra 
to be completely arbitrary functions of $\l$ and fitting for them directly 
from the data.
Unfortunately, this only works for the above-mentioned rank 2 
assumption about coherence for \planck, since the number of components
cannot exceed half of the number of channels.
We have restricted the foreground power spectra, frequency spectra and
frequency correlations to be fairly smooth functions, since all such 
functions measured to date have been fairly dull and featureless.

\subsection{Outlook}

% In conclusion,
A large number of papers have now 
painted a rosy picture of the future of
cosmology, with CMB experiments measuring cosmological parameters to 
unprecedented accuracy over the next decade.
In this paper, we have tried quite hard to spoil this picture,
using foreground models with hundreds of harmful parameters and 
pushing them to limits of physical plausibility and current constraints.
Although we have found that great care needs to be taken in 
the foreground removal phase of the data analysis to avoid 
potentially perilous pitfalls, 
we have failed to tarnish the overall picture with more than a few
minor blemishes, degrading the accuracy on certain measurements 
by small factors. Although much work certainly remains to be done on
the foreground problem, this is cause for cautious optimism.

\bigskip
{\bf Acknowledgements:}
We thank John Bahcall, Lloyd Knox, David Spergel, 
Alexei Tchepurnov, Philip Tegmark, Matias Zaldarriaga 
and an anonymous referee for helpful comments, Luigi Toffolatti for kindly
providing his source count model data, and 
Uro\v s Seljak \& Matias Zaldarriaga for 
their CMBFAST 
code\footnote{http://www.sns.ias.edu/$\sim$matiasz/CMBFAST/cmbfast.html}, 
which was used for generating our CMB power spectra. 
We thank the Alfred P. Sloan Foundation for funding the foreground
workshop where this work was first presented, and the workshop
participants for useful comments and suggestions.
This work was supported by
NASA though grant NAG5-6034 and 
Hubble Fellowship HF-01084.01-96A from STScI, operated by AURA, Inc. 
under NASA contract NAS5-26555. 
D.J.E.\ is supported by a Frank and Peggy Taplin Membership at the IAS,
W.H.\ by the W.M.\ Keck Foundation,
and D.J.E.\ and W.H.\ by NSF-9513835.

%%%%%%%%%%%%%%%%%%%%%% REFERENCES: %%%%%%%%%%%%%%%%%%%%%%%%%

\bigskip
\bigskip
\bigskip
This paper is available with figures and fortran code from 
{\it h t t p://www.sns.ias.edu/$\tilde{~}$max/foregrounds.html}

%\onecolumn

% \begin{figure}[tbh] 
% \centerline{\epsfxsize=18cm\epsffile{channels_T.ps}}
% \caption{\label{CTfig}\footnotesize%
% Our unpolarized foreground power spectrum models are shown at 30, 100 and
% 217 GHz for the optimistic (OPT), middle-of-the-road (MID) and
% pessimistic (PESS) scenarios described in the text. 
% }
% \end{figure}
% 
% 
% 
% \begin{figure}[tbh] 
% \centerline{\epsfxsize=18cm\epsffile{channels_X.ps}}
% \caption{\label{CXfig}\footnotesize%
% Same as for \fig{CTfig}, but for the temperature-polarization
% cross power spectrum.
% }
% \end{figure}
% 
% 
% \begin{figure}[tbh] 
% \centerline{\epsfxsize=18cm\epsffile{channels_E.ps}}
% \caption{\label{CEfig}\footnotesize%
% Same as for \fig{CTfig}, but for the E-polarization
% power spectrum.
% The B-spectra are similar.
% }
% \end{figure}

% \begin{figure*}[tbh] 
% \centerline{\epsfxsize=18cm\epsffile{cleaned_planck.ps}}
% \caption{\label{CleanedPlanckFig}\footnotesize%
% Power spectra of the noise and residual foregrounds in the
% three cleaned maps is shown for Planck.
% The B-spectra are similar to those
% shown for E-polarization.}
% \end{figure*}
% 
% \begin{figure*}[tbh] 
% \centerline{\epsfxsize=18cm\epsffile{cleaned_map.ps}}
% \caption{\label{CleanedMAPfig}\footnotesize%
% Same as previous figure, but for MAP.
% }
% \end{figure*}
% 

\ed

% TO DO IF TIME AND ENERGY:

% Figure 666 shows how 
% this criterion
% partitions CMB experiments into two classes: those for which 
% internal cleaning suffices and those which need external point 
% source data to reach their full potential.

PERHAPS EXPLAIN OUR MODEL FOR THE CROSS-CORRELATION BETWEEN T \& E
AND BETWEEN DIFFERENT FOREGROUNDS BETTER.

%%%  Can we try the perfectly correlated case?}

* Add patchy reionization?